\documentclass[twocolumn]{aastex631}

\usepackage{enumitem}
\usepackage{amsmath}
\usepackage{latexsym}
\usepackage{graphicx}
\usepackage{epsf}
\usepackage{CJK}
\usepackage{wasysym}
\usepackage{natbib}

\newcommand{\ha}{H$\alpha$}
\newcommand{\hb}{H$\beta$}

\newcommand{\myr}{\mathrm{Myr}}

\newcommand{\triplelens}{A2744-QSO1}
\newcommand{\jwst}{\textit{JWST}}

\received{MMDDYY}
\revised{MMDDYY}
\accepted{MMDDYY}

\submitjournal{ApJ}

\shorttitle{Triply Imaged Little Red Dot}
\shortauthors{Ma et al}

\graphicspath{{./}{figures/}}

\begin{document}
\begin{CJK*}{UTF8}{gbsn}

\title{UNCOVER: 404 Error -- Models Not Found for the Triply Imaged Little Red Dot A2744-QSO1}

\correspondingauthor{Yilun Ma}
\author[0000-0002-0463-9528]{Yilun Ma (马逸伦)}
\affiliation{Department of Astrophysical Sciences, Princeton University, Princeton, NJ 08544, USA}
\email{yilun@princeton.edu}

\author[0000-0002-5612-3427]{Jenny E. Greene}
\affiliation{Department of Astrophysical Sciences, Princeton University, Princeton, NJ 08544, USA}

\author[0000-0003-4075-7393]{David J. Setton}
\affiliation{Department of Astrophysical Sciences, Princeton University, Princeton, NJ 08544, USA}

\author[0000-0002-3216-1322]{Marta Volonteri}
\affiliation{Institut d'Astrophysique de Paris, Sorbonne Universit\'{e}, CNRS, UMR 7095, 98 bis bd Arago, F-75014 Paris, France}

\author[0000-0001-6755-1315]{Joel Leja}
\affiliation{Department of Astronomy \& Astrophysics, The Pennsylvania State University, University Park, PA 16802, USA}
\affiliation{Institute for Computational \& Data Sciences, The Pennsylvania State University, University Park, PA 16802, USA}
\affiliation{Institute for Gravitation and the Cosmos, The Pennsylvania State University, University Park, PA 16802, USA}

\author[0000-0001-9269-5046]{Bingjie Wang (王冰洁)}
\affiliation{Department of Astronomy \& Astrophysics, The Pennsylvania State University, University Park, PA 16802, USA}
\affiliation{Institute for Computational \& Data Sciences, The Pennsylvania State University, University Park, PA 16802, USA}
\affiliation{Institute for Gravitation and the Cosmos, The Pennsylvania State University, University Park, PA 16802, USA}


\author[0000-0001-5063-8254]{Rachel Bezanson}
\affiliation{Department of Physics and Astronomy and PITT PACC, University of Pittsburgh, Pittsburgh, PA 15260, USA}

\author[0000-0003-2680-005X]{Gabriel Brammer}
\affiliation{Cosmic Dawn Center (DAWN), Niels Bohr Institute, University of Copenhagen, Jagtvej 128, K{\o}benhavn N, DK-2200, Denmark}

\author[0000-0002-7031-2865]{Sam E. Cutler}
\affiliation{Department of Astronomy, University of Massachusetts, Amherst, MA 01003, USA}

\author[0000-0001-8460-1564]{Pratika Dayal}
\affiliation{Kapteyn Astronomical Institute, University of Groningen, P.O. Box 800, 9700 AV Groningen, The Netherlands}

\author[0000-0002-8282-9888]{Pieter van Dokkum}
\affiliation{Department of Astronomy, Yale University, New Haven, CT 06511, USA}

\author[0000-0001-6278-032X]{Lukas J. Furtak}
\affiliation{Physics Department, Ben-Gurion University of the Negev, P.O. Box 653, Be'er-Sheva 84105, Israel}

\author[0000-0002-3254-9044]{Karl Glazebrook}
\affiliation{Centre for Astrophysics and Supercomputing, Swinburne University of Technology, PO Box 218, Hawthorn, VIC 3122, Australia}

\author[0000-0003-4700-663X]{Andy D. Goulding}
\affiliation{Department of Astrophysical Sciences, Princeton University, Princeton, NJ 08544, USA}

\author[0000-0002-2380-9801]{Anna de Graaff}
\affiliation{Max-Planck-Institut f\"ur Astronomie, K\"onigstuhl 17, D-69117, Heidelberg, Germany}

\author[0000-0002-5588-9156]{Vasily Kokorev}
\affiliation{Department of Astronomy, The University of Texas at Austin, Austin, TX 78712, USA}

\author[0000-0002-2057-5376]{Ivo Labbe}
\affiliation{Centre for Astrophysics and Supercomputing, Swinburne University of Technology, Melbourne, VIC 3122, Australia}

\author[0000-0002-9651-5716]{Richard Pan}
\affiliation{Department of Physics and Astronomy, Tufts University, 574 Boston Ave, Medford, MA 02155, USA}

\author[0000-0002-0108-4176]{Sedona H. Price}
\affiliation{Department of Physics and Astronomy and PITT PACC, University of Pittsburgh, Pittsburgh, PA 15260, USA}

\author[0000-0003-1614-196X]{John R. Weaver}
\affiliation{Department of Astronomy, University of Massachusetts, Amherst, MA 01003, USA}

\author[0000-0003-2919-7495]{Christina C. Williams}
\affiliation{NSF National Optical-Infrared Astronomy Research Laboratory, 950 North Cherry Avenue, Tucson, AZ 85719, USA}

\author[0000-0001-7160-3632]{Katherine E. Whitaker}
\affiliation{Department of Astronomy, University of Massachusetts, Amherst, MA 01003, USA}
\affiliation{Cosmic Dawn Center (DAWN), Niels Bohr Institute, University of Copenhagen, Jagtvej 128, K{\o}benhavn N, DK-2200, Denmark}

\author[0000-0002-0350-4488]{Adi Zitrin}
\affiliation{Department of Physics, Ben-Gurion University of the Negev, P.O. Box 653, Be'er-Sheva 84105, Israel}

\renewcommand{\thefootnote}{\fnsymbol{footnote}}
\footnotetext[1]{\textcolor{black}{The error code 404 in the title represents that no satisfying model is found for \triplelens. In computer network communication, the code ``404" refers to the error occurring when the server cannot find the requested information. }}
\renewcommand{\thefootnote}{\arabic{footnote}}

\begin{abstract}
\textit{JWST} has revealed an abundance of compact, red objects at $z\approx5-8$ dubbed ``little red dots" (LRDs), whose SEDs display a faint blue UV continuum followed by a steep rise in the optical. Despite extensive study of their characteristic V-shaped SEDs, the nature of LRDs remains unknown. We present a new analysis of the NIRSpec/PRISM spectrum of A2744-QSO1, a triply imaged LRD at $z=7.04$ from the UNCOVER survey. The spectrum shows a strong Balmer break and broad Balmer emission lines, both of which are difficult to explain with models invoking exclusively AGN or stellar contributions. Our fiducial model decomposes the spectrum into a post-starburst galaxy dominating the UV-optical continuum and a reddened AGN being sub-dominant at all wavelength and contributing at $\sim20\%$ level. However, \textcolor{black}{this model} infers a stellar mass of $M_\star\approx 4\times10^9\,\mathrm{M_\odot}$ within a radius of $r_\mathrm{e}<30\,$pc, driving its central density to the highest among observations to date. This high central density could be explained if A2744-QSO1 is the early-forming core of a modern-day massive elliptical galaxy that later puffed up via the inside-out growth channel. The models also necessitate an unusually steep dust \textcolor{black}{extinction} law to preserve the strong break strength, though this steepness may be explained by a deficit of large dust grains. It is also probable that these challenges reflect our ignorance of A2744-QSO1's true nature. Future variability and reverberation mapping studies could help disentangle the galaxy and AGN contribution to the continuum, and deeper redder observations could also unveil the dust properties in LRDs.
\end{abstract}


\section{Introduction}\label{sec:intro}

Before the launch of \textit{James Webb Space Telescope} (\textit{JWST}; \citealt{Gardner2023}), galaxies and active galactic nuclei (AGN) were often discovered by rest-frame ultraviolet (UV) selection methods and/or drop-out techniques at $z\gtrsim4$ \citep[e.g.,][]{Fan2001, Bouwens2015}. However, these UV-selection methods tend to be biased towards preferentially selecting the most massive and luminous black holes (BHs) in the early universe \citep{Fan2023}. Thanks to its high sensitivity and spatial resolution, \textit{JWST} has revolutionized our understanding of black holes in the high-redshift universe by enabling detections of fainter sources, instead of solely the brightest AGN and quasars \citep{Goulding2023, Harikane2023, Maiolino2023}. 

One of the most intriguing results from \textit{JWST} is the discovery of a population of compact red sources at high redshift \citep{Furtak2023, Labbe2023uncoverLRD, Akins2024, Barro2024, Kocevski2024, Kokorev2024PhotLRD}. Because of their size and color, these sources are dubbed little red dots (LRDs; \citealt{Matthee2024}). The LRDs are identified from photometric samples for their V-shaped photometric spectral energy distribution (SED), which consists of a faint blue ultraviolet (UV) continuum followed by a steep red rise in continuum flux into the rest-frame optical wavelengths. 

Motivated by the compact sizes ($r_\mathrm{e}\approx100\,\mathrm{pc}$; \citealt{Labbe2023uncoverLRD, Akins2024}) of these objects, photometric SED modeling analyses have often invoked an AGN component on top of a galaxy SED \citep[e.g.,][]{Furtak2023, Barro2024, Perez-Gonzalez2024}. Indeed, in follow-up spectroscopic programs, nearly 80\% of the photometrically selected LRDs show broad Balmer and Paschen emission lines with $\mathrm{FWHM}>2000\,\mathrm{km\,s^{-1}}$ typically seen in type-1 AGN \citep[e.g.,][]{Kocevski2023, Greene2024, Furtak2024Nature, Matthee2024, Wang2024BRD}. The common presence of broad lines corroborates the potential existence of accreting massive BHs in LRDs. If one takes the BH mass estimates inferred from emission line scaling relations at face value, LRDs would host overmassive BHs when compared to their host galaxies ($M_\mathrm{BH}/M_\star\approx1\%$; \citealt{Kokorev2023LRDz8, Furtak2024Nature, Wang2024RUBIESz78MassGal}). These BH mass measurements also imply that the BHs in LRDs accrete at high Eddington ratios \citep{Furtak2024Nature, Greene2024}. Many other BH seeding and growth mechanisms such as primordial origins \citep{Dayal2024} and prolonged accretion in high-density environments \citep{Inayoshi&Ichikawa2024} have also been proposed for these intriguing objects. 

Another important point is that LRDs have a surprisingly high number density of roughly $10^{-5}\,\mathrm{Mpc^{-3}\,mag^{-1}}$ for $z>4$ \citep{Greene2024, Matthee2024, Kocevski2024}. Both \cite{Akins2024} and \cite{Greene2024} find that LRDs are $\sim100$ times more common than one would expect if simply extrapolating the UV luminosity functions of high-redshift quasars to match the UV luminosity of LRDs; they also account for 1\% of the UV-selected galaxies at $z\approx4-6$. Moreover, \cite{Kocevski2024} find that the number density of LRDs photometrically selected by rest-frame spectral slopes drop significantly at $z\lesssim4$. This could suggest that LRDs may mark a previously unknown and poorly understood episode of BH growth prevalent in the early universe.

Nevertheless, despite extensive theoretical and observational studies on these intriguing sources, the nature of LRDs is still uncertain. Many models can fit \textcolor{black}{the shape of the rest-frame UV-optical SED well.} The UV can be described by both unobscured scattered light from the central BH and/or a dust-free stellar population, whereas the red optical can be described by either a reddened AGN or an attenuated stellar component \citep{Killi2023, Greene2024, Perez-Gonzalez2024, Wang2024BRD, Williams2024}. \textcolor{black}{\cite{Li2024} also attempt to model the UV-optical photometric SED with an AGN reddened by a gray extinction law. However, spectroscopic observations reveal that LRDs consistently exhibit continuum inflection at the Balmer limit (or even sometimes strong Balmer breaks, see \citealt{Furtak2024Nature}, \citealt{Labbe2024Monster}, \citealt{Wang2024BRD}, and this work), which are hard to be explained with dust alone. These single-component and composite models, particularly when an AGN is involved, are also hard to reconcile with observations beyond the UV-optical SED. The flattening of the rest-frame near-infrared SED \citep{Williams2024} in LRDs, their X-ray non-detections \citep{Akins2024, Ananna2024, Yue2024}, their lack of a rising mid-infrared continuum associated with the dusty torus typically observed in reddened AGNs \citep{Akins2024, Wang2024BRD}, as well as lack of variability in many LRDs \citep{KokuboHarikane2024, Tee2024, Zhang2024}} are often interpreted as significant challenges to the AGN interpretation, leading to the hypothesis of a stellar origin of LRDs. However, such explanations cannot be easily reconciled with the uncomfortably efficient early formation history \citep{Wang2024RUBIESz78MassGal} and the existence of broad lines \citep{Maiolino2024LRDGeometry}; nonetheless, the feasibility of such a scenario has been suggested by \cite{Baggen2024BLnotAGN}. \textcolor{black}{The ALMA 1.2\,mm dust continuum non-detection in LRDs \citep{Fujimoto2023uncoverALMA, Akins2024} also challenges the stellar origin of their UV-optical SED from dust-reddened galaxies, although hotter-than-usual dust temperatures could resolve this issue \citep{Casey2024}.} The X-ray non-detections \textcolor{black}{and weak variability} have also been interpreted as evidence for super-Eddington AGN \citep[e.g.,][]{Lambrides2024, PacucciNarayan2024, Zhang2024}, but the exact SED shape for such accretors is highly uncertain and poorly understood \citep[e.g.,][]{Kubota&Done2019}. 

The challenges in understanding the SED of LRDs are only complicated further by their faintness ($-20\,\lesssim M_\mathrm{UV}\lesssim-16$). Fortunately, foreground galaxy clusters can serve as gravitational lenses to efficiently obtain high-SNR spectra of LRDs and conduct detailed modeling of their SED shapes. In this work, we present \textcolor{black}{a} spectral analysis that decomposes the UV-optical continuum emission of \triplelens{}. This source is a triply imaged LRD at $z=7.04$ in the Ultradeep NIRSpec and NIRCam ObserVations before the Epoch of Reionization (UNCOVER) survey that displays a strong break feature at rest-frame $\sim3600\,\mathrm{\AA}$ \citep{Furtak2023, Furtak2024Nature}. Section~\ref{sec:source} describes the observations, the data, and our knowledge of the properties of \triplelens{} to date. In Section~\ref{sec:analysis}, we present our spectral analysis with multiple models to decompose the continuum emission of \triplelens. Lastly, we discuss some of the challenges in the modeling process in Section~\ref{sec:discussion}. Throughout this work, we assume a cosmology of $\Omega_\mathrm{m,0}=0.3$, $\Omega_\mathrm{\Lambda,0}=0.7$, and $H_0=70\,\mathrm{km\,s^{-1}\,Mpc^{-1}}$. All magnitudes are expressed in AB magnitudes.

\section{Data and Source Description}\label{sec:source}
\begin{figure*}
    \centering
    \includegraphics[width=\textwidth]{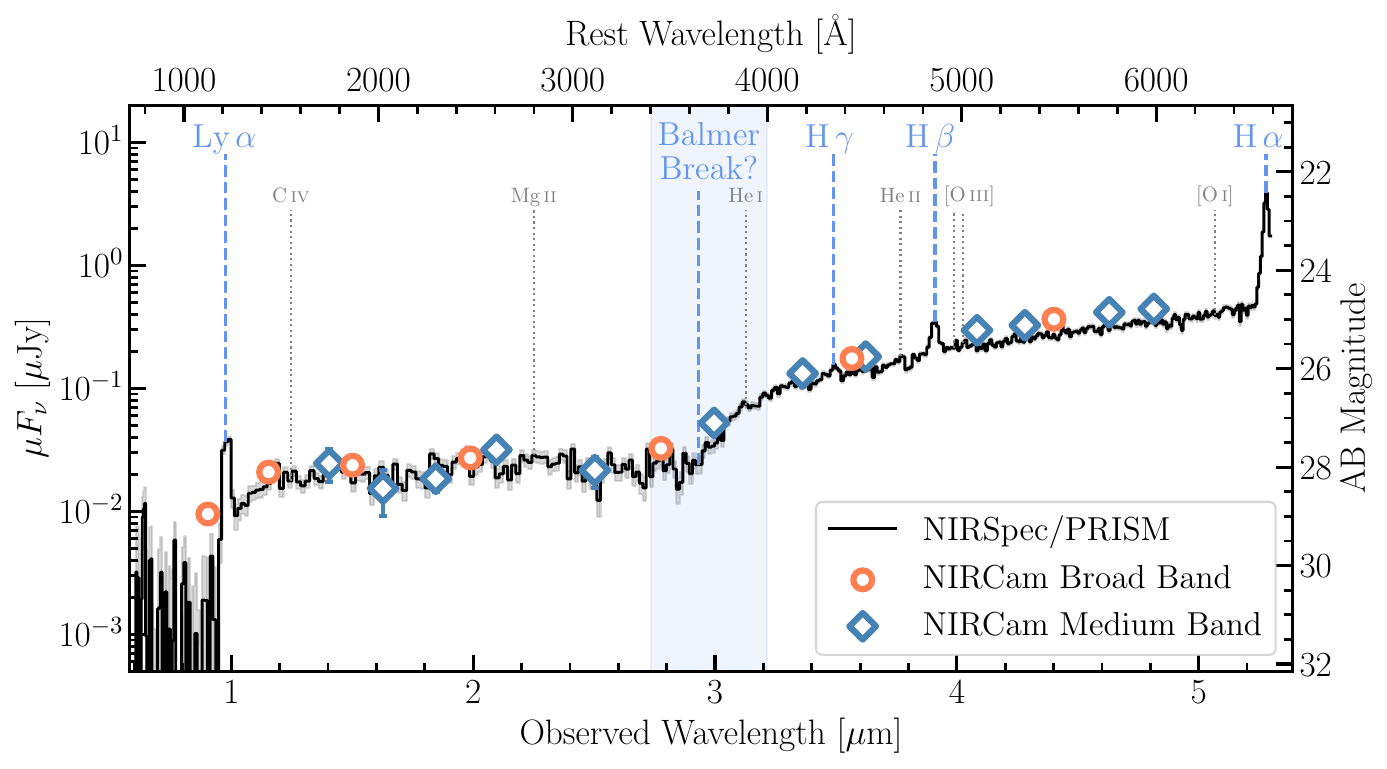}
    \caption{The NIRSpec/PRISM spectrum of image A (MSA13123) of \triplelens{} with no lensing correction applied is shown in black. The shaded gray area around the flux is the $1\sigma$-range of the observed spectrum. The circles and diamonds are photometry from the NIRCam broad-band and medium-band filters, respectively \citep{Suess2024MEGASCIENCE}. Strong hydrogen emission lines and the break are labelled in blue. Helium and metal lines are labelled in gray. }
    \label{fig:spec_with_phot}
\end{figure*}

\subsection{Observations}

The UNCOVER survey \citep{Bezanson2022uncover} is a \jwst{} Cycle 1 Treasury program. It was designed to exploit the foreground galaxy cluster Abell\,2744 \citep[$z=0.308$;][]{Lotz2017} as a gravitational lens to obtain deep NIRCam broad-band imaging, photometrically detecting sources with $m_\mathrm{F444W}\lesssim 30\,\mathrm{mag}$ for characterization \citep{Weaver2024}. In addition, medium-band imaging from observations taken in Cycle 2 are also available for most of the 45\,arcmin$^2$-field as part of the Medium Bands, Mega Science survey \citep{Suess2024MEGASCIENCE}. 

\triplelens{} was observed as part of the NIRSpec/PRISM follow-up campaign of the UNCOVER survey, which includes 668 unique targets \citep{Price2024Spectra}. The spectroscopic observations used a 2-POINT-WITH-NIRCam-SIZE2 dither pattern and nodded with a 3 shutter slitlet pattern with the apertures at a position angle of 44.56 degrees. The three images of \triplelens{} were included in five (A), three (B), and four (C) of the seven micro-shutter array (MSA) mask configurations, respectively. The total exposure times for the three images are 16.7 hours (A), 9.4 hours (B), and 11.7 hours (C), respectively. 

\subsection{Data Reduction}
\cite{Price2024Spectra} provide the full detailed description of the reduction process of the NIRSpec/PRISM spectrum, so we only give a brief summary here. We begin with level 1 products from MAST\footnote{Available from: \href{https://dx.doi.org/10.17909/8k5c-xr27}{https://dx.doi.org/10.17909/8k5c-xr27}} and run the \textit{JWST} \texttt{jwst} \citep{Bushouse2024JWSTpipeline} stage 1 pipeline with \texttt{msaexp} \citep{Brammer2023a_msaexp} and \texttt{grizli} \citep{Brammer2023b_grizli}; snowballs are corrected in this process. We then correct for the $1/f$ noise and the bias offset. World coordinate system (WCS) assignments, 2D slit extraction, slit flat-fielding, vignetting correction, and photometric calibration are carried out in the \texttt{jwst} stage 2 pipeline. We subtract the local background by taking the difference of frames at different nodding positions. The background-subtracted exposures are then drizzle resampled onto a common grid before aligning and stacking. Finally, we modify the optimal extraction algorithm by \cite{Horne1986} to account for the wavelength-dependent PRISM resolution and extract the final 1D spectrum. We do not combine the spectra of all three images and only utilize the one \textcolor{black}{with the highest SNR}, image A (MSA~13123), in UNCOVER/Mega Science Data Release 4 (DR4) for the analysis. This is to avoid possible spectral shape changes among the spectra of the different gravitationally lensed images due to potential time variability \citep{Golubchik2024, KokuboHarikane2024, Zhang2024} \textcolor{black}{as well as the data quality issues resulted from the contamination from a nearby foreground galaxy light (image B) or merely lower SNR (image C). Using a single spectrum allows us to take the high-SNR data as it is and explore the models without any post-reduction manipulation (see more discussion in Section~\ref{sec:overall_settings}).} The final 1D-spectrum is offset from the photometry by a roughly constant factor of $\sim1.1$, which is likely due to imperfect slit loss corrections, but the continuum shape is not affected. 

In this work, we adopt a lensing magnification of $\mu=5.4\pm0.1$, corresponding to the image covered by MSA13123 \citep[image A;][]{Furtak2024Nature} to de-magnify the spectrum in our modeling procedures. Taking the magnification into account, the effective exposure time on our spectrum is approximately 90 hours. The tangential magnification, used for size constraints, is $\mu_{\mathrm{t}}=3.21\pm0.05$. The magnifications are computed at the position of image A and the redshift of the source from the latest UNCOVER strong-lensing model of Abell\,2744. The parametric lens model of the cluster was initially constructed in \cite{Furtak2023LensingModel} using the \citet{Zitrin2015SLcode} analytic method and \texttt{v2.0} has been updated with \textit{JWST} spectroscopic redshifts in \citet{Price2024Spectra}. The model is publicly available in DR4 on the UNCOVER website\footnote{\href{https://jwst-uncover.github.io/DR4.html}{https://jwst-uncover.github.io/DR4.html}}.

\subsection{A2744-QSO1}
\triplelens{} is the first LRD discovered in the UNCOVER field, selected for its compact size and V-shaped photometric SED shape \citep{Furtak2023}. It is also the LRD with the best size constraint due to its image multiplicity and the strong lensing magnification in the UNCOVER field. The three images of the object are all consistent with point sources across all the NIRCam filters. An upper limit of $r_\mathrm{e}<30\,\mathrm{pc}$ is thus estimated based on the lensing magnification and the point spread function in the F115W band \citep{Furtak2023, Furtak2024Nature}. Given that some LRDs show signatures of extended morphology in the UV \citep[e.g.,][]{Kokorev2024LRDz4, Labbe2024Monster}, it is likely that this size upper limit probes the physical extent of the stellar populations in \triplelens. Yet, we also note that spatially extended scattered light has been observed in low-redshift AGN as well \citep[e.g.,][]{Zakamska2005, Zakamska2006}. Indeed, in order to model the photometric SED of \triplelens{} and to explain its compactness, \cite{Furtak2023} prefer a model where the UV continuum originates from the unobscured scattered AGN light and the optical continuum is produced by a reddened AGN. 

The low-resolution NIRSpec/PRISM spectrum of \triplelens, shown in Figure~\ref{fig:spec_with_phot}, confirms the source redshift at $z_\mathrm{spec}=7.04$ \citep{Furtak2024Nature}. The spectrum reveals broad Balmer emission lines with $\mathrm{FWHM}\approx2500\,\mathrm{km\,s^{-1}}$ \citep{Furtak2024Nature} --- such line widths are typically seen in type-1 AGNs, corroborating the potential existence of a BH. The authors also measure line luminosities of $L_\mathrm{H\alpha} = 7.1\pm0.3\times10^{41}\,\mathrm{erg\,s^{-1}}$ and $L_\mathrm{H\beta}=9.5\pm0.3\times10^{40}\,\mathrm{erg\,s^{-1}}$; in contrast, the metal lines are very weak. Based on the line width and luminosity, \cite{Furtak2024Nature} estimate that \triplelens{} hosts a BH with $M_\mathrm{BH}=4_{-1}^{+2}\times10^7\,\mathrm{M_\odot}$ that is accreting at 30\,\% of its Eddington luminosity, although one should be cautious about inferring $M_\mathrm{BH}$ using standard scaling relations at such high redshift \citep{King2024, Lupi2024}, particularly when we still lack an understanding of the SED of LRDs. Equally prominent in the PRISM spectrum in Figure~\ref{fig:spec_with_phot} is a discontinuity and change in continuum slope at $\sim3\,\mathrm{\mu m}$ observed frame (referred as ``the break" hereafter). It is worth noting that the break coincides with 3600$-$4000\,\AA{} in the rest-frame where continuous Balmer absorption occurs \citep{Setton2024LRD3600}, hinting at a continuum signature of the galaxy host. Assuming the highest surface density observed to date \citep{Baggen2023, Vanzella2023}, \cite{Furtak2024Nature} \textcolor{black}{estimate} a fiducial stellar mass upper limit of $M_\star<1.4\times10^9\,\mathrm{M_\odot}$. Nevertheless, the authors do not conduct a detailed analysis of the continuum to support their inferred stellar mass upper limit. 

\section{modeling}\label{sec:analysis}
Given all of the information mentioned above, we aim to search for a self-consistent model capable of \textcolor{black}{reproducing the rest-frame UV-optical continuum shape of \triplelens.} In this section, we present several SED fits that we perform on the de-magnified spectrum of \triplelens. The SED models are produced using combinations of AGN continua and synthesized high-resolution spectra of composite stellar populations. We first make a few general comments applicable to all the modeling procedures in this work before delving into the different options to model the spectrum of \triplelens. 

\subsection{Overall Settings}\label{sec:overall_settings}
Throughout our analysis, we conduct the spectral fitting with {\tt\string PyMultiNest} \citep{Buchner2014, Feroz2008, Feroz2009, Feroz2019}, a nested sampling code. Such an algorithm is particularly useful when multiple modes potentially exist in the posterior distributions. 

We model the continuum emission of \triplelens{} without emission lines. The strong emission lines have significant broad components and may also harbor narrow components, both of which can vary in width and fractional contribution from one line to another --- this is difficult to disentangle at the low resolution of the PRISM spectra and is outside of the scope of this work. Additionally, \cite{Furtak2024Nature} and \cite{Greene2024} already provide an extensive analysis and discussion of the emission lines of \triplelens{} and of the LRDs in the UNCOVER field. Since the metal emission lines are very weak for \triplelens{}, as shown in Figure~\ref{fig:spec_with_phot}, modeling the continuum alone is much easier than in the case of some other LRDs \citep[e.g.,][]{Labbe2024Monster}. Therefore, we mask at the positions of strong broad emission lines, i.e., \ha, \hb, and H\,$\gamma$, to avoid complications in the modeling process. We also mask $\lambda_\mathrm{rest}<1400\,\mathrm{\AA}$ since the physical origin of this UV emission remains ambiguous. The flux drop just redward of the Ly$\alpha$ emission line may be due to the damping wing of Ly$\alpha$ absorption in the circumgalactic medium \textcolor{black}{given that the high $A_V\approx2-4$ inferred in both simulated and observed LRDs \citep{Killi2023, Volonteri2024, Wang2024BRD} would translates to a column density of $N_\mathrm{H}\approx10^{21-22}\,\mathrm{cm^{-2}}$}. The roll over could also be attributed to the two-photon continuum \citep{Cameron2024, Katz2024}, although we do not detect the recombination edge of the hydrogen nebular continuum that is usually associated with it. 

\begin{figure*}
    \centering
    \gridline{
    \fig{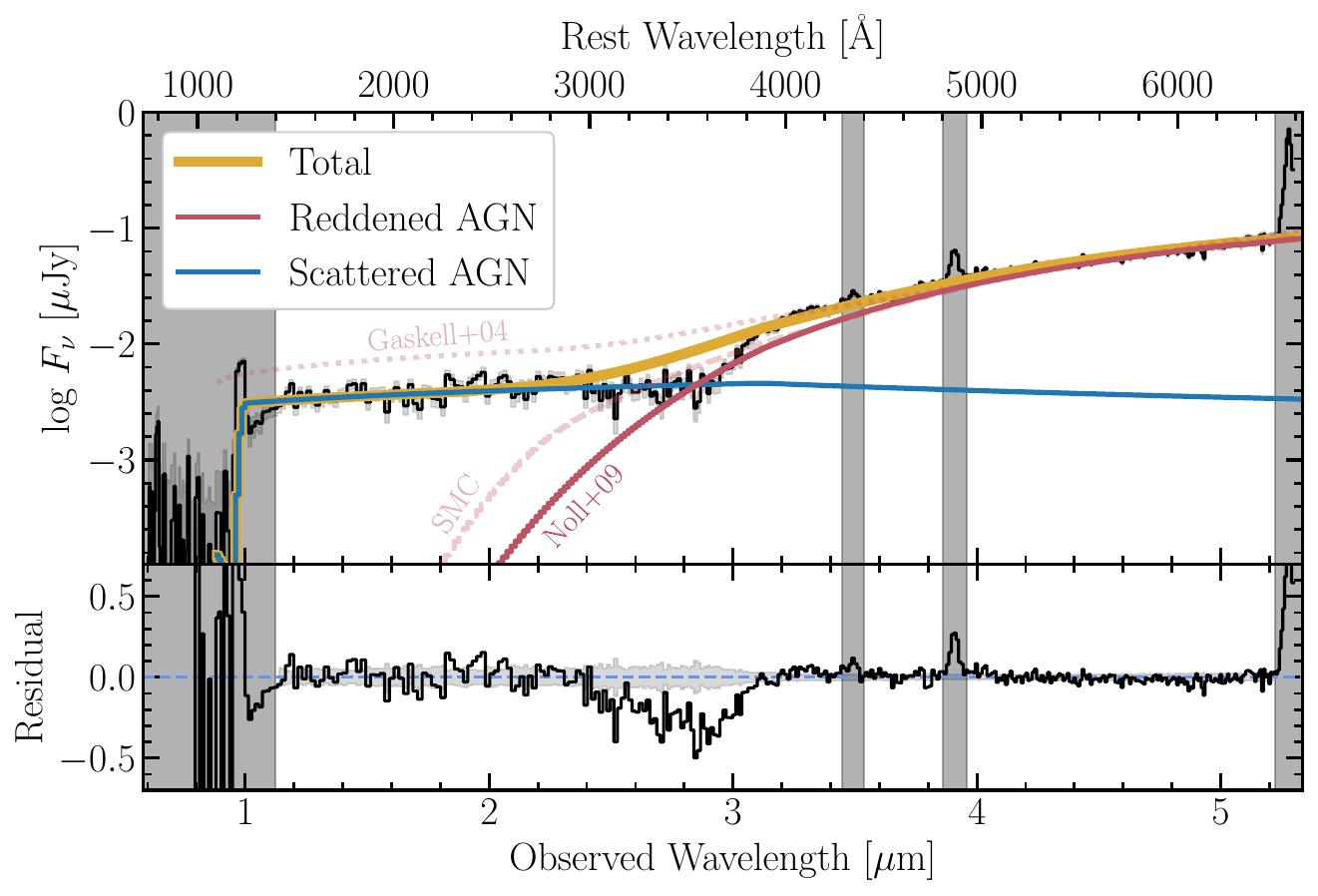}{\columnwidth}{(a) AGN-only Model}
    \fig{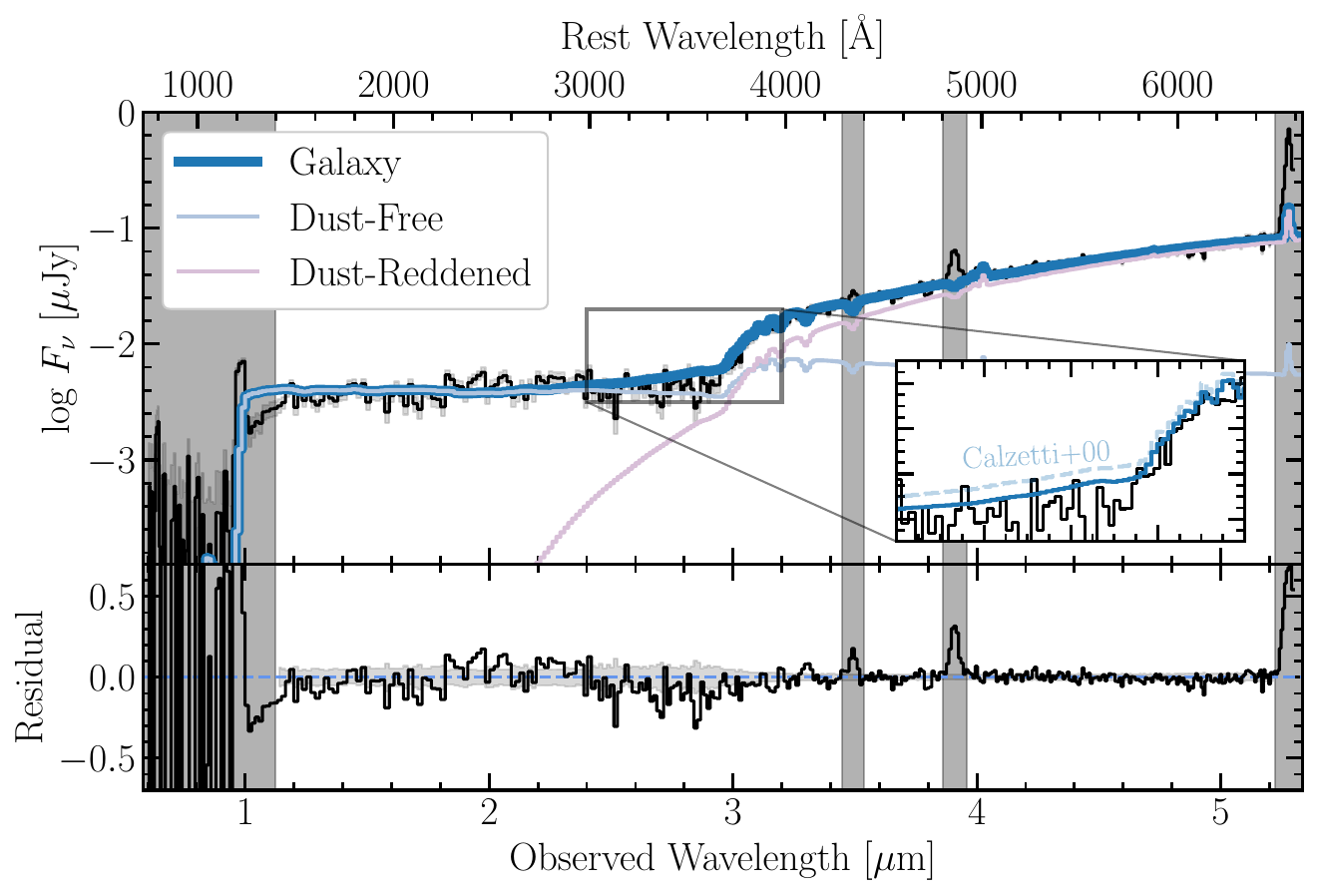}{\columnwidth}{(b) Galaxy-only Model}
    }
    \gridline{
    \fig{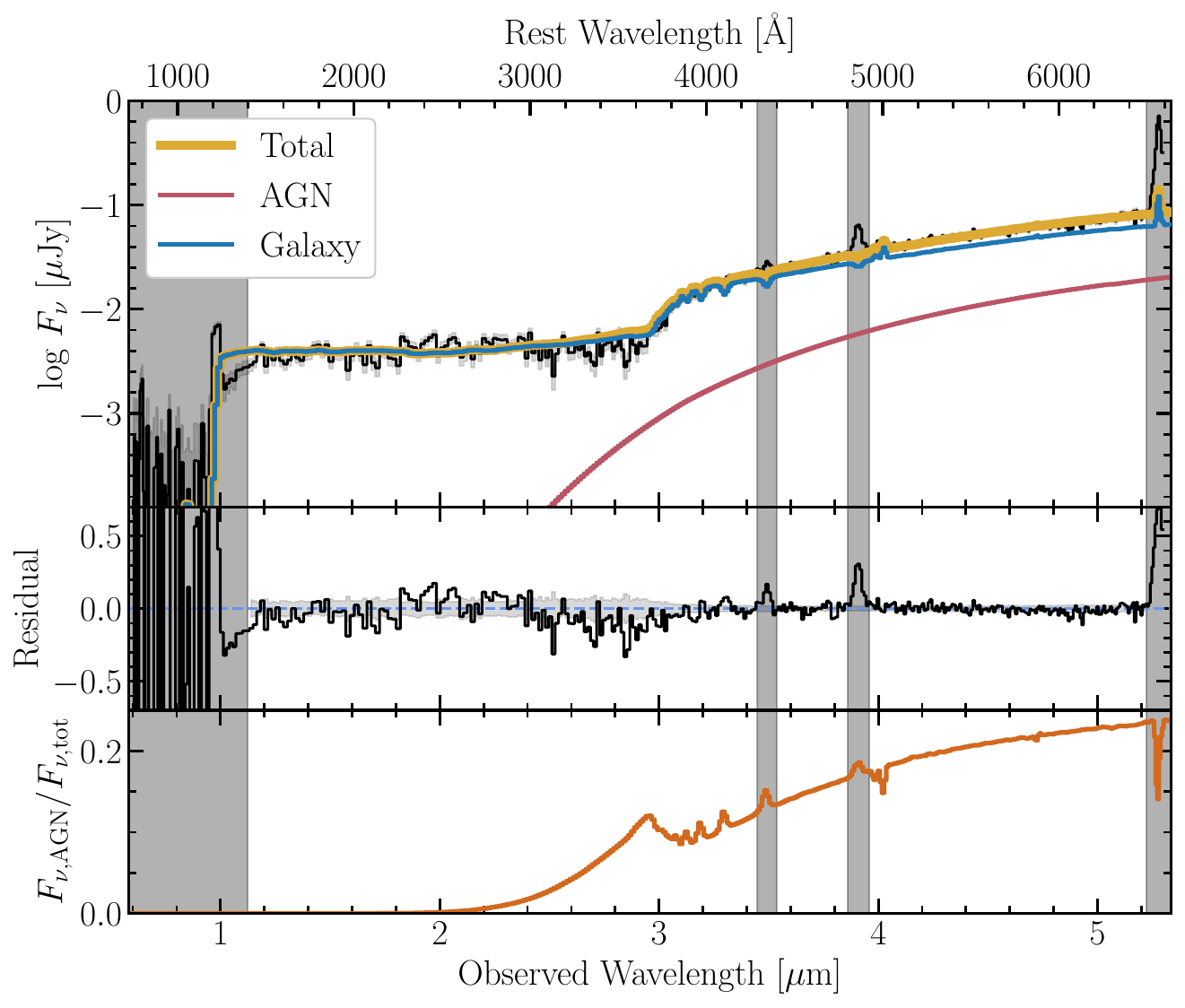}{\columnwidth}{(c) AGN+Galaxy Model}
    \fig{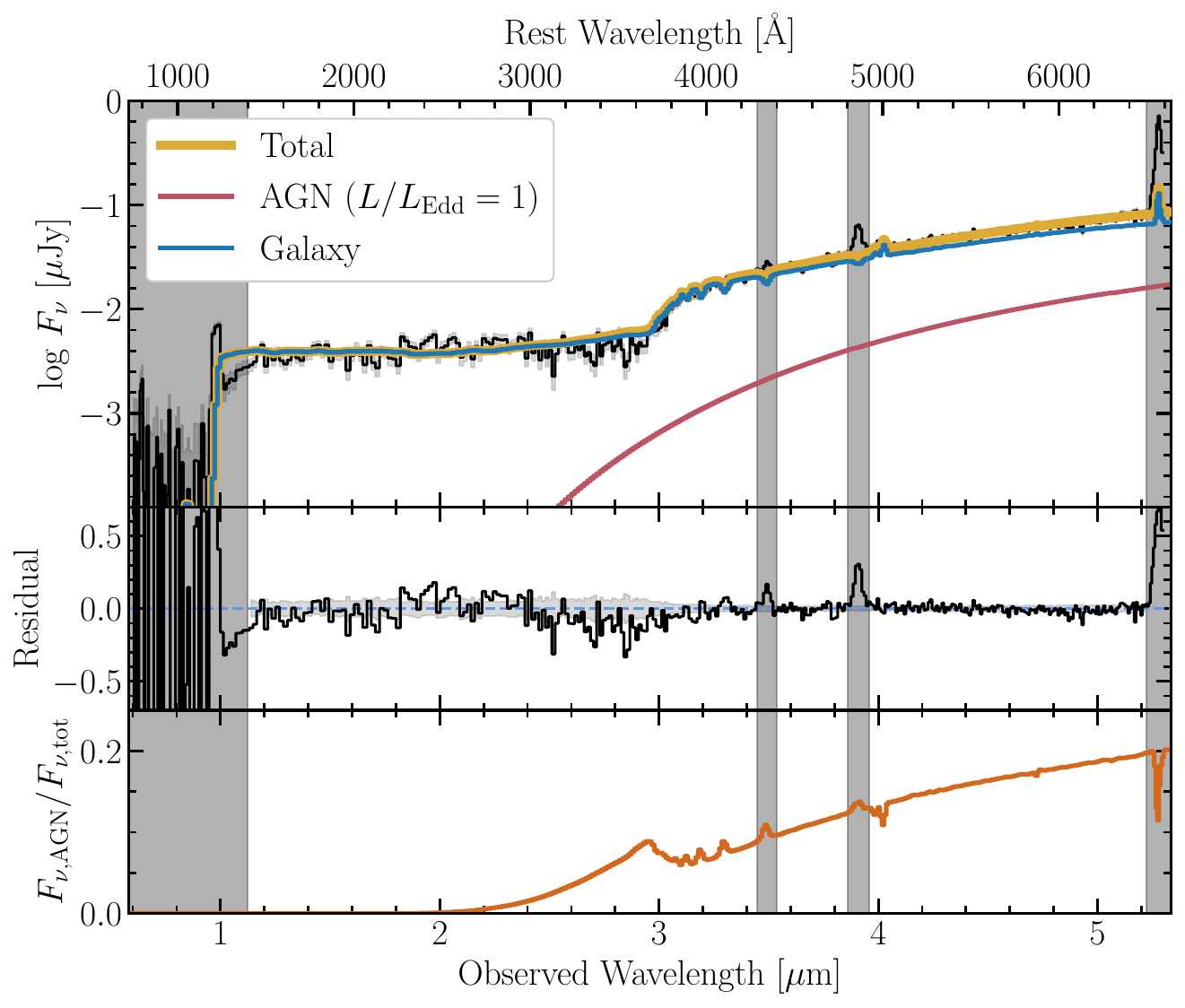}{\columnwidth}{(d) Eddington AGN+Galaxy Model}
    }
    \caption{The best-fit models for the four different scenarios proposed in this work. We plot the de-magnified spectra here. The solid red curve in each panel represents the reddened AGN continuum assuming the \cite{Noll2009dustlaw} dust law. The dashed red curves in the AGN-only model are the best-fit reddened AGN component assuming the SMC dust law \citep{Gordon2003} and the \cite{Gaskell2004} gray extinction law. The solid blue curves are starlight from the galaxy except in the AGN-only model, in which it represents the scattered unattenuated AGN continuum. \textcolor{black}{The light blue and pink curves in panel (b) are the decomposed dust-free and dust-reddened stellar population components in the galaxy model, respectively.} The vertical bands are wavelength ranges that are masked in the fitting procedures. The dashed blue curve in the galaxy-only model is the best-fit assuming a \cite{Calzetti2000} dust law. For the two AGN+galaxy composite models, we also show the fraction of AGN contribution to the total continuum model as a function of wavelength in a third panel as the orange curve.}
    \label{fig:best_fit_models}
\end{figure*}

Next, we proceed to consider the spectral resolution. The model spectra output from stellar population synthesis codes are of much higher resolution than those observed by the NIRSpec/PRISM. The published resolution curve by \cite{Jakobsen2022} and that on the \jwst{} user documentation webpage are calibrated using uniformly illuminated slits. \cite{deGraaff2024LSF1.3} show that point sources would yield up to a factor of $\sim2$ higher spectral resolution. Since \triplelens{} is unresolved throughout all the NIRCam bands as we mention in Section~\ref{sec:source}, this is the appropriate regime in our case. After accounting for the additional broadening in the reduction and coadding processes, we conservatively increase the published spectral resolution by a factor of 1.3 at all wavelengths for simplicity and convolve the model spectra to this resolution before comparing it to the observed data, similar to the treatment by \cite{Greene2024}. Using the more realistic wavelength- and size-dependent correction factor mostly affects the line width measurements rather than continuum modeling. We also test other constant factors ranging from 1 to 1.8 and confirm that the continuum shape, including the break, is not sensitive to the specific value of this correction factor. The model and observed spectra are compared in $\log F_\nu$ versus wavelength (as shown in Figure~\ref{fig:spec_with_phot} and Figure~\ref{fig:best_fit_models}). 

Lastly, we note that similar spectral decomposition analysis of LRDs has been carried out in the literature \citep[e.g.,][]{Killi2023, Wang2024BRD, Wang2024RUBIESz78MassGal}. Our modeling procedures are similar to those works in many respects (see descriptions in the following sections). \textcolor{black}{For instance, \cite{Wang2024BRD, Wang2024RUBIESz78MassGal}} conduct a polynomial calibration of the spectroscopic continuum to the photometry along with error inflation in the modeling processes to account for slit losses and other data quality/calibration issues. These steps may also leave room to potentially mitigate shortcomings in the models themselves. However, since the spectrum of image A of \triplelens{} has a very high $\mathrm{SNR}\approx20$ per pixel, and the medium band photometry of \triplelens{} confirms the spectral shape \citep[and Figure~\ref{fig:spec_with_phot}]{Furtak2024Nature}, we do not adopt \textcolor{black}{these} steps, but instead take the data as they are for our modeling analysis.

\begin{deluxetable*}{cclc}
    \centering
    \tablecaption{Free Parameters and Their Priors in the modeling\label{tab:prior}}
    \tablehead{\colhead{Parameter} & \colhead{Unit} & \colhead{Description} & \colhead{Prior}} 
    \startdata
    $\log M_\star$ & $M_\odot$ & Total stellar Mass & [7, 12] \\
    $\log \tau$ & Gyr & $e$-folding timescale in a delayed-$\tau$ star formation history & [$-3.3$, 1] \\
    $t_\mathrm{start}$ & Gyr & Starting time of the delayed-$\tau$ star formation history & [0, $t_\mathrm{universe}$] \\
    $A_V^\text{gal}$ & mag & $V$-band attenuation of the stars (also seen by the AGN) & [0, 5] \\
    $\delta$ & --- & Multiplicative power law index of the \cite{Noll2009dustlaw} dust law & [$-1$, 0.4] \\
    $f_\mathrm{nodust}$ & \textcolor{black}{\%} &Fraction of stars not experiencing any dust attenuation & \textcolor{black}{[0, 100]} \\[5pt]
    \hline
    $\log \lambda L_{\lambda,3000}$ & $\mathrm{erg\,s^{-1}}$ &Luminosity of the unreddened AGN continuum at rest-frame 3000\,\AA & [43, 47] \\
    $A_V^\mathrm{AGN}$ & mag & $V$-band extinction of the AGN (not seen by the stars) & [0, 5] \\
    $f_\mathrm{scat}$ & \textcolor{black}{\%} & Fraction of light scattered from the unreddened AGN continuum & \textcolor{black}{[0, 100]} \\ 
    $\log M_\mathrm{BH}$ & $M_\odot$ & Black hole mass & [6, 9]
    \enddata
    \tablecomments{The table presents all free parameters invoked in this work. A given model does not necessarily involve all the free parameters listed above. All priors are uniform within the ranges. The age of the universe $t_\mathrm{universe}$ at $z=7.04$ is 0.745\,Gyr under the concordance cosmology assumed in this work.}
\end{deluxetable*}

\begin{deluxetable*}{lc|rrrr}
    \centering
    \tablecaption{Best-Fit Parameters in Different Models\label{tab:result}}
    \tablehead{
    \colhead{Parameter} & \colhead{Unit} &\colhead{AGN-only} & \colhead{Galaxy-only} & \colhead{AGN+Galaxy} & \colhead{EAGN+Galaxy}
    }
    \startdata
    $\chi_\nu^2$ & --- & 4.32 & 2.74 & 2.81 & 2.85 \\[1pt]
    \hline
    $\log M_\star$ & $\mathrm{M_\odot}$ & \nodata & $9.71_{-0.03}^{+0.03}$ & $9.56_{-0.03}^{+0.02}$ & $9.59_{-0.03}^{+0.02}$ \\
    $\log\tau$ & Gyr & \nodata & $-1.94_{-0.05}^{+0.05}$ & $-1.88_{-0.03}^{+0.02}$ & $-1.88_{-0.03}^{+0.02}$ \\
    $t_\mathrm{start}$ & Gyr & \nodata & $0.64_{-0.01}^{+0.01}$ & $0.63_{-0.01}^{+0.01}$ & $0.63_{-0.01}^{+0.01}$ \\
    $A_V^\mathrm{gal}$ & mag & \nodata & $2.12_{-0.02}^{+0.02}$ & $2.02_{-0.02}^{+0.02}$ & $2.00_{-0.02}^{+0.01}$ \\
    $\delta$ & --- & $<-1$ & $<-1$ & $<-1$ & $<-1$ \\
    $f_\mathrm{nodust}$ & \% & \nodata & $1.85_{-0.09}^{+0.12}$ & $2.84_{-0.10}^{+0.12}$ & $2.67_{-0.09}^{+0.09}$ \\[5pt]
    \hline
    $\log\lambda L_{\lambda,3000}$ & $\mathrm{erg\,s^{-1}}$ & $44.42_{-0.02}^{+0.02}$ & \nodata & $43.99_{-0.03}^{+0.03}$ & \nodata \\ 
    $A_V^\mathrm{AGN}$ & mag & $2.08_{-0.01}^{+0.01}$ & \nodata & $0.50_{-0.09}^{+0.09}$ & $0.19_{-0.08}^{+0.09}$ \\
    $f_\mathrm{scat}$ & \textcolor{black}{\%} & $1.13_{-0.02}^{+0.02}$ & \nodata & \nodata & \nodata \\[5pt]
    \hline
    $\log M_\mathrm{BH}$ & $\mathrm{M_\odot}$ & $^\dagger7.60_{-0.12}^{+0.18}$ & $^\dagger7.60_{-0.12}^{+0.18}$ & $^\dagger7.60_{-0.12}^{+0.18}$ & $6.17_{-0.03}^{+0.03}$ 
    \enddata
    \tablecomments{The best-fit parameters and their uncertainties are presented. Any entry with ``$\cdots$" means the parameter is not involved in the modeling. The limit on the dust law slope $\delta$ indicates that the parameter is pushing against the prior boundaries. We use ``$\dagger$" to mark out the BH masses inferred by \cite{Furtak2024Nature} using standard emission line scaling relations.}
\end{deluxetable*}

\subsection{AGN-Only Model}\label{sec:agn_only}
The broad Balmer emission lines displayed in the spectrum of \triplelens{} as well as its compact morphology are the major arguments for the presence of an AGN \citep{Furtak2024Nature, Greene2024}. Somewhat similar populations of reddened broad-line AGN are known at lower redshift \citep[e.g.,][]{Glikman2012, Banerji2015}. Therefore, we use the preferred AGN-only model from \cite{Furtak2023} as a starting point and test its validity in explaining the spectral shape. In brief, this model attributes the optical flux to a reddened AGN continuum and the UV flux to unobscured, scattered light from the AGN. 

We adopt the broken power law continuum empirically derived by \cite{Temple2021} from a population of type-1 quasars at $0<z<5$ to be the intrinsic continuum of the AGN. Following the parametrization of \cite{Temple2021}, the scaling of the continuum level is controlled by the unextincted luminosity at rest-frame 3000\,\AA, $\lambda L_{\lambda,3000}$. The continuum is then reddened using the \cite{Noll2009dustlaw} curve parameterized by $A_V$ and a multiplicative power law slope $\delta$. \textcolor{black}{We ignore the 2175\,\AA\ bump in our setup --- this is because at the typical LRD extinction of $A_V\approx2$ \citep{Volonteri2024, Wang2024BRD, Wang2024RUBIESz78MassGal}, the reddened continuum is always significantly suppressed and does not contribute to the observed flux at the 2175\,\AA\ (see the best-fit models in Figure~\ref{fig:best_fit_models}). Under such a setup, we do not have the ability to detect the existence of the UV bump.} Given that the AGN is only observed along a single  line of sight, it is true that an extinction curve instead of an attenuation curve is more appropriate in the modeling. Yet, this parameterization we choose is flexible enough to cover slopes ranging from a \cite{Calzetti2000} extinction law to the steep SMC curve \citep{Gordon2003}. \textcolor{black}{Since the dust physics is still uncertain at such high redshift, we adopt a broad prior on the dust law slopes ($\delta\in[-1, 0.4]$, also see Table~\ref{tab:prior}) that is the same as the one used by \cite{Wang2024BRD}. This prior range ensures that both gray extinction curves (i.e., smaller $A_\mathrm{UV}/A_V$, \textcolor{black}{$\delta\approx0.4$}) and dust laws steeper than that of SMC \textcolor{black}{($\delta\approx-0.5$)} can be sampled. The prior upper bound is set to avoid an unphysically flat extinction curve.} The scattered UV continuum is modeled as the unobscured AGN continuum scaled down by the scattering fraction $f_\mathrm{scat}$. This AGN-only model includes four free parameters: the intrinsic luminosity $\lambda L_{3000}$, the scattering fraction $f_\mathrm{scat}$, the dust reddening $A_V$, and dust law slope $\delta$. \textcolor{black}{It is true that both $\lambda L_{3000}$ and $A_V$ would scale the modeled continuum and would be correlated, but $A_V$ also affects the shape of the reddened continuum, making sure that the two parameters are not entirely degenerate.}

We show in Figure~\ref{fig:best_fit_models}a that the best-fit AGN-only model in this scenario does not well reproduce the observed spectral shape, with a reduced chi-square value of $\chi_\nu=4.27$. The summation of the reddened continuum and the scattered intrinsic power law produces a transition that is too smooth to fit the data around rest-frame 3600\,\AA, i.e., the model overproduces flux at the break. We note that the dust law slope is pushing against the prior range (see Table~\ref{tab:result} \textcolor{black}{and Figure~\ref{fig:corner_agn_only}}), suppressing the UV flux as much as allowed. It is also steeper than that of the Small Magellanic Cloud (SMC), which is well suited to describe extinction in more typical AGN \citep[e.g.,][]{Hopkins2004}. In fact, when we \textcolor{black}{explicitly} use the SMC extinction law to fit the data, the reddened AGN component already overproduces flux at the break, as we show with a dashed curve in Figure~\ref{fig:best_fit_models}a. \textcolor{black}{Similarly, a gray \cite{Gaskell2004} extinction curve cannot account for the break strength either, as we show with the dotted curve in Figure~\ref{fig:best_fit_models}a. }

Therefore, we disfavor such an AGN-only model even though it is preferred for the photometric SED \citep{Furtak2023}, because the combination of an unobscured power law and a reddened one cannot reproduce the strong break well, a feature that is only revealed with the addition of spectroscopy. It is possible that either stellar light contributes to the continuum in addition to an AGN, or that the intrinsic AGN continuum and/or dust extinction curve of \triplelens{} does not resemble those of typical AGNs.

\subsection{Galaxy-Only Model}\label{sec:galaxy_only}

\begin{figure}
    \centering
    \includegraphics[width=\columnwidth]{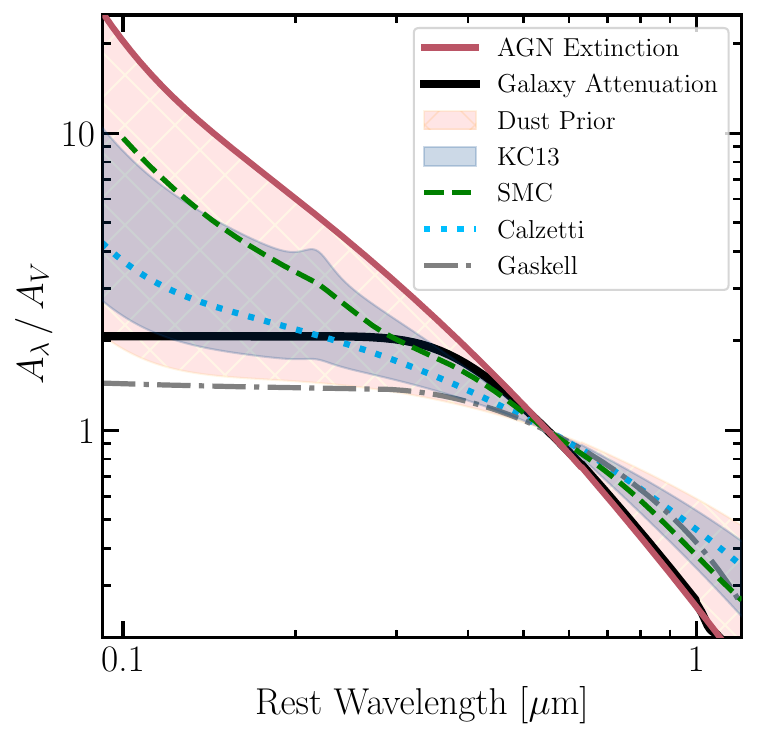}
    \caption{The possible dust attenuation laws are presented. \textcolor{black}{The dark blue shaded area shows the ranges empirically measured by \cite{KriekConroy2013}}. The red hatched shade shows the dust law priors in the fitting procedures. The solid curve (red) is the dust extinction law and the effective galaxy attenuation law (black) that the fits prefer, where the former also pushes against the prior boundary.} 
    \label{fig:dust}
\end{figure} 

Given that the AGN-only model fails to reproduce the observed break strength, we next seek models that produce strong breaks. The break feature has been associated with a Balmer break \citep{Wang2024BRD,Wang2024RUBIESz78MassGal}, where it has been found that a galaxy-only model produces a statistically acceptable fit to the continuum. Thus, we naturally ask the question whether the continuum of \triplelens{} is solely produced by galaxy light, which could easily show a Balmer break in the spectra.

We generate the galaxy SED with the flexible stellar population synthesis code (FSPS; \citealt{Conroy2009, Conroy2010}) using the MIST isochorones \citep{Dotter2016, Choi2016} and the MILES library of stellar spectrum templates \citep{Sanchez-Blazquez2006, Falcon-Barroso2011}. We also assume a simple delayed-$\tau$ star formation history, parameterized by the $e$-folding timescale $\tau$ and the starting time $t_\mathrm{start}$, and a \cite{Chabrier2003} initial mass function (IMF). The nebular continuum is also included using the modeling by \cite{Byler2017} to account for the potential recombination and two-photon continua present at UV wavelengths. Other weak nebular lines may also contribute to the model flux given the PRISM resolution, particularly in the break region. We assume a stellar metallicity $\log(Z/Z_\odot)=-1$ to reflect the limited time available for multiple episodes of chemical enrichment in the first billion years after the Big Bang. This is supported by previous modeling; \cite{Wang2024BRD,Wang2024RUBIESz78MassGal} infer similar stellar metallicities for other LRD-like objects at similar and lower redshifts using a similar setup as ours. \cite{Gallazzi2005} also indicates $-1\lesssim\log(Z/Z_\odot)\lesssim-0.5$ for local galaxies with stellar mass between $10^9\,\mathrm{M_\odot}$ and $10^{10}\,\mathrm{M_\odot}$, which turns out to be our estimate, too (see discussion below). The extremely weak metal emission lines in \triplelens{} that hint at low gas-phase metallicity also resonate with this assumed low stellar metallicity. We choose the dust law to be in the form of the \cite{Noll2009dustlaw} curve. Lastly, we always allow a fraction of the stars to be entirely unattenuated --- this is parameterized by $f_\mathrm{nodust}\equiv M_{\star,A_V=0}/M_\star$, where $M_\star$ is the total stellar mass. In total, this model consists of six free parameters: $M_\star$, $\tau$, $t_\mathrm{start}$, $A_V^\mathrm{gal}$, $\delta$, and $f_\mathrm{nodust}$; their respective priors are presented in Table~\ref{tab:prior}. 

We present the best-fit galaxy-only model in the top right panel of Figure~\ref{fig:best_fit_models}b, the best-fit parameters in Table~\ref{tab:result} \textcolor{black}{and their posterior distributions in Figure~\ref{fig:corner_galaxy_only}}. If the entire continuum were to originate from stellar light, it would require a heavily reddened stellar population. The recent starburst has formed virtually all of the observed stellar mass ($t_\mathrm{age}\approx100\,\myr$ and $\tau\approx 10\,\myr$) in this model. The galaxy would also be quite massive ($M_\star\approx5\times10^9\,\mathrm{M_\odot}$), which, when combined with its compact size of $r_\mathrm{e}<30\,\mathrm{pc}$, implies a value that is a factor of at least five larger than the upper limit on stellar mass derived in \cite{Furtak2024Nature} by assuming an empirical maximum stellar density \citep{Baggen2023, Vanzella2023}. Furthermore, in order for the stellar light not to overproduce any flux in the break, the dust law is required to be steeper than the \cite{Calzetti2000} and SMC curves and pushes against the prior boundary. We show in the inset of Figure~\ref{fig:best_fit_models}b that indeed, a grayer \cite{Calzetti2000} dust law does not produce the break strength as well as our steep dust law does. \textcolor{black}{Even with such a steep dust extinction curve, Figure~\ref{fig:best_fit_models}b shows that because the flux from the reddened stellar population still does not drop off sufficiently fast, the addition of the unobscured continuum makes the modeled break weaker than the observed one, very similar to the case in the AGN-only model.} We note that models with more free parameters that assume a non-parametric star formation history are still unable to reproduce the break without a steep dust law (B. Wang, priv. comm., 2024), suggesting that our results are not sensitive to our choice of star formation history. However, there is an unattenuated stellar component responsible for the UV continuum. Thus, although the underlying extinction law is steep ($A_\mathrm{1500}/A_V\approx9.4$), the effective attenuation law to the stellar population has a UV-optical slope of $A_\mathrm{1500}/A_V\approx2.1$ when accounting for the unobscured stars (see Figure~\ref{fig:dust}). This UV-optical slope is not surprising for an object at optical attenuation of $A_V\approx2$ and consistent with the trend found at low redshift where attenuation laws are grayer (flatter slope between UV and optical) at higher $A_V$ \citep{Salim&Narayanan2020}. The phenomenon can be attributed to partially unobscured UV flux ranging from simple prescriptions as used in our model to more complex star-dust geometries \citep[e.g.,][]{Narayanan2018a, Salim&Narayanan2020, Trayford2020}. 

Both the steep extinction law and the high stellar density may hint at the improbability of this model. Yet, nullifying the model requires stronger evidence. The fact that we have not seen objects with such high densities and/or steep dust law slopes does not mean they do not exist. Another significant challenge that the galaxy-only model faces is the deficiency of ionizing photons to produce the observed \ha{} luminosity. We first de-redden the best-fit galaxy-only model and remove the IGM absorption from the model. We then convert the number of photons between rest-frame 10\,\AA{} and 912\,\AA{} in the intrinsic (unattenuated) galaxy SED into an \ha{} luminosity assuming case-B recombination \citep{OsterbrockFerland2006}. This fiducial \ha{} luminosity computed from the best-fit galaxy-only model is $\sim100$ times smaller than the observed value measured by \cite{Furtak2024Nature} even without applying a reddening correction to the line luminosity. The presence of an AGN, as implied by the broad lines, would likely be required for the addition of more ionizing photons. Nonetheless, it is plausible to achieve high $L_\mathrm{H\alpha}$ with an O-star-dominated stellar population, but such a galaxy SED would contradict the observed strong Balmer break, which necessitates late-B and A-type stars within our modeling scheme. Beyond our setup, it is possible that the Balmer break has a non-stellar origin \citep[e.g.,][]{Inayoshi&Maiolino2024}, and we discuss this possibility in Section~\ref{sec:other_models}. Therefore, given the challenges faced by this galaxy-only model, we conclude that the presence of an AGN is nearly inevitable.

\subsection{AGN + Galaxy Model}\label{sec:agn_galaxy}
Perhaps a more sensible and appropriately flexible model would include both the AGN and the galaxy. This would be a natural choice, given the strong break and the broad Balmer lines. Such a composite model has been shown to fit the data of other LRDs (and LRD-like objects) with Balmer breaks \citep{Wang2024BRD, Wang2024RUBIESz78MassGal}. In addition, the AGN could also serve as a reservoir of ionizing photons to produce the observed \ha{} luminosity.  

We model the galaxy as described in Section~\ref{sec:galaxy_only}. The AGN is also modeled as in Section~\ref{sec:agn_only} using the empirical \cite{Temple2021} SED. In this composite model, the scattered light component is replaced with unobscured stellar light, although we stress that the UV origin still remains ambiguous. Both scattered AGN/galaxy light and unobscured star formation remain open possibilities \citep[e.g.,][]{Killi2023, Labbe2023uncoverLRD, Greene2024, Matthee2024} as we already demonstrate in the AGN-only and galaxy-only models. Adding any additional components in the UV would yield significant degeneracy in the fitting process as the UV slope due to star formation and that of an AGN is very similar \citep{Greene2024}. The AGN continuum is also reddened by galactic dust ($A_V^\mathrm{gal}$) in addition to its own circumnuclear dust ($A_V^\mathrm{AGN}$), i.e. $A_{V,\mathrm{tot}}^\mathrm{AGN} = A_V^\mathrm{AGN}+A_V^\mathrm{gal}$. For simplicity in the model, we assume that the AGN and the galaxy share the same dust law slope. The model is similar to \cite{Wang2024BRD}, who model an LRD at $z\approx3$, although the authors use two different dust laws and a non-parametric star formation history in their analysis. This composite model includes eight free parameters: the stellar mass $M_\star$, $e$-folding timescale $\tau$ of star formation, starting time $t_\mathrm{start}$ of star formation, galaxy reddening $A_V^\mathrm{gal}$, AGN luminosity $\lambda L_{\lambda,3000}$, AGN dust extinction $A_V^\mathrm{AGN}$, and dust law slope $\delta$, and the fraction $f_\mathrm{nodust}$ of stars that are unobscured by dust. We show the best-fit parameters in Table~\ref{tab:result} \textcolor{black}{and their posterior distributions in Figure~\ref{fig:corner_agn_galaxy}.}

As shown in Figure~\ref{fig:best_fit_models}c, the best-fit model in the AGN+Galaxy scenario is consistent with a galaxy-dominated continuum, which is quite similar to the galaxy-only model described in Section~\ref{sec:galaxy_only}, but explains the discrepant \ha{} emission. The UV flux is entirely attributed to unobscured stellar light. A dusty 100-Myr-old galaxy is responsible for producing the observed break and most of the optical continuum, similar to the galaxy-only model described earlier. The reddened AGN only starts to contribute non-negligibly to the continuum emission redward of $\lambda_\mathrm{rest}\approx5000\,\mathrm{\AA}$ but never dominates anywhere in the NIRSpec coverage. However, \textcolor{black}{by extrapolating the unobscured AGN SED to rest-frame 10\,\AA{} and integrating up to rest-frame 912\,\AA, we estimate that} this model does predict an extincted \ha{} luminosity that is a factor of $\sim2$ higher than the observed value. Moreover, due to the AGN contribution in the rest-frame optical, the inferred stellar mass is lowered by a factor of $\sim1.5$ as listed in Table~\ref{tab:result}, but $M_\star\approx4\times10^9\,\mathrm{M_\odot}$ is still rather massive for its compact size. We note that Balmer absorption lines are present in the model even at PRISM resolution, but observational noise and emission line infilling could easily wash out their presence in the observed spectrum. 

While the joint model relieves some tension, the dust law slope remains steep and pushing against the prior \textcolor{black}{(see Figure~\ref{fig:corner_agn_galaxy})} such that the extincted near-UV photons from either the dust-obscured stars or the AGN do not wash out the break strength in the model. The dust law predicts a Balmer decrement of 11.4 for the broad line components, while the observed value is $7.4\pm0.4$ using the combined broad and narrow emission line \citep{Furtak2024Nature}. However, we stress that kinematic decomposition of the emission lines is challenging at the PRISM resolution \citep{Greene2024}, so the measured Balmer decrement does not necessarily reflect the true extinction experienced by the AGN. Since the steep extinction law slope is already in place in the galaxy-only fit, and given the AGN+galaxy model is also galaxy-dominated, the steep extinction law for the AGN is possibly inherited from there. Of course, there is no reason a priori that the AGN and galaxy should share the same shape for their underlying extinction/attenuation laws; we will further discuss other extinction curves in Section~\ref{sec:dust}. 

Regardless, the failure of our AGN-only model in Section~\ref{sec:agn_only} already shows that if the \textcolor{black}{reddened} AGN were to dominate anywhere in the optical continuum of \triplelens, its extinction law would need to be as steep as our prior allows, if not more (i.e., $\delta<-1$ or $A_\mathrm{1500}/A_V\gtrsim9$; also see Figure~\ref{fig:best_fit_models}a). This is necessary for the reddened continuum to drop fast enough to preserve the break strength. Otherwise, any grayer dust extinction curve would lead to a less extincted UV continuum and force the AGN to be even more sub-dominant to preserve the break. In that case, we would once again face the deficiency of ionizing photons to explain the broad line luminosities as well as high equivalent width of the broad lines, because the fit would approach a galaxy-only one. 

\textcolor{black}{In addition, there is another combination of AGN+galaxy composite model, where the UV continuum is dominated by the scattered unobscured AGN continuum and the optical is dominated by a dust-reddened galaxy with $A_V\approx2$ \citep[e.g.,][]{Kocevski2023}. Here, we only draw analogy from the AGN-only model to briefly describe its inability to reproduce the spectral shape of \triplelens. In the AGN-only model (Figure~\ref{fig:best_fit_models}a), we already show that if an unobscured AGN continuum dominates the UV, it remains dominant at rest-frame 3600\,\AA\ as well. The summation of such a UV continuum and a reddened stellar population would once again overproduce the flux at the break region, very similar to the AGN-only case. Indeed, actual fitting yields a reduced $\chi^2$ of 3.02. Such a model would also lead to an extremely high \ha\ equivalent width. Thus, we disfavor such a model for explaining the UV-optical continuum of \triplelens.}

\subsection{Eddington AGN + Galaxy Model}\label{sec:eddington_agn}

\begin{figure}
    \centering
    \includegraphics[width=\columnwidth]{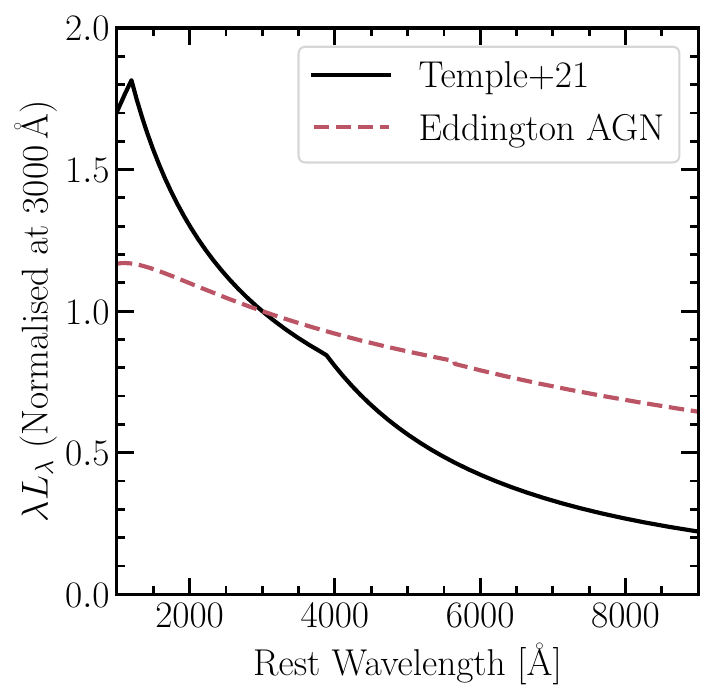}
    \caption{The two AGN SEDs are compared in this figure. The solid black curve is the empirical SED from type-1 quasars from SDSS obtained by \cite{Temple2021}. The red dashed curve is an AGN SED with assumed $L_\mathrm{bol}/L_\mathrm{Edd}=1$ and $M_\mathrm{BH}=10^{6.22}\,\mathrm{M_\odot}$. }
    \label{fig:agn_sed}
\end{figure}

The previous AGN+galaxy model works well, but it still requires uncomfortably high mean stellar density and steep extinction slope. This is because the continuum is mostly dominated by galaxy and both the reddened stellar population and the AGN must not over-produce flux near $\sim3000\,$\AA. It is possible that an intrinsically redder AGN continuum could alleviate the steepness of the dust law while allowing the AGN to dominate at redder wavelengths, thus lowering the stellar mass. AGN with high Eddington ratios then become great candidates to naturally provide a intrinsically redder continuum when compared to typical AGN accreting at sub-Eddington luminosities \citep{Volonteri2017, Kubota&Done2019}. \cite{Furtak2024Nature} suggest that \triplelens{} is one such BH with a high Eddington ratio. By adopting a BH mass of $10^{7.3}\,\mathrm{M_\odot}$ measured from the emission lines and scaling relations, the authors conclude that \textcolor{black}{\triplelens{} is} accreting at 30\% of Eddington luminosity. \textcolor{black}{The X-ray non-detections of LRDs \citep{Akins2024, Ananna2024, Yue2024}, including \triplelens{} with a $2\sigma$ upper limit of $L_\mathrm{2-10\,keV}<10^{43.68}\,\mathrm{erg\,s^{-1}}$ \citep{Ananna2024}, is another line of evidence consistent with high Eddington ratios \citep{Lambrides2024}.} Both numerical simulations and observations at lower redshifts have shown that \textcolor{black}{AGN} and quasars accreting at high fractions of the Eddington limit could become X-ray weak \citep{Ma2024, PacucciNarayan2024} due to a lack of hot gas to up-scatter the UV photons and/or the disk photosphere obscuring \textcolor{black}{the X-ray emitting region \citep{Jiang2019}}. \cite{Greene2024} also suggest that near- or super-Eddington accretion rates may help explain the high number density of the LRD population. 

Motivated by these points and the need for an intrinsically red continuum to preserve the break, we adopt the AGN SED from \cite{Volonteri2017}. Their model is inspired by the \cite{ShakuraSunyaev1973} disk solution and parameterized by the BH mass $M_\mathrm{BH}$ and Eddington ratio $L_\mathrm{bol}/L_\mathrm{Edd}$. Under this parameterisation, the peak position of the big blue bump as well as the UV-optical slope are variable: the SED assumes the peak temperature scaling from \cite{Thomas2016}, normalizing the free parameters to obtain a good match with quasar bolometric corrections in the relevant luminosity and $M_\mathrm{BH}$ range. For this exercise, we have further modified the SED to have a redder intrinsic SED when the Eddington ratio is high, mimicking the effect of radiation trapping in super-Eddington sources \citep{Begelman1979}. This is achieved by increasing the radius corresponding to the peak temperature. Since it is uncertain whether local scaling relations can be applied at high redshift, the previously measured BH mass and the Eddington ratio of \triplelens{} \citep{Furtak2024Nature} may not accurately reflect reality. We therefore assume that \triplelens{} is Eddington-limited ($L_\mathrm{bol}/L_\mathrm{Edd}=1$) in this exercise to break the degeneracy in the modeling when both $M_\mathrm{BH}$ and $L_\mathrm{bol}/L_\mathrm{Edd}$ are set free, since both are involved in scaling the full spectrum. The goal is to test whether a redder (physically motivated) AGN continuum than \cite{Temple2021} could alleviate the steep dust law problem and have AGN dominate the rest-frame optical flux. We show the comparison between the two SEDs in Figure~\ref{fig:agn_sed}. Except for the intrinsic AGN SED, the setup is the same as described in Section~\ref{sec:agn_galaxy}. The model also totals 8 parameters, whose best-fit values are presented in Table~\ref{tab:result} \textcolor{black}{and parameter posterior distributions in Figure~\ref{fig:corner_agn_galaxy_edd}}.  

We show in Figure~\ref{fig:best_fit_models}d that qualitatively, this setup yields the same model as the regular AGN+Galaxy setup, where the entire UV-optical continuum is dominated by a massive dust-reddened post-starburst galaxy, and the AGN is subdominant across all wavelengths. The fit also infers a BH mass that is nearly 1\,dex lower than the estimation from scaling relations. We note that since the size of the broad line region may change at high Eddington ratio \citep[e.g.,][]{Lupi2024}, BH masses estimated from emission line scalings may not be accurate. Furthermore, given that the best-fit model in this case is still galaxy-dominated and given our ignorance of the intrinsic SED of the AGN, our derived BH properties should be taken with a grain of salt. This model involving the Eddington-limited AGN is virtually indistinguishable from the standard AGN+galaxy model. The steep-dust-slope challenge is not alleviated by the redder AGN model, either \textcolor{black}{(see Figure~\ref{fig:best_fit_models}d and Figure~\ref{fig:corner_agn_galaxy_edd})}; the best-fit remains galaxy-dominated, and the steepness of the dust law is necessary to preserve the strong break strength, just as in the galaxy-only model described in Section~\ref{sec:galaxy_only}. More importantly, the extincted \ha{} luminosity predicted by this Eddington AGN+galaxy model is roughly half of observed value. Since this model also under-supplies the photon budget needed to explain the high $L_\mathrm{H\alpha}$ of \triplelens{}, we disfavor this model as well. 

\section{Discussion}\label{sec:discussion}
We have presented four different options to describe the continuum emission of \triplelens{} in the previous section. The AGN-only model fails to produce the strong Balmer break feature and is therefore disfavored. The galaxy-only model and Eddington AGN+galaxy model yield reasonable fits to the continuum shape but do not provide enough UV photons to ionize the ISM to produce the observed broad line luminosity --- for this reason, these models are also unsatisfactory. We therefore take the regular AGN+Galaxy model to be the fiducial model, where the UV-optical continuum is dominated by a massive post-starburst galaxy and the reddened AGN begins to contribute non-negligibly redward of H$\beta$. 

Nonetheless, challenges still exist in this composite model. The best-fit stellar mass is so large compared to \triplelens's compact size that the derived stellar surface density is among the largest seen in current observations. Also, in order for the model match the flux just blueward of the break, dust extinction laws steeper than that of the SMC are required. Given these challenges, perhaps the most likely scenario is that we are missing some physical elements in all of our modeling. In this section, we discuss some of these issues and the implications of galaxy and BH growth below. 

\subsection{Galaxy and BH Properties}

\begin{figure*}
    \centering
    \gridline{
    \fig{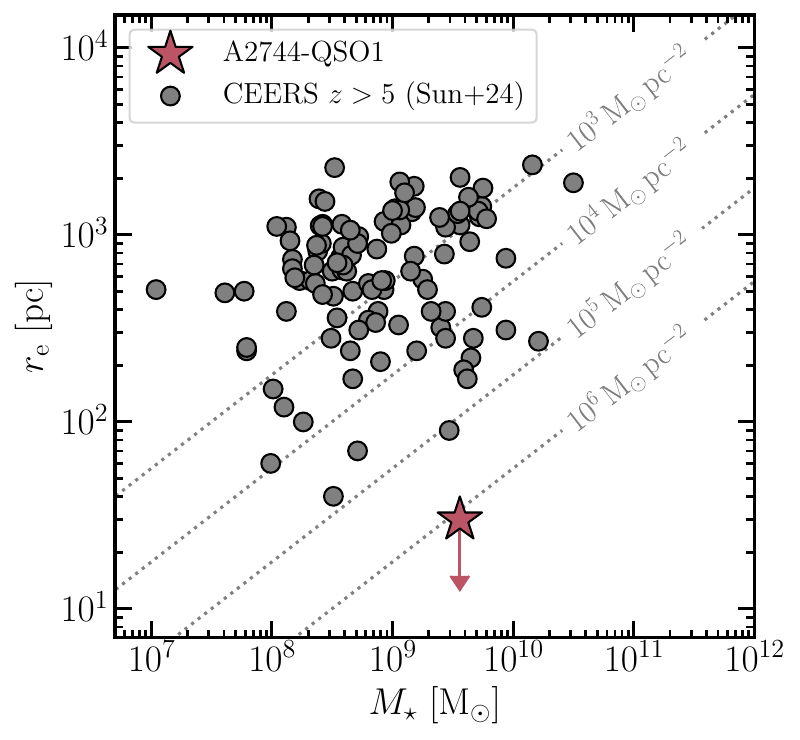}{\columnwidth}{~}
    \fig{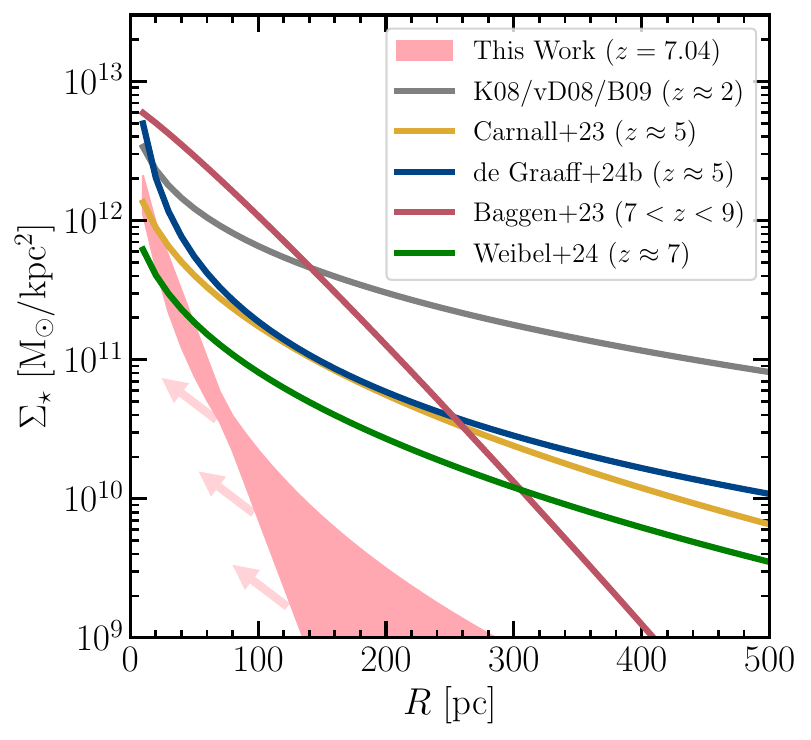}{\columnwidth}{~}}
    \caption{\textbf{Left:} The stellar mass and size of \triplelens{} (red star) is plotted against a sample of $z>5$ galaxies \citep{Sun2024} represented by the gray circles. The dashed lines are lines of constant surface mass density. \textbf{Right:} The assumed density profile of \triplelens{} (range due to assuming $n=1$ and $n=4$) is compared against other compact objects at various redshifts (\citealt{vanDokkum2008MassiveQ}, see also \citealt{Kriek2008} and \citealt{Bezanson2009}); \citealt{Baggen2023, Carnall2023, deGraaff2024, Weibel2024}) using the $r_\mathrm{e}=30\,\mathrm{pc}$ upper limit. The arrows represent the direction of the changes in the curve assuming sizes smaller than 30\,pc.}
    \label{fig:mass_size_density}
\end{figure*}

\begin{figure}
    \centering
    \includegraphics[width=\columnwidth]{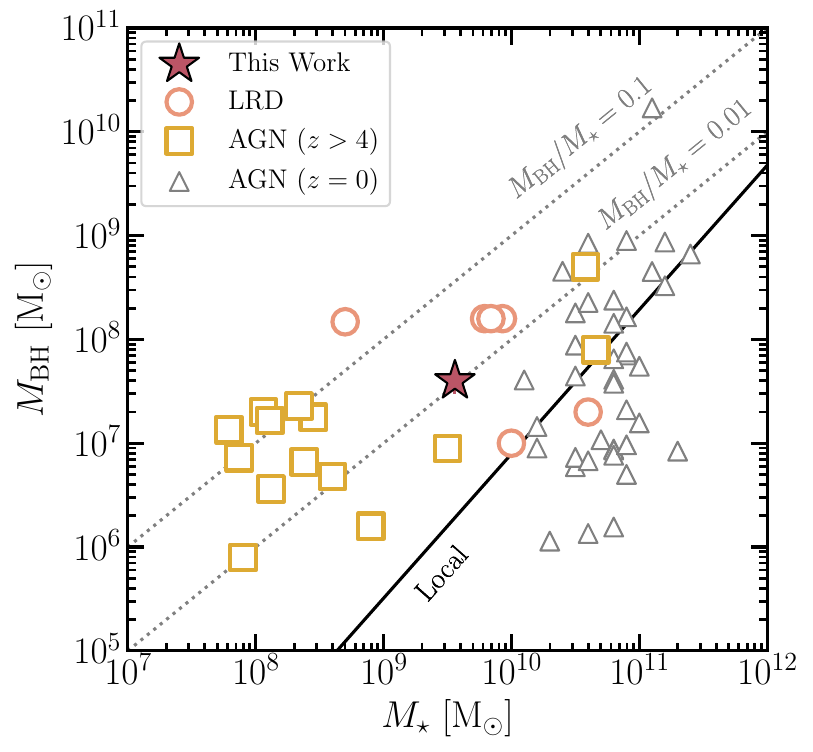}
    \caption{The stellar mass and BH mass of \triplelens{} is plotted against various samples. The BH mass of \triplelens{} is taken from \cite{Furtak2024Nature} since our inferred $M_\mathrm{BH}$ from the Eddington AGN+galaxy model assumes $L/L_\mathrm{Edd}=1$ and is highly uncertain and systematics-dominated. The solid star shows our measurement of the stellar mass from our fiducial AGN+Galaxy model. The circles are LRDs from \cite{Fujimoto2022}, \cite{Kokorev2023LRDz8}, \cite{Wang2024RUBIESz78MassGal}, and \cite{Kokorev2024LRDz4}. The high-$z$ AGN comparison sample are the yellow squares and are from \cite{Carnall2023}, \cite{Larson2023}, \cite{Maiolino2023}, \cite{Chisholm2024}, \cite{Juodzbalis2024}, and \cite{Maiolino2024GNz11}. The local comparison sample (gray triangles) and scaling relation (solid black line) are from \cite{Greene2020}.}
    \label{fig:mbh_mgal}
\end{figure}

We first discuss the galaxy and BH properties taking our fiducial AGN+galaxy model at face value. This composite model infers a stellar mass of $\sim4\times10^9\,\mathrm{M_\odot}$ within an extremely small effective radius of $30\,\mathrm{pc}$, comparable to the size of a star cluster. We show in the left panel of Figure~\ref{fig:mass_size_density} that this mass corresponds to $\langle\Sigma_\star\rangle\approx 10^6\,\mathrm{M_\odot\,pc^{-2}}$ and is among the densest in a sample of high-redshift systems. In this fit, all UV light originates from the galaxy; if the AGN were to contribute in the UV, the inferred stellar density would be reduced. However, although the average density seems extreme, we show on the right panel of the same figure that the central density (assuming S\'{e}rsic index $n=1$ to $n=4$) of \triplelens{} is similar to but does not exceed the high central density found in many other compact massive quiescent galaxies at $z\gtrsim5$ \citep{Carnall2023, deGraaff2024}. Compact massive quiescent galaxies have been found at lower redshift as well \citep[e.g.,][]{vanDokkum2008MassiveQ, Szomoru2012, Wright2024}. These galaxies tend to show an extended wing on scales of $\gtrsim200\,\mathrm{pc}$ in their density profile due to the high S\'ersic indices (see the $z\approx2$ curve in Figure~\ref{fig:mass_size_density} as an example) when compared with our 30\,pc-object. We may thus picture their predecessors to be objects like \triplelens{} that later grow by minor mergers in the outskirts of the systems \citep{Bezanson2009, Naab2009, vanDokkum2010, Newman2013a, Suess2023MinorMergerInQG}. This scenario is particularly plausible given that some LRDs show a preference for overdense regions \citep{Labbe2024Monster, Lin2024, Matthee2024LRDoverdensity}. We mention that observations with adaptive optics on future 20-30\,m level ground-based telescopes could potentially resolve such a compact morphology and provide more insight on the light profile of LRDs. 

Moreover, \triplelens{} also has one of the strongest breaks observed at $z\gtrsim5$ even when other LRDs are included. \cite{Weibel2024} measure a Balmer break strength of $3.33\pm0.15$ for the \triplelens, as defined by \cite{Wang2024BRD} to be the ratio of the median fluxes in rest-frame 3620--3630\,\AA{} and 4000--4100\,\AA. In order to reproduce the strong break, our model necessitates the galaxy host of \triplelens{} to be an A-star-dominated massive post-starburst galaxy with all of its stellar population forming nearly simultaneously. Such a star formation history on a star cluster size scale would require that feedback from supernova explosions does not kick in and disrupt this intense starburst. Indeed, \cite{Dekel2023} propose that massive galaxies can efficiently form via feedback-free starbursts if the metallicity is low and the gas density is high, which seems to be suitable assumptions for \triplelens{}.

We now focus on the BH in \triplelens, which we estimate to have a intrinsic luminosity of $\lambda L_{\lambda,3000}\approx 10^{44}\,\mathrm{erg\,s^{-1}}$ in our fiducial model. Taking $M_\mathrm{BH}=4\times10^7\,\mathrm{M_\odot}$ derived from the H$\alpha$ emission line \citep{Furtak2024Nature}, we show in Figure~\ref{fig:mbh_mgal} that our fiducial model yields $M_\mathrm{BH}/M_\star\approx 1\%$. This ratio is consistent with other LRDs with estimates for both masses. Like many high-$z$ AGN \citep[e.g.][]{Maiolino2023}, the BH in \triplelens{} is overmassive compared to the local population \citep{Greene2020}. Comparably extreme $M_\mathrm{BH}/M_\star$ have been observed before, even locally as we show in Figure~\ref{fig:mbh_mgal}. These extreme ratios could reflect an evolution in the intrinsic $M_\mathrm{BH}$-$M_\star$ relation, but it is also possible that these overmassive BHs in LRDs only represent the scatter in a redshift-invariant scaling relation due to selection biases and measurement uncertainties \citep{Lauer2007, Li2024MstarMBH}. The latter is especially feasible given that some LRDs do fall onto the local relation \citep{Fujimoto2022, Kokorev2023LRDz8}. Alternatively, tidal stripping in an overdense environment, which the LRDs may prefer \citep[e.g.,][]{Matthee2024LRDoverdensity} as we mentioned before, could also result in both compact morphology and an overmassive BHs \citep{Volonteri2008, Barber2016, Voggel2019}.

The high black-hole-to-galaxy mass ratio of \triplelens{} and the LRD population \citep[e.g.,][]{Kokorev2023LRDz8, Wang2024RUBIESz78MassGal} may shed light on seeding mechanisms of early BHs, especially given their high number density at $z\gtrsim4$ \citep{Pacucci2023}. Furthermore, given the compact size and the consequently implied high stellar density of \triplelens{}, seeding mechanisms in high-density environments would be relevant \citep[e.g.,][]{Devecchi&Volonteri2009, Alexander&Natarajan2014, Reinoso2023, Dayal2024}. However, we re-iterate here that considering the inability of our model to resolve issues like the dust law and the validity of local scaling relations at high redshift, the BH and galaxy masses must be taken with a grain of salt. As shown in the Eddington AGN+Galaxy fit, the AGN continuum could be entirely buried beneath the galaxy. This scenario leads to a smaller BH mass and larger galaxy mass, and would bring the mass ratio even closer to the local relation as well.

\subsection{The Role of Dust}\label{sec:dust}

\begin{figure}
    \centering
    \includegraphics[width=\columnwidth]{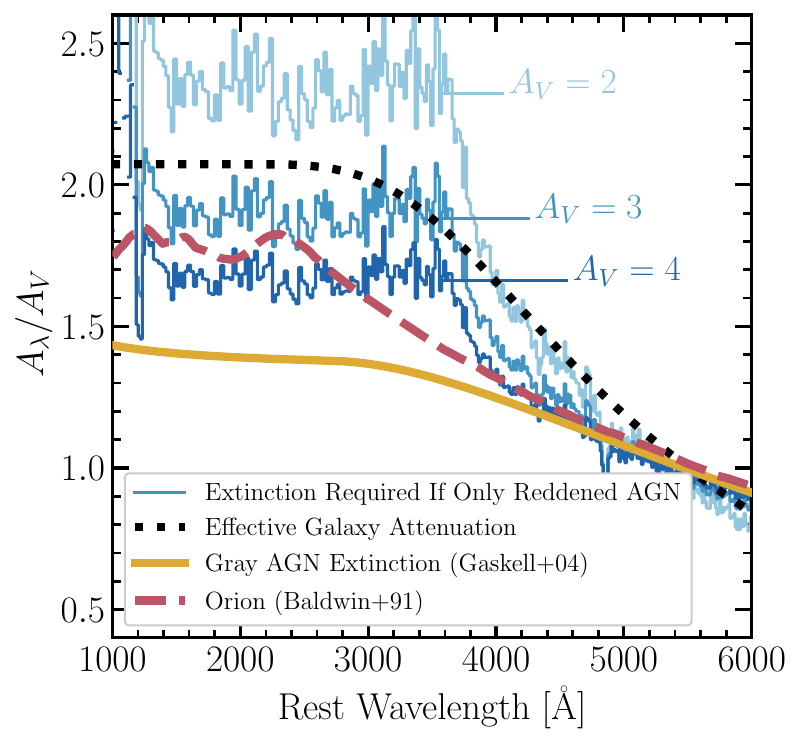}
    \caption{The empirical extinction laws (shown in blue) one would obtain if the entire UV-optical continuum of \triplelens{} were dominated by a reddened AGN power law is compared against the gray \citep{Gaskell2004} extinction curve (yellow solid curve) and the extinction curve found in the Orion Nebula \citep[red dashed curve;][]{Baldwin1991}. The different shades of blue represent different levels of assumed extinction. The black dotted curve is the effective galaxy attenuation curve already shown in Figure~\ref{fig:dust}.} 
    \label{fig:only_agn_dust}
\end{figure} 

In addition to the high stellar density of \triplelens, another uncomfortable component of the fiducial model is the necessity for both the AGN and galaxy to have a steep dust law beyond the SMC one, as shown in Figure~\ref{fig:dust} and discussed in Section~\ref{sec:agn_only} and \ref{sec:agn_galaxy}. We stress here that the UV flattening in the effective galaxy attenuation curve in Figure~\ref{fig:dust} only reflects our choice to model the UV emission as unobscured stellar light; if we were to use the AGN scattered light to describe the UV, this effective attenuation law would not show such a flattening behavior. Regardless how we model the UV, the required dust law redward of the break, which is independent of this imposed star-dust geometry, is inferred to be quite steep as well. This steepness is strictly required such that the summation of the reddened galaxy/AGN component in the optical and the unobscured component in the UV preserves the strong break. We first discuss the steepness under the context of our modeling framework, and then discuss whether any other dust extinction laws beyond our modeling scheme could explain the spectral shape of \triplelens. 

\textcolor{black}{In all the models, the inferred dust extinction curve has an unusually steep UV-to-optical slope of $A_{1500}/A_V\approx9$ (or $R_V\approx1.8$), if not steeper given our limited dust slope prior, although similarly steep dust laws have been observed at lower redshift \citep{Cikota2018, Markov2024}. In contrast, at comparable redshifts $z\approx6-7$ as \triplelens, \cite{Markov2024} also find that the dust laws of star-forming galaxies tend to have a shallower slope of $A_{1500}/A_V\approx2-3$.} A steep dust extinction curve could be due to a size distribution tilted towards small grains. Na\"{i}vely, our finding is consistent with the simple picture that lower metallicity environments lead to steeper dust laws, echoing the weak metal emission lines in the observed spectrum. Since dust grains tend to be smaller when metallicity is low, they lead to more extinction at shorter wavelengths and hence steepen the extinction curve, as we observe in the SMC \citep{Gordon2003, Dubois2024}. In addition, large grains may be preferentially destroyed by AGN shocks \citep{DraineSalpeter1979, Jones1994}, \textcolor{black}{resulting in the steepening of the extinction law in AGN environments \citep[e.g.,][]{Hall2002, Jiang2013}.} Given the compact size of \triplelens, it is not hard for shocks from a few supernovae or the AGN itself, if any, to shape the dust properties of the entire system. \textcolor{black}{The compactness and the presence of an AGN could potentially be the reasons for the inconsistency between our inferred steep extinction laws in LRDs and the shallower dust extinction laws in other high-redshift galaxies mentioned before. In the latter case, dust sputtering could just take place more gradually on timescales of 0.5-1\,Gyr \citep{Langeroodi2024, Markov2024} without the AGN shocking and efficiently destroying the larger grains}. Thus, the steep dust law could be explained physically by the dust size distribution \textcolor{black}{favoring small grains.} 

It is also true that dust extinction laws with entirely different shapes beyond our parameterization do exist. For instance, gray extinction curves, which are usually flat at short wavelength \citep[e.g.,][]{Maiolino2001, Czerny2004, Gaskell2004} usually found in composite AGN spectra have been invoked to explain the V-shaped SED of LRDs \citep{Killi2023, Li2024}. \cite{Killi2023} are able to model the NIRSpec/PRISM spectrum of a $z\approx4.5$ LRD using the \cite{Gaskell2004} extinction curve, yielding an optical continuum dominated by a reddened AGN and a star-forming galaxy responsible for the UV flux. Seeking a more general model to explain the SED shape from NIRCam and MIRI photometry, \cite{Li2024} also utilize a gray extinction curve similar to that found in the Orion nebula \citep{Baldwin1991}, using the absence of small grains to explain the V-shaped SED and the flattening of the NIR flux. With this extinction curve, the authors claim that the UV-optical continuum of LRDs is entirely dominated by a dust-reddened AGN. However, we show in Figure~\ref{fig:best_fit_models}a that the gray \citep{Gaskell2004} extinction law entirely misses the shape of the strong break in \triplelens{}. Moreover, we show in Figure~\ref{fig:only_agn_dust} the dust extinction laws one would empirically infer from the data if the observed spectrum of \triplelens{} where entirely dominated by a reddened AGN power-law at different assumed values of $A_V$. The inferred three-sloped dust extinction curve would not resemble the two-sloped gray extinction laws like the \cite{Gaskell2004} and the Orion ones due to the break feature. \textcolor{black}{In contrast, the effective galaxy attenuation law shown in both Figure~\ref{fig:dust} and Figure~\ref{fig:only_agn_dust} does resemble a gray dust law despite its steeper slope in the optical and higher UV attenuation than the \cite{Gaskell2004} and Orion curves \textcolor{black}{as proposed by \cite{Li2024}}. However, we reiterate that the gray shape of the attenuation law is purely due to our assumed star-dust geometry where the UV is produced by unobscured stellar light. The underlying extinction law is much steeper than any gray extinction laws presented.} Therefore, we do not consider that gray extinction laws to be sufficient to explain the V-shape of \triplelens{} and its strong break simultaneously. 

Just as with the BH and galaxy masses, the dust law also has its own uncertainty --- this may mark our ignorance of the dust properties in the LRDs and/or the early universe. The difficulty for our model to resolve the steepness of the dust law is also suggestive of the incompleteness of our model. Uncertainties in fitting the dust laws are exacerbated by our ignorance about the underlying continuum shape. However, although modeling photometric SEDs could be successful with a pre-chosen extinction curve, it is when spectroscopic data becomes available that modeling the break, particularly its strength, becomes challenging as is the case for \triplelens{} and other LRDs with similarly strong breaks \citep{Labbe2024Monster, Wang2024BRD}. 

To further understand dust properties in \triplelens{} and the larger LRD population, it is crucial to obtain deeper and redder data from ALMA. Currently, ALMA observations of \triplelens{} yield no significant detection in the 1.2~mm continuum as expected from the 20-30\,K cold dust heated by stellar light, with a $3\sigma$ upper limit of 0.1\,mJy \citep{Labbe2023uncoverLRD, Fujimoto2023uncoverALMA, Furtak2024Nature} --- the prediction from our galaxy model exceeds this limit. Given the compact size of \triplelens{}, hotter dust would reconcile the nondetection by shifting the peak of the dust thermal emission to shorter wavelengths \citep{Casey2024}. If deep observations covering rest-frame sub-millimeter wavelengths are available, we may be able to constrain the dust properties such as their temperature as well as the dust extinction expectations.

\subsection{Alternative Models}\label{sec:other_models}

All of our discussion above is predicated on the assumption that some combination of the power-law AGN+galaxy+attenuation model can describe the truth. \textcolor{black}{We note that these models are not perfect, either. Putting aside the resulting high stellar density from these models, the residuals in Figure~\ref{fig:best_fit_models} also show that all four models still consistently produce weaker Balmer breaks than than the observed one even when the dust extinction curves are required to be unusually steep. Plausibly, we should acknowledge the fact that we may not understand the physical origin of the spectral shape of \triplelens{}, and more exotic models may be invoked. }

Under our scheme, the strong break necessitates an A-star-dominated stellar population, and the broad line luminosity and line width requires an AGN to photoionize the circumnuclear gas. Having AGN continuum explicitly present in the model also avoids the line equivalent width from reaching infinity (relative to the AGN continuum). As suggested by \cite{Maiolino2024LRDGeometry}, the BLR clouds could be very compact and dense with a large covering factor around the accretion disk. Thus, it is possible that only the disk continuum is obscured, and broad line emission gets out. In this case, we effectively have a galaxy-dominated continuum and the AGN produces broad lines with high equivalent widths. However, as noted by \cite{KokuboHarikane2024}, this geometry is rather contrived and does not match the high number density of LRDs based on variability studies of Sloan quasars \citep{Sesar2007}. 

On this note, we re-explore the galaxy-only model, assuming an AGN is entirely absent. Particularly, \cite{Baggen2024BLnotAGN} propose that the broad, high-luminosity Balmer emission lines present in LRDs' spectrum may be achieved solely by stellar populations without an AGN. The authors point out that the broad line width may simply reflect the velocity dispersion of the ionized gas tracing the stellar distribution in such a compact dense system, and the presence of an AGN may not be needed. Indeed, if we take the 30\,pc-upper limit on the effective radius and the stellar mass derived from our galaxy-only model at face value, we would obtain a lower limit of $\mathrm{FWHM} \gtrsim 1400\,\mathrm{km\,s^{-1}}$ for the emission line width, consistent with the observed value for \triplelens{} \citep{Furtak2024Nature, Greene2024}. Such high velocity dispersion may also provides an explanation for the lack of observed absorption features at the break region as the PRISM resolution could smear out these broad absorption lines, although observational noise and emission line infilling could already explain the smoothness of the observed spectrum as we mention in Section~\ref{sec:galaxy_only}. Without photoionization from the AGN, the high $L_\mathrm{H\alpha}$ may also be achieved with shocks or collisional excitation in the potentially high-density gas due to the compact size of \triplelens{} \citep[and references therein]{Draine1993, Stasinska&Izotov2001, Baggen2024BLnotAGN}. Future modeling of these mechanisms is needed to confirm their viability. Moreover, a top-heavy initial mass function resulting in a stellar population favoring massive stars could provide sufficient ionizing photons \citep[and references therein]{Schaerer2024a, vanDokkum&Conroy2024, Baggen2024BLnotAGN}. Massive stars stripped of their envelopes in binaries may also provide sufficiently hard ionizing spectrum to reproduce the strong emission lines without significantly changing the SED shape redder than the Lyman break \citep{Gotberg2020}. 

It is also possible that the entire UV-optical continuum is dominated by an exotic AGN whose SED shape is not well characterized --- this is particularly plausible given the potential high Eddington ratio of \triplelens{} and the entire LRD population. For instance, \cite{Thompson2005} provide a disk solution with accretion rate near the Eddington limit and whose disk scale height drops at the location of $T\approx10^4\,\mathrm{K}$ because of an opacity gap from hydrogen recombination at that temperature. Naturally, this scenario would result in a two-piece continuum (with appropriate dust reddening) because the inner and outer disk have different structures. Also, the strong break could be produced by continuous Balmer absorption by dense gas near the disk photosphere as well \citep{Inayoshi&Maiolino2024} --- this is analogous to the star-like disk SED proposed by \cite{LaorDavis2011}. Nonetheless, detailed modeling of the AGN SEDs with high Eddington ratios, similar to those down by \cite{Kubota&Done2019}, is required in the future in the context of LRDs. 

Regardless of the exact shape of the intrinsic SED, if the continuum emission from \triplelens{} is dominated by an AGN, it should show signatures of variability on timescales of months to years in the rest frame. For sources like \triplelens{} that has three (or generally, multiple) gravitationally lensed images, every single epoch of observation corresponds to three (multiple) epochs in the rest-frame, making the light curve sampling much more efficient \citep{Golubchik2024}. If any variability signature is detected, this would confirm the AGN nature of \triplelens{} and enable reverberation mapping with \textit{JWST} to obtain a more accurate and precise BH mass measurement. Furthermore, high-SNR medium- or high-resolution spectroscopy may reveal absorption lines in the break, thus confirming the stellar origin of the break \citep[e.g.,][]{Kokorev2024LRDz4} without seeking for any non-traditional models to explain the full continuum shape.

\section{Summary}

In this work, we conduct spectral decomposition of the NIRSpec/PRISM spectrum of the triply imaged \triplelens, one of the first LRDs discovered by JWST \citep{Furtak2023, Furtak2024Nature} with a strong break feature at rest-frame $3600$-$4000$\,\AA. Thanks to the strong magnification due to gravitational lensing of the foreground galaxy cluster, the spectrum we obtain has a high SNR of $\sim20$ per pixel, and its spectral shape can be confirmed by both the broad band and medium band photometry. The consistency of \triplelens{} being point-like across all three lensed images in all photometric bands allows us to set a firm size upper limit of $r_\mathrm{e}<30\,\mathrm{pc}$ for \triplelens. In this light, we attempt to model the UV-optical SED and describe the peculiar spectral shape of \triplelens\ with a combination of AGN and stellar populations. Our fiducial model includes a composite of galaxy light that dominates the continuum and a significantly reddened AGN component that is entirely sub-dominant. However, the need to reproduce the strong break strength and the bright emission lines leads us to a specific (contrived) combination of physical parameters, including an uncomfortably high stellar density and unusually steep extinction law. Therefore, we conclude that we do not find any model, including the fiducial one, to be fully satisfactory, and the true nature of \triplelens{} remains unconfirmed. 

Our analysis demonstrates that while the observed spectral shape can be well reproduced by our models, many challenges to physically interpret the models are still present. In addition to the difference in the details of model setups, more generally, it is indeed true that we embark on a different approach from previous attempts of LRD modeling \citep[e.g.,][]{Wang2024BRD, Wang2024RUBIESz78MassGal} by trusting the shape of the continuum measurement given the high SNR. However, these emergent challenges to physically interpret the models are consistent across multiple approaches despite their differences in approach. Thus, we encourage both theoretical modeling and deep long-wavelength follow-up observations for \triplelens{} (and the larger LRD population) to unveil their enigmatic nature.

\section*{Acknowledgements}

This work is based in part on observations made with the NASA/ESA/CSA \textit{James Webb Space Telescope}. The data presented in this work was obtained from the Mikulski Archive for Space Telescopes (MAST) at the Space Telescope Science Institute (STScI), which is operated by the Association of Universities for Research in Astronomy, Inc., under NASA contract NAS 5-03127 for \textit{JWST}. These observations are associated with \textit{JWST} Cycle 1 GO program \#2561 and Cycle 2 GO program \#4111. The specific observations analyzed can be accessed at \dataset[10.17909/8k5c-xr27]{\doi{10.17909/8k5c-xr27}}. Support for program JWST-GO2561 was provided by NASA through a grant from the STScI, which is operated by the Associations of Universities for Research in Astronomy, Incorporated, under NASA contract NAS5-26555.

P.D. acknowledges support from the NWO grant 016.VIDI.189.162 (``ODIN") and warmly thanks the European Commission's and University of Groningen's CO-FUND Rosalind Franklin program. 

K.G. acknowledges support from Australian Research Council Laureate Fellowship FL180100060.

A.Z. acknowledges support by grant No.~2020750 from the United States-Israel Binational Science Foundation (BSF) and grant No.~2109066 from the United States National Science Foundation (NSF), by the Israel Science Foundation Grant No.~864/23, and by the Ministry of Science \& Technology, Israel. 

The work of CCW is supported by NOIRLab, which is managed by the Association of Universities for Research in Astronomy (AURA) under a cooperative agreement with the National Science Foundation.  

\appendix
\section{Posterior Distributions of the MultiNest Modeling}
\restartappendixnumbering

In this appendix, we show the parameters' posterior distributions output from MultiNest for the four different models presented in Section~\ref{sec:analysis} and Figure~\ref{fig:best_fit_models}. 

\begin{figure}[h]
    \centering
    \includegraphics[width=0.5\columnwidth]{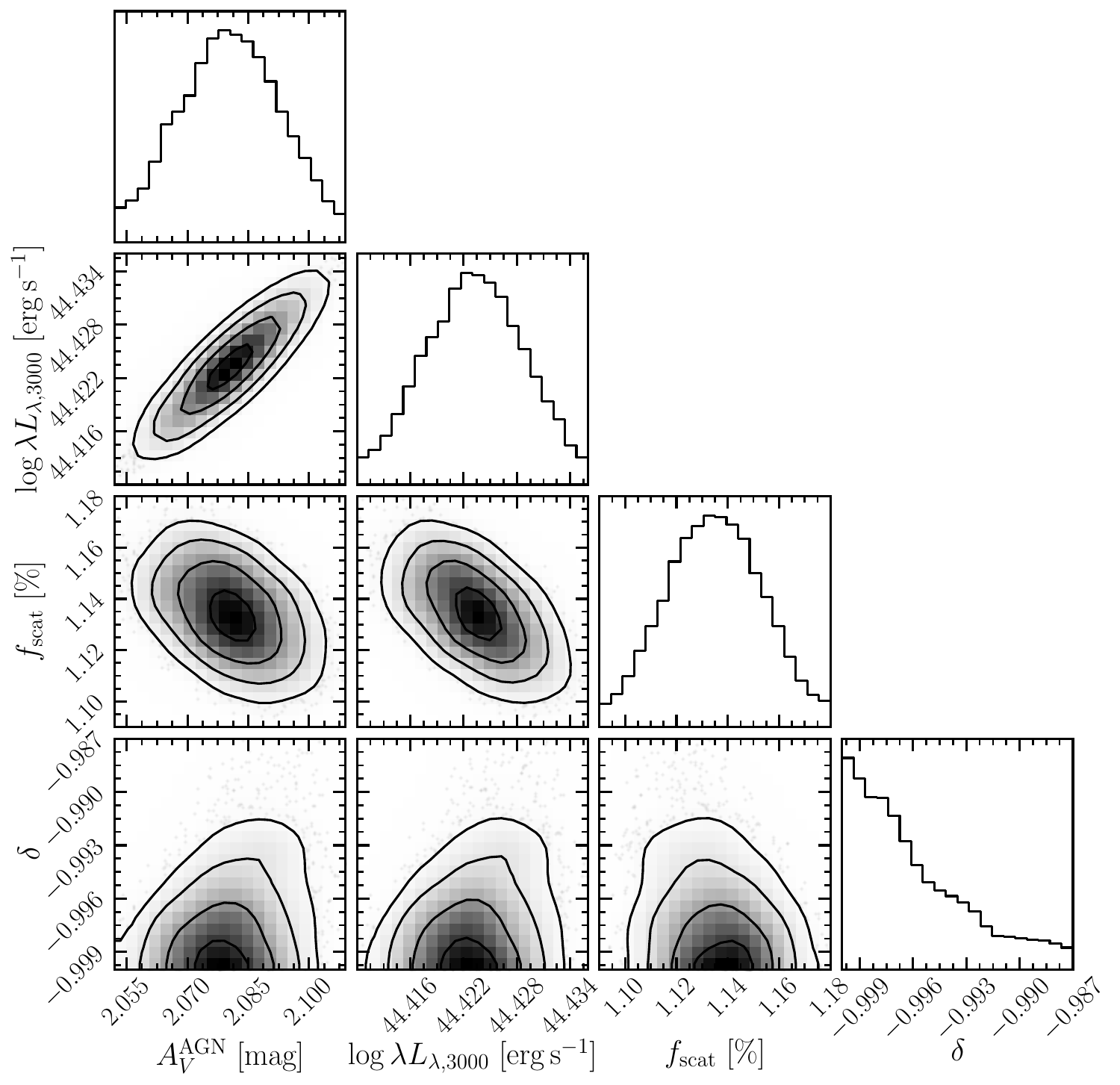}
    \caption{The posterior distributions for the four parameters involved in the AGN-only model.} 
    \label{fig:corner_agn_only}
\end{figure} 

\begin{figure}[h]
    \centering
    \includegraphics[width=0.7\columnwidth]{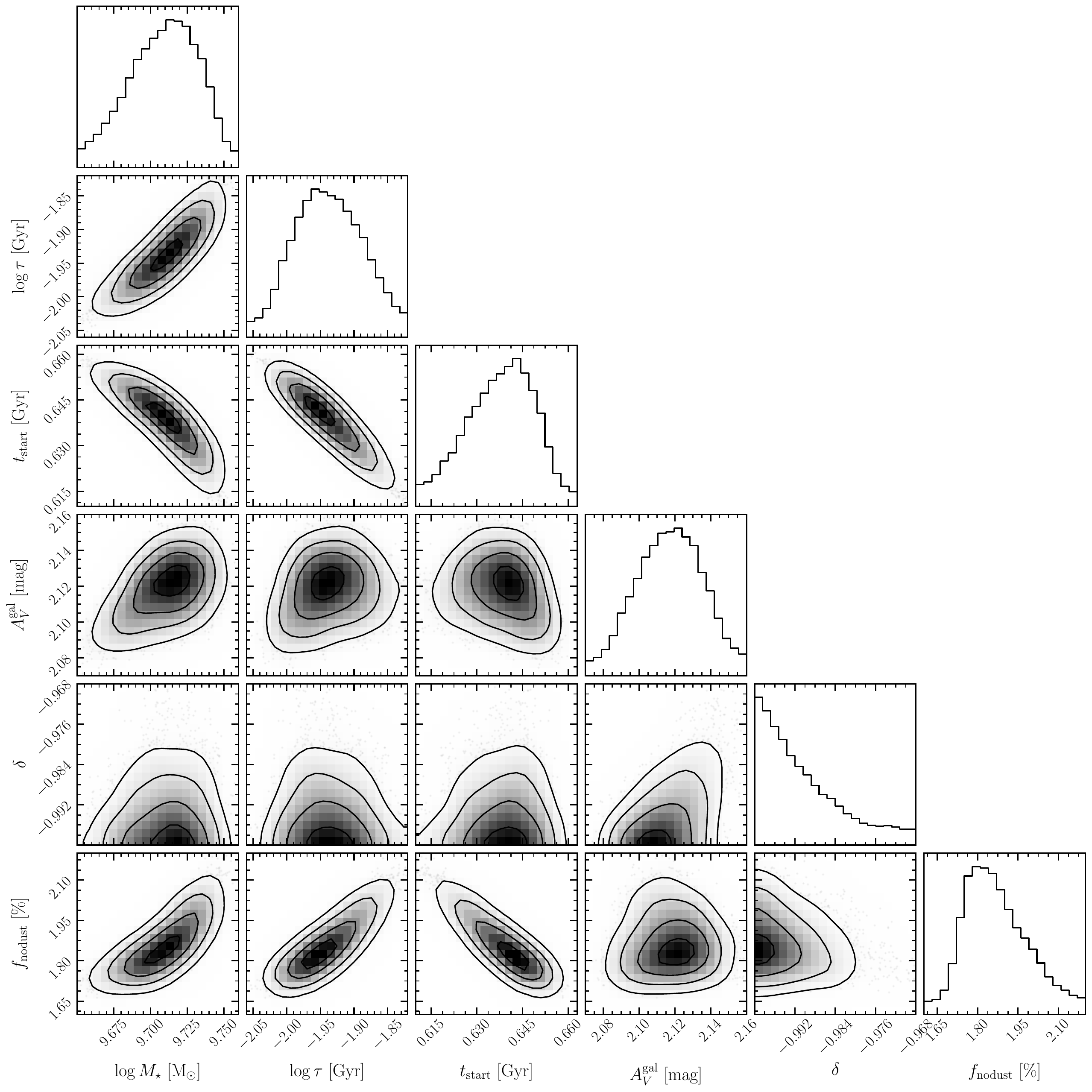}
    \caption{The posterior distributions for the four parameters involved in the galaxy-only model.} 
    \label{fig:corner_galaxy_only}
\end{figure} 

\begin{figure}[h]
    \centering
    \includegraphics[width=0.8\columnwidth]{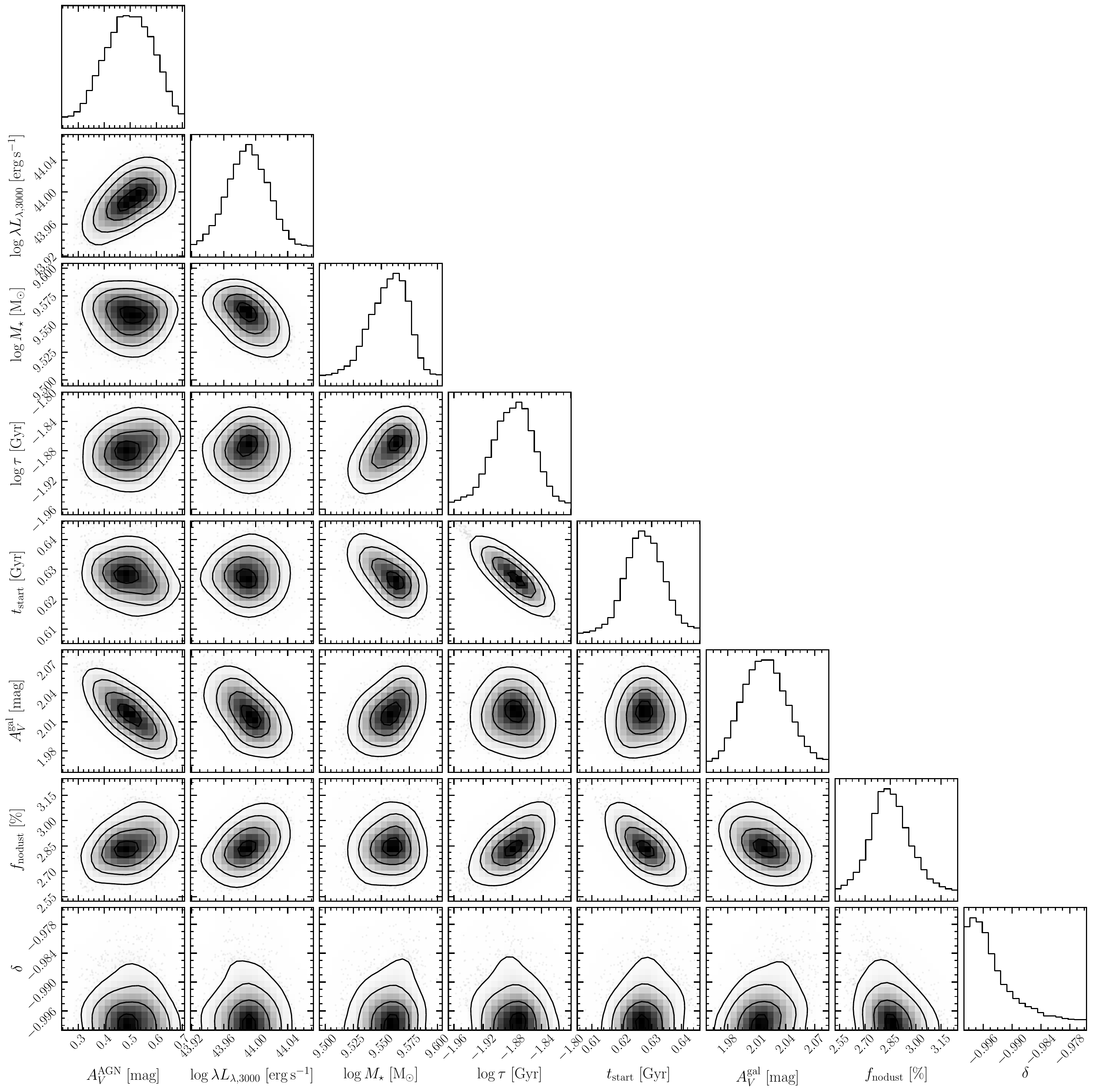}
    \caption{The posterior distributions for the four parameters involved in the AGN+Galaxy model.} 
    \label{fig:corner_agn_galaxy}
\end{figure}

\begin{figure}[h]
    \centering
    \includegraphics[width=0.8\columnwidth]{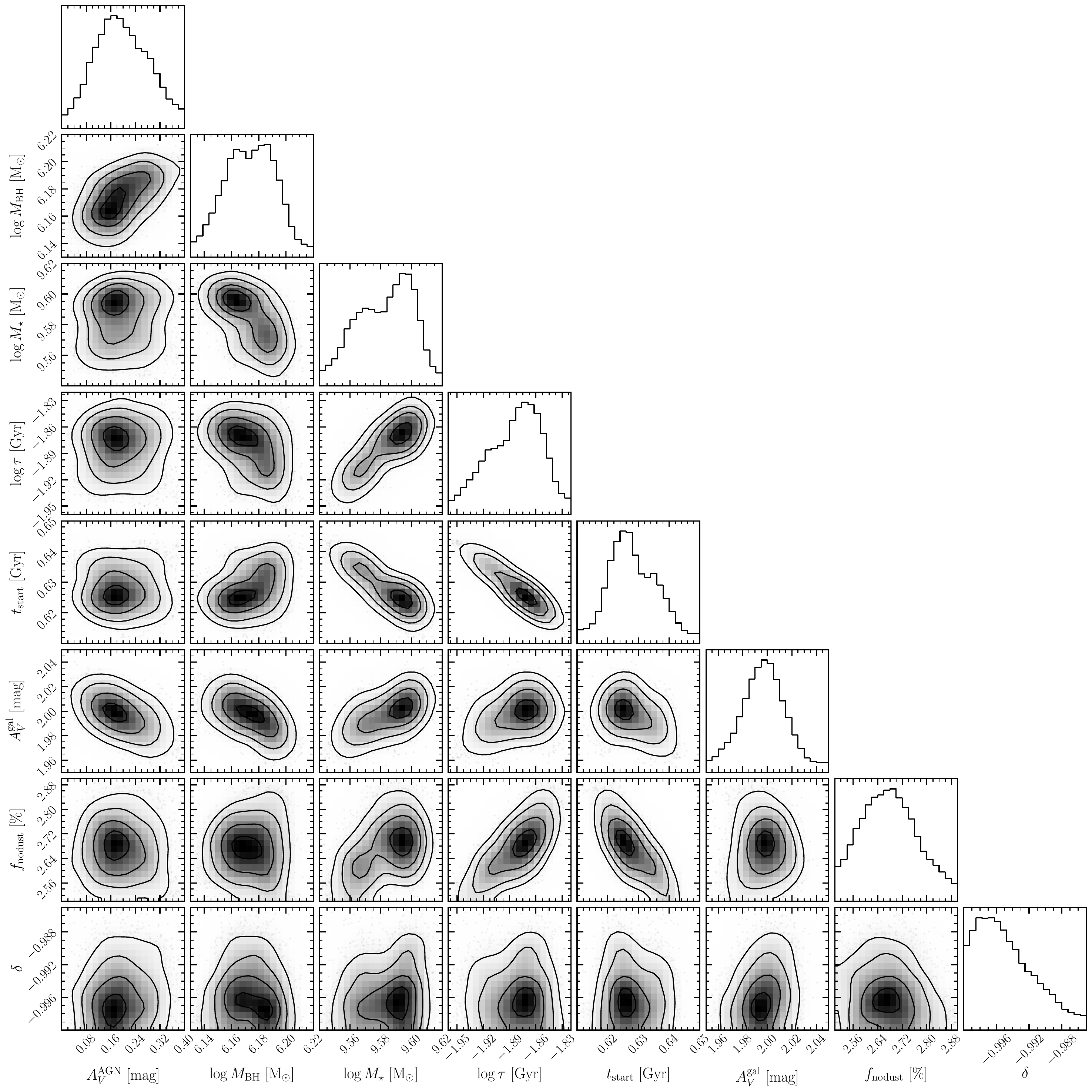}
    \caption{The posterior distributions for the four parameters involved in the Eddington AGN+Galaxy model.} 
    \label{fig:corner_agn_galaxy_edd}
\end{figure}

\bibliography{ref}{}

\begin{thebibliography}{}
\expandafter\ifx\csname natexlab\endcsname\relax\def\natexlab#1{#1}\fi
\providecommand{\url}[1]{\href{#1}{#1}}
\providecommand{\dodoi}[1]{doi:~\href{http://doi.org/#1}{\nolinkurl{#1}}}
\providecommand{\doeprint}[1]{\href{http://ascl.net/#1}{\nolinkurl{http://ascl.net/#1}}}
\providecommand{\doarXiv}[1]{\href{https://arxiv.org/abs/#1}{\nolinkurl{https://arxiv.org/abs/#1}}}

\bibitem[{{Akins} {et~al.}(2024){Akins}, {Casey}, {Lambrides}, {Allen}, {Andika}, {Brinch}, {Champagne}, {Cooper}, {Ding}, {Drakos}, {Faisst}, {Finkelstein}, {Franco}, {Fujimoto}, {Gentile}, {Gillman}, {Gozaliasl}, {Harish}, {Hayward}, {Hirschmann}, {Ilbert}, {Kartaltepe}, {Kocevski}, {Koekemoer}, {Kokorev}, {Liu}, {Long}, {McCracken}, {McKinney}, {Onoue}, {Paquereau}, {Renzini}, {Rhodes}, {Robertson}, {Shuntov}, {Silverman}, {Tanaka}, {Toft}, {Trakhtenbrot}, {Valentino}, \& {Zavala}}]{Akins2024}
{Akins}, H.~B., {Casey}, C.~M., {Lambrides}, E., {et~al.} 2024, arXiv e-prints, arXiv:2406.10341, \dodoi{10.48550/arXiv.2406.10341}

\bibitem[{{Alexander} \& {Natarajan}(2014)}]{Alexander&Natarajan2014}
{Alexander}, T., \& {Natarajan}, P. 2014, Science, 345, 1330, \dodoi{10.1126/science.1251053}

\bibitem[{{Ananna} {et~al.}(2024){Ananna}, {Bogd{\'a}n}, {Kov{\'a}cs}, {Natarajan}, \& {Hickox}}]{Ananna2024}
{Ananna}, T.~T., {Bogd{\'a}n}, {\'A}., {Kov{\'a}cs}, O.~E., {Natarajan}, P., \& {Hickox}, R.~C. 2024, arXiv e-prints, arXiv:2404.19010, \dodoi{10.48550/arXiv.2404.19010}

\bibitem[{{Baggen} {et~al.}(2023){Baggen}, {van Dokkum}, {Labb{\'e}}, {Brammer}, {Miller}, {Bezanson}, {Leja}, {Wang}, {Whitaker}, {Suess}, \& {Nelson}}]{Baggen2023}
{Baggen}, J. F.~W., {van Dokkum}, P., {Labb{\'e}}, I., {et~al.} 2023, \apjl, 955, L12, \dodoi{10.3847/2041-8213/acf5ef}

\bibitem[{{Baggen} {et~al.}(2024){Baggen}, {van Dokkum}, {Brammer}, {de Graaff}, {Franx}, {Greene}, {Labb{\'e}}, {Leja}, {Maseda}, {Nelson}, {Rix}, {Wang}, \& {Weibel}}]{Baggen2024BLnotAGN}
{Baggen}, J. F.~W., {van Dokkum}, P., {Brammer}, G., {et~al.} 2024, arXiv e-prints, arXiv:2408.07745, \dodoi{10.48550/arXiv.2408.07745}

\bibitem[{{Baldwin} {et~al.}(1991){Baldwin}, {Ferland}, {Martin}, {Corbin}, {Cota}, {Peterson}, \& {Slettebak}}]{Baldwin1991}
{Baldwin}, J.~A., {Ferland}, G.~J., {Martin}, P.~G., {et~al.} 1991, \apj, 374, 580, \dodoi{10.1086/170146}

\bibitem[{{Banerji} {et~al.}(2015){Banerji}, {Alaghband-Zadeh}, {Hewett}, \& {McMahon}}]{Banerji2015}
{Banerji}, M., {Alaghband-Zadeh}, S., {Hewett}, P.~C., \& {McMahon}, R.~G. 2015, \mnras, 447, 3368, \dodoi{10.1093/mnras/stu2649}

\bibitem[{{Barber} {et~al.}(2016){Barber}, {Schaye}, {Bower}, {Crain}, {Schaller}, \& {Theuns}}]{Barber2016}
{Barber}, C., {Schaye}, J., {Bower}, R.~G., {et~al.} 2016, \mnras, 460, 1147, \dodoi{10.1093/mnras/stw1018}

\bibitem[{{Barro} {et~al.}(2024){Barro}, {P{\'e}rez-Gonz{\'a}lez}, {Kocevski}, {McGrath}, {Trump}, {Simons}, {Somerville}, {Yung}, {Arrabal Haro}, {Akins}, {Bagley}, {Cleri}, {Costantin}, {Davis}, {Dickinson}, {Finkelstein}, {Giavalisco}, {G{\'o}mez-Guijarro}, {Hathi}, {Hirschmann}, {Holwerda}, {Huertas-Company}, {Kartaltepe}, {Koekemoer}, {Lucas}, {Papovich}, {Pirzkal}, {Seill{\'e}}, {Tacchella}, {Wuyts}, {Wilkins}, {de la Vega}, {Yang}, \& {Zavala}}]{Barro2024}
{Barro}, G., {P{\'e}rez-Gonz{\'a}lez}, P.~G., {Kocevski}, D.~D., {et~al.} 2024, \apj, 963, 128, \dodoi{10.3847/1538-4357/ad167e}

\bibitem[{{Begelman}(1979)}]{Begelman1979}
{Begelman}, M.~C. 1979, \mnras, 187, 237, \dodoi{10.1093/mnras/187.2.237}

\bibitem[{{Bezanson} {et~al.}(2009){Bezanson}, {van Dokkum}, {Tal}, {Marchesini}, {Kriek}, {Franx}, \& {Coppi}}]{Bezanson2009}
{Bezanson}, R., {van Dokkum}, P.~G., {Tal}, T., {et~al.} 2009, \apj, 697, 1290, \dodoi{10.1088/0004-637X/697/2/1290}

\bibitem[{{Bezanson} {et~al.}(2022){Bezanson}, {Labbe}, {Whitaker}, {Leja}, {Price}, {Franx}, {Brammer}, {Marchesini}, {Zitrin}, {Wang}, {Weaver}, {Furtak}, {Atek}, {Coe}, {Cutler}, {Dayal}, {van Dokkum}, {Feldmann}, {Forster Schreiber}, {Fujimoto}, {Geha}, {Glazebrook}, {de Graaff}, {Greene}, {Juneau}, {Kassin}, {Kriek}, {Khullar}, {Maseda}, {Mowla}, {Muzzin}, {Nanayakkara}, {Nelson}, {Oesch}, {Pacifici}, {Pan}, {Papovich}, {Setton}, {Shapley}, {Smit}, {Stefanon}, {Taylor}, \& {Williams}}]{Bezanson2022uncover}
{Bezanson}, R., {Labbe}, I., {Whitaker}, K.~E., {et~al.} 2022, arXiv e-prints, arXiv:2212.04026, \dodoi{10.48550/arXiv.2212.04026}

\bibitem[{{Bouwens} {et~al.}(2015){Bouwens}, {Illingworth}, {Oesch}, {Trenti}, {Labb{\'e}}, {Bradley}, {Carollo}, {van Dokkum}, {Gonzalez}, {Holwerda}, {Franx}, {Spitler}, {Smit}, \& {Magee}}]{Bouwens2015}
{Bouwens}, R.~J., {Illingworth}, G.~D., {Oesch}, P.~A., {et~al.} 2015, \apj, 803, 34, \dodoi{10.1088/0004-637X/803/1/34}

\bibitem[{{Brammer}(2023{\natexlab{a}})}]{Brammer2023a_msaexp}
{Brammer}, G. 2023{\natexlab{a}}, {msaexp: NIRSpec analyis tools}, 0.6.17,  Zenodo, \dodoi{10.5281/zenodo.7299500}

\bibitem[{{Brammer}(2023{\natexlab{b}})}]{Brammer2023b_grizli}
---. 2023{\natexlab{b}}, {grizli}, 1.9.11,  Zenodo, \dodoi{10.5281/zenodo.1146904}

\bibitem[{{Buchner} {et~al.}(2014){Buchner}, {Georgakakis}, {Nandra}, {Hsu}, {Rangel}, {Brightman}, {Merloni}, {Salvato}, {Donley}, \& {Kocevski}}]{Buchner2014}
{Buchner}, J., {Georgakakis}, A., {Nandra}, K., {et~al.} 2014, \aap, 564, A125, \dodoi{10.1051/0004-6361/201322971}

\bibitem[{{Bushouse} {et~al.}(2024){Bushouse}, {Eisenhamer}, {Dencheva}, {Davies}, {Greenfield}, {Morrison}, {Hodge}, {Simon}, {Grumm}, {Droettboom}, {Slavich}, {Sosey}, {Pauly}, {Miller}, {Jedrzejewski}, {Hack}, {Davis}, {Crawford}, {Law}, {Gordon}, {Regan}, {Cara}, {MacDonald}, {Bradley}, {Shanahan}, {Jamieson}, {Teodoro}, {Williams}, \& {Pena-Guerrero}}]{Bushouse2024JWSTpipeline}
{Bushouse}, H., {Eisenhamer}, J., {Dencheva}, N., {et~al.} 2024, {JWST Calibration Pipeline}, 1.14.0,  Zenodo, \dodoi{10.5281/zenodo.10870758}

\bibitem[{{Byler} {et~al.}(2017){Byler}, {Dalcanton}, {Conroy}, \& {Johnson}}]{Byler2017}
{Byler}, N., {Dalcanton}, J.~J., {Conroy}, C., \& {Johnson}, B.~D. 2017, \apj, 840, 44, \dodoi{10.3847/1538-4357/aa6c66}

\bibitem[{{Calzetti} {et~al.}(2000){Calzetti}, {Armus}, {Bohlin}, {Kinney}, {Koornneef}, \& {Storchi-Bergmann}}]{Calzetti2000}
{Calzetti}, D., {Armus}, L., {Bohlin}, R.~C., {et~al.} 2000, \apj, 533, 682, \dodoi{10.1086/308692}

\bibitem[{{Cameron} {et~al.}(2024){Cameron}, {Katz}, {Witten}, {Saxena}, {Laporte}, \& {Bunker}}]{Cameron2024}
{Cameron}, A.~J., {Katz}, H., {Witten}, C., {et~al.} 2024, \mnras, \dodoi{10.1093/mnras/stae1547}

\bibitem[{{Carnall} {et~al.}(2023){Carnall}, {McLure}, {Dunlop}, {McLeod}, {Wild}, {Cullen}, {Magee}, {Begley}, {Cimatti}, {Donnan}, {Hamadouche}, {Jewell}, \& {Walker}}]{Carnall2023}
{Carnall}, A.~C., {McLure}, R.~J., {Dunlop}, J.~S., {et~al.} 2023, \nat, 619, 716, \dodoi{10.1038/s41586-023-06158-6}

\bibitem[{{Casey} {et~al.}(2024){Casey}, {Akins}, {Kokorev}, {McKinney}, {Cooper}, {Long}, {Franco}, \& {Manning}}]{Casey2024}
{Casey}, C.~M., {Akins}, H.~B., {Kokorev}, V., {et~al.} 2024, arXiv e-prints, arXiv:2407.05094, \dodoi{10.48550/arXiv.2407.05094}

\bibitem[{{Chabrier}(2003)}]{Chabrier2003}
{Chabrier}, G. 2003, \pasp, 115, 763, \dodoi{10.1086/376392}

\bibitem[{{Chisholm} {et~al.}(2024){Chisholm}, {Berg}, {Endsley}, {Gazagnes}, {Richardson}, {Lambrides}, {Greene}, {Finkelstein}, {Flury}, {Guseva}, {Henry}, {Hutchison}, {Izotov}, {Marques-Chaves}, {Oesch}, {Papovich}, {Saldana-Lopez}, {Schaerer}, \& {Stephenson}}]{Chisholm2024}
{Chisholm}, J., {Berg}, D.~A., {Endsley}, R., {et~al.} 2024, arXiv e-prints, arXiv:2402.18643, \dodoi{10.48550/arXiv.2402.18643}

\bibitem[{{Choi} {et~al.}(2016){Choi}, {Dotter}, {Conroy}, {Cantiello}, {Paxton}, \& {Johnson}}]{Choi2016}
{Choi}, J., {Dotter}, A., {Conroy}, C., {et~al.} 2016, \apj, 823, 102, \dodoi{10.3847/0004-637X/823/2/102}

\bibitem[{{Cikota} {et~al.}(2018){Cikota}, {Hoang}, {Taubenberger}, {Patat}, {Mazzei}, {Cox}, {Zelaya}, {Cikota}, {Tomasella}, {Benetti}, \& {Rodeghiero}}]{Cikota2018}
{Cikota}, A., {Hoang}, T., {Taubenberger}, S., {et~al.} 2018, \aap, 615, A42, \dodoi{10.1051/0004-6361/201731395}

\bibitem[{{Conroy} \& {Gunn}(2010)}]{Conroy2010}
{Conroy}, C., \& {Gunn}, J.~E. 2010, \apj, 712, 833, \dodoi{10.1088/0004-637X/712/2/833}

\bibitem[{{Conroy} {et~al.}(2009){Conroy}, {Gunn}, \& {White}}]{Conroy2009}
{Conroy}, C., {Gunn}, J.~E., \& {White}, M. 2009, \apj, 699, 486, \dodoi{10.1088/0004-637X/699/1/486}

\bibitem[{{Czerny} {et~al.}(2004){Czerny}, {Li}, {Loska}, \& {Szczerba}}]{Czerny2004}
{Czerny}, B., {Li}, J., {Loska}, Z., \& {Szczerba}, R. 2004, \mnras, 348, L54, \dodoi{10.1111/j.1365-2966.2004.07590.x}

\bibitem[{{Dayal}(2024)}]{Dayal2024}
{Dayal}, P. 2024, arXiv e-prints, arXiv:2407.07162, \dodoi{10.48550/arXiv.2407.07162}

\bibitem[{{de Graaff} {et~al.}(2024{\natexlab{a}}){de Graaff}, {Rix}, {Carniani}, {Suess}, {Charlot}, {Curtis-Lake}, {Arribas}, {Baker}, {Boyett}, {Bunker}, {Cameron}, {Chevallard}, {Curti}, {Eisenstein}, {Franx}, {Hainline}, {Hausen}, {Ji}, {Johnson}, {Jones}, {Maiolino}, {Maseda}, {Nelson}, {Parlanti}, {Rawle}, {Robertson}, {Tacchella}, {{\"U}bler}, {Williams}, {Willmer}, \& {Willott}}]{deGraaff2024LSF1.3}
{de Graaff}, A., {Rix}, H.-W., {Carniani}, S., {et~al.} 2024{\natexlab{a}}, \aap, 684, A87, \dodoi{10.1051/0004-6361/202347755}

\bibitem[{{de Graaff} {et~al.}(2024{\natexlab{b}}){de Graaff}, {Setton}, {Brammer}, {Cutler}, {Suess}, {Labbe}, {Leja}, {Weibel}, {Maseda}, {Whitaker}, {Bezanson}, {Boogaard}, {Cleri}, {De Lucia}, {Franx}, {Greene}, {Hirschmann}, {Matthee}, {McConachie}, {Naidu}, {Oesch}, {Price}, {Rix}, {Valentino}, {Wang}, \& {Williams}}]{deGraaff2024}
{de Graaff}, A., {Setton}, D.~J., {Brammer}, G., {et~al.} 2024{\natexlab{b}}, arXiv e-prints, arXiv:2404.05683, \dodoi{10.48550/arXiv.2404.05683}

\bibitem[{{Dekel} {et~al.}(2023){Dekel}, {Sarkar}, {Birnboim}, {Mandelker}, \& {Li}}]{Dekel2023}
{Dekel}, A., {Sarkar}, K.~C., {Birnboim}, Y., {Mandelker}, N., \& {Li}, Z. 2023, \mnras, 523, 3201, \dodoi{10.1093/mnras/stad1557}

\bibitem[{{Devecchi} \& {Volonteri}(2009)}]{Devecchi&Volonteri2009}
{Devecchi}, B., \& {Volonteri}, M. 2009, \apj, 694, 302, \dodoi{10.1088/0004-637X/694/1/302}

\bibitem[{{Dotter}(2016)}]{Dotter2016}
{Dotter}, A. 2016, \apjs, 222, 8, \dodoi{10.3847/0067-0049/222/1/8}

\bibitem[{{Draine} \& {McKee}(1993)}]{Draine1993}
{Draine}, B.~T., \& {McKee}, C.~F. 1993, \araa, 31, 373, \dodoi{10.1146/annurev.aa.31.090193.002105}

\bibitem[{{Draine} \& {Salpeter}(1979)}]{DraineSalpeter1979}
{Draine}, B.~T., \& {Salpeter}, E.~E. 1979, \apj, 231, 438, \dodoi{10.1086/157206}

\bibitem[{{Dubois} {et~al.}(2024){Dubois}, {Rodr{\'\i}guez Montero}, {Guerra}, {Trebitsch}, {Han}, {Beckmann}, {Yi}, {Lewis}, \& {Jang}}]{Dubois2024}
{Dubois}, Y., {Rodr{\'\i}guez Montero}, F., {Guerra}, C., {et~al.} 2024, \aap, 687, A240, \dodoi{10.1051/0004-6361/202449784}

\bibitem[{{Falc{\'o}n-Barroso} {et~al.}(2011){Falc{\'o}n-Barroso}, {S{\'a}nchez-Bl{\'a}zquez}, {Vazdekis}, {Ricciardelli}, {Cardiel}, {Cenarro}, {Gorgas}, \& {Peletier}}]{Falcon-Barroso2011}
{Falc{\'o}n-Barroso}, J., {S{\'a}nchez-Bl{\'a}zquez}, P., {Vazdekis}, A., {et~al.} 2011, \aap, 532, A95, \dodoi{10.1051/0004-6361/201116842}

\bibitem[{{Fan} {et~al.}(2023){Fan}, {Ba{\~n}ados}, \& {Simcoe}}]{Fan2023}
{Fan}, X., {Ba{\~n}ados}, E., \& {Simcoe}, R.~A. 2023, \araa, 61, 373, \dodoi{10.1146/annurev-astro-052920-102455}

\bibitem[{{Fan} {et~al.}(2001){Fan}, {Narayanan}, {Lupton}, {Strauss}, {Knapp}, {Becker}, {White}, {Pentericci}, {Leggett}, {Haiman}, {Gunn}, {Ivezi{\'c}}, {Schneider}, {Anderson}, {Brinkmann}, {Bahcall}, {Connolly}, {Csabai}, {Doi}, {Fukugita}, {Geballe}, {Grebel}, {Harbeck}, {Hennessy}, {Lamb}, {Miknaitis}, {Munn}, {Nichol}, {Okamura}, {Pier}, {Prada}, {Richards}, {Szalay}, \& {York}}]{Fan2001}
{Fan}, X., {Narayanan}, V.~K., {Lupton}, R.~H., {et~al.} 2001, \aj, 122, 2833, \dodoi{10.1086/324111}

\bibitem[{{Feroz} \& {Hobson}(2008)}]{Feroz2008}
{Feroz}, F., \& {Hobson}, M.~P. 2008, \mnras, 384, 449, \dodoi{10.1111/j.1365-2966.2007.12353.x}

\bibitem[{{Feroz} {et~al.}(2009){Feroz}, {Hobson}, \& {Bridges}}]{Feroz2009}
{Feroz}, F., {Hobson}, M.~P., \& {Bridges}, M. 2009, \mnras, 398, 1601, \dodoi{10.1111/j.1365-2966.2009.14548.x}

\bibitem[{{Feroz} {et~al.}(2019){Feroz}, {Hobson}, {Cameron}, \& {Pettitt}}]{Feroz2019}
{Feroz}, F., {Hobson}, M.~P., {Cameron}, E., \& {Pettitt}, A.~N. 2019, The Open Journal of Astrophysics, 2, 10, \dodoi{10.21105/astro.1306.2144}

\bibitem[{{Fujimoto} {et~al.}(2022){Fujimoto}, {Brammer}, {Watson}, {Magdis}, {Kokorev}, {Greve}, {Toft}, {Walter}, {Valiante}, {Ginolfi}, {Schneider}, {Valentino}, {Colina}, {Vestergaard}, {Marques-Chaves}, {Fynbo}, {Krips}, {Steinhardt}, {Cortzen}, {Rizzo}, \& {Oesch}}]{Fujimoto2022}
{Fujimoto}, S., {Brammer}, G.~B., {Watson}, D., {et~al.} 2022, \nat, 604, 261, \dodoi{10.1038/s41586-022-04454-1}

\bibitem[{{Fujimoto} {et~al.}(2023){Fujimoto}, {Bezanson}, {Labbe}, {Brammer}, {Price}, {Wang}, {Weaver}, {Fudamoto}, {Oesch}, {Williams}, {Dayal}, {Feldmann}, {Greene}, {Leja}, {Whitaker}, {Zitrin}, {Cutler}, {Furtak}, {Pan}, {Chemerynska}, {Kokorev}, {Miller}, {Atek}, {van Dokkum}, {Juneau}, {Kassin}, {Khullar}, {Marchesini}, {Maseda}, {Nelson}, {Setton}, \& {Smit}}]{Fujimoto2023uncoverALMA}
{Fujimoto}, S., {Bezanson}, R., {Labbe}, I., {et~al.} 2023, arXiv e-prints, arXiv:2309.07834, \dodoi{10.48550/arXiv.2309.07834}

\bibitem[{{Furtak} {et~al.}(2023{\natexlab{a}}){Furtak}, {Zitrin}, {Plat}, {Fujimoto}, {Wang}, {Nelson}, {Labb{\'e}}, {Bezanson}, {Brammer}, {van Dokkum}, {Endsley}, {Glazebrook}, {Greene}, {Leja}, {Price}, {Smit}, {Stark}, {Weaver}, {Whitaker}, {Atek}, {Chevallard}, {Curtis-Lake}, {Dayal}, {Feltre}, {Franx}, {Fudamoto}, {Marchesini}, {Mowla}, {Pan}, {Suess}, {Vidal-Garc{\'\i}a}, \& {Williams}}]{Furtak2023}
{Furtak}, L.~J., {Zitrin}, A., {Plat}, A., {et~al.} 2023{\natexlab{a}}, \apj, 952, 142, \dodoi{10.3847/1538-4357/acdc9d}

\bibitem[{{Furtak} {et~al.}(2023{\natexlab{b}}){Furtak}, {Zitrin}, {Weaver}, {Atek}, {Bezanson}, {Labb{\'e}}, {Whitaker}, {Leja}, {Price}, {Brammer}, {Wang}, {Marchesini}, {Pan}, {Dayal}, {van Dokkum}, {Feldmann}, {Fujimoto}, {Franx}, {Khullar}, {Nelson}, \& {Mowla}}]{Furtak2023LensingModel}
{Furtak}, L.~J., {Zitrin}, A., {Weaver}, J.~R., {et~al.} 2023{\natexlab{b}}, \mnras, 523, 4568, \dodoi{10.1093/mnras/stad1627}

\bibitem[{{Furtak} {et~al.}(2024){Furtak}, {Labb{\'e}}, {Zitrin}, {Greene}, {Dayal}, {Chemerynska}, {Kokorev}, {Miller}, {Goulding}, {de Graaff}, {Bezanson}, {Brammer}, {Cutler}, {Leja}, {Pan}, {Price}, {Wang}, {Weaver}, {Whitaker}, {Atek}, {Bogd{\'a}n}, {Charlot}, {Curtis-Lake}, {van Dokkum}, {Endsley}, {Feldmann}, {Fudamoto}, {Fujimoto}, {Glazebrook}, {Juneau}, {Marchesini}, {Maseda}, {Nelson}, {Oesch}, {Plat}, {Setton}, {Stark}, \& {Williams}}]{Furtak2024Nature}
{Furtak}, L.~J., {Labb{\'e}}, I., {Zitrin}, A., {et~al.} 2024, \nat, 628, 57, \dodoi{10.1038/s41586-024-07184-8}

\bibitem[{{Gallazzi} {et~al.}(2005){Gallazzi}, {Charlot}, {Brinchmann}, {White}, \& {Tremonti}}]{Gallazzi2005}
{Gallazzi}, A., {Charlot}, S., {Brinchmann}, J., {White}, S. D.~M., \& {Tremonti}, C.~A. 2005, \mnras, 362, 41, \dodoi{10.1111/j.1365-2966.2005.09321.x}

\bibitem[{{Gardner} {et~al.}(2023){Gardner}, {Mather}, {Abbott}, {Abell}, {Abernathy}, {Abney}, {Abraham}, {Abraham}, {Abul-Huda}, {Acton}, {Adams}, {Adams}, {Adler}, {Adriaensen}, {Aguilar}, {Ahmed}, {Ahmed}, {Ahmed}, {Albat}, {Albert}, {Alberts}, {Aldridge}, {Allen}, {Allen}, {Altenburg}, {Altunc}, {Alvarez}, {{\'A}lvarez-M{\'a}rquez}, {Alves de Oliveira}, {Ambrose}, {Anandakrishnan}, {Andersen}, {Anderson}, {Anderson}, {Anderson}, {Anderson}, {Aprea}, {Archer}, {Arenberg}, {Argyriou}, {Arribas}, {Artigau}, {Arvai}, {Atcheson}, {Atkinson}, {Averbukh}, {Aymergen}, {Bacinski}, {Baggett}, {Bagnasco}, {Baker}, {Balzano}, {Banks}, {Baran}, {Barker}, {Barrett}, {Barringer}, {Barto}, {Bast}, {Baudoz}, {Baum}, {Beatty}, {Beaulieu}, {Bechtold}, {Beck}, {Beddard}, {Beichman}, {Bellagama}, {Bely}, {Berger}, {Bergeron}, {Bernier}, {Bertch}, {Beskow}, {Betz}, {Biagetti}, {Birkmann}, {Bjorklund}, {Blackwood}, {Blazek}, {Blossfeld}, {Bluth}, {Boccaletti}, {Boegner}, {Bohlin}, {Boia}, {B{\"o}ker}, {Bonaventura}, {Bond},
  {Bosley}, {Boucarut}, {Bouchet}, {Bouwman}, {Bower}, {Bowers}, {Bowers}, {Boyce}, {Boyer}, {Boyer}, {Boyer}, {Boyer}, {Bradley}, {Brady}, {Brandl}, {Brannen}, {Breda}, {Bremmer}, {Brennan}, {Bresnahan}, {Bright}, {Broiles}, {Bromenschenkel}, {Brooks}, {Brooks}, {Brown}, {Brown}, {Brown}, {Bruce}, {Bryson}, {Bujanda}, {Bullock}, {Bunker}, {Bureo}, {Burt}, {Bush}, {Bushouse}, {Bussman}, {Cabaud}, {Cale}, {Calhoon}, {Calvani}, {Canipe}, {Caputo}, {Cara}, {Carey}, {Case}, {Cesari}, {Cetorelli}, {Chance}, {Chandler}, {Chaney}, {Chapman}, {Charlot}, {Chayer}, {Cheezum}, {Chen}, {Chen}, {Cherinka}, {Chichester}, {Chilton}, {Chittiraibalan}, {Clampin}, {Clark}, {Clark}, {Clark}, {Claybrooks}, {Cleveland}, {Cohen}, {Cohen}, {Col{\'o}n}, {Coleman}, {Colina}, {Comber}, {Comeau}, {Comer}, {Conde Reis}, {Connolly}, {Conroy}, {Contos}, {Contreras}, {Cook}, {Cooper}, {Cooper}, {Correia}, {Correnti}, {Cossou}, {Costanza}, {Coulais}, {Cox}, {Coyle}, {Cracraft}, {Crew}, {Curtis}, {Cusveller}, {Da Costa Maciel}, {Dailey},
  {Daugeron}, {Davidson}, {Davies}, {Davis}, {Davis}, {Day}, {de Chambure}, {de Jong}, {De Marchi}, {Dean}, {Decker}, {Delisa}, {Dell}, {Dellagatta}, {Dembinska}, {Demosthenes}, {Dencheva}, {Deneu}, {DePriest}, {Deschenes}, {Dethienne}, {Detre}, {Diaz}, {Dicken}, {DiFelice}, {Dillman}, {Disharoon}, {Dixon}, {Doggett}, {Dominguez}, {Donaldson}, {Doria-Warner}, {Santos}, {Doty}, {Douglas}, {Doyon}, {Dressler}, {Driggers}, {Driggers}, {Dunn}, {DuPrie}, {Dupuis}, {Durning}, {Dutta}, {Earl}, {Eccleston}, {Ecobichon}, {Egami}, {Ehrenwinkler}, {Eisenhamer}, {Eisenhower}, {Eisenstein}, {El Hamel}, {Elie}, {Elliott}, {Elliott}, {Engesser}, {Espinoza}, {Etienne}, {Etxaluze}, {Evans}, {Fabreguettes}, {Falcolini}, {Falini}, {Fatig}, {Feeney}, {Feinberg}, {Fels}, {Ferdous}, {Ferguson}, {Ferrarese}, {Ferreira}, {Ferruit}, {Ferry}, {Filippazzo}, {Firre}, {Fix}, {Flagey}, {Flanagan}, {Fleming}, {Florian}, {Flynn}, {Foiadelli}, {Fontaine}, {Fontanella}, {Forshay}, {Fortner}, {Fox}, {Framarini}, {Francisco}, {Franck}, {Franx},
  {Franz}, {Friedman}, {Friend}, {Frost}, {Fu}, {Fullerton}, {Gaillard}, {Galkin}, {Gallagher}, {Galyer}, {Garc{\'\i}a Mar{\'\i}n}, {Gardner}, {Garland}, {Garrett}, {Gasman}, {G{\'a}sp{\'a}r}, {Gastaud}, {Gaudreau}, {Gauthier}, {Geers}, {Geithner}, {Gennaro}, {Gerber}, {Gereau}, {Giampaoli}, {Giardino}, {Gibbons}, {Gilbert}, {Gilman}, {Girard}, {Giuliano}, {Gkountis}, {Glasse}, {Glassmire}, {Glauser}, {Glazer}, {Goldberg}, {Golimowski}, {Gonzaga}, {Gordon}, {Gordon}, {Goudfrooij}, {Gough}, {Graham}, {Grau}, {Green}, {Greene}, {Greene}, {Greenfield}, {Greenhouse}, {Greve}, {Greville}, {Grimaldi}, {Groe}, {Groebner}, {Grumm}, {Grundy}, {G{\"u}del}, {Guillard}, {Guldalian}, {Gunn}, {Gurule}, {Gutman}, {Guy}, {Guyot}, {Hack}, {Haderlein}, {Hagan}, {Hagedorn}, {Hainline}, {Haley}, {Hami}, {Hamilton}, {Hammann}, {Hammel}, {Hanley}, {Hansen}, {Hardy}, {Harnisch}, {Harr}, {Harris}, {Hart}, {Hartig}, {Hasan}, {Hashim}, {Hashimoto}, {Haskins}, {Hawkins}, {Hayden}, {Hayden}, {Healy}, {Hecht}, {Heeg}, {Hejal}, {Helm},
  {Hengemihle}, {Henning}, {Henry}, {Henry}, {Henshaw}, {Hernandez}, {Herrington}, {Heske}, {Hesman}, {Hickey}, {Hilbert}, {Hines}, {Hinz}, {Hirsch}, {Hitcho}, {Hodapp}, {Hodge}, {Hoffman}, {Holfeltz}, {Holler}, {Hoppa}, {Horner}, {Howard}, {Howard}, {Huber}, {Hunkeler}, {Hunter}, {Hunter}, {Hurd}, {Hurst}, {Hutchings}, {Hylan}, {Ignat}, {Illingworth}, {Irish}, {Isaacs}, {Jackson}, {Jaffe}, {Jahic}, {Jahromi}, {Jakobsen}, {James}, {James}, {James}, {Jamieson}, {Jandra}, {Jayawardhana}, {Jedrzejewski}, {Jeffers}, {Jensen}, {Joanne}, {Johns}, {Johnson}, {Johnson}, {Johnson}, {Johnson}, {Johnson}, {Johnson}, {Johnstone}, {Jollet}, {Jones}, {Jones}, {Jones}, {Jones}, {Jones}, {Jordan}, {Jordan}, {Jue}, {Jurkowski}, {Justis}, {Justtanont}, {Kaleida}, {Kalirai}, {Kalmanson}, {Kaltenegger}, {Kammerer}, {Kan}, {Kanarek}, {Kao}, {Karakla}, {Karl}, {Kassin}, {Kauffman}, {Kavanagh}, {Kelley}, {Kelly}, {Kendrew}, {Kennedy}, {Kenny}, {Keski-Kuha}, {Keyes}, {Khan}, {Kidwell}, {Kimble}, {King}, {King}, {Kinzel}, {Kirk},
  {Kirkpatrick}, {Klaassen}, {Klingemann}, {Klintworth}, {Knapp}, {Knight}, {Knollenberg}, {Knutsen}, {Koehler}, {Koekemoer}, {Kofler}, {Kontson}, {Kovacs}, {Kozhurina-Platais}, {Krause}, {Kriss}, {Krist}, {Kristoffersen}, {Krogel}, {Krueger}, {Kulp}, {Kumari}, {Kwan}, {Kyprianou}, {Labador}, {Labiano}, {Lafreni{\`e}re}, {Lagage}, {Laidler}, {Laine}, {Laird}, {Lajoie}, {Lallo}, {Lam}, {LaMassa}, {Lambros}, {Lampenfield}, {Lander}, {Langston}, {Larson}, {Larson}, {LaVerghetta}, {Law}, {Lawrence}, {Lee}, {Lee}, {Lee}, {Leisenring}, {Leveille}, {Levenson}, {Levi}, {Levine}, {Lewis}, {Lewis}, {Lewis}, {Libralato}, {Lidon}, {Liebrecht}, {Lightsey}, {Lilly}, {Lim}, {Lim}, {Ling}, {Link}, {Link}, {Lipinski}, {Liu}, {Lo}, {Lobmeyer}, {Logue}, {Long}, {Long}, {Long}, {Long}, {L{\'o}pez-Caniego}, {Lotz}, {Love-Pruitt}, {Lubskiy}, {Luers}, {Luetgens}, {Luevano}, {Lui}, {Lund}, {Lundquist}, {Lunine}, {L{\"u}tzgendorf}, {Lynch}, {MacDonald}, {MacDonald}, {Macias}, {Macklis}, {Maghami}, {Maharaja}, {Maiolino},
  {Makrygiannis}, {Malla}, {Malumuth}, {Manjavacas}, {Marini}, {Marrione}, {Marston}, {Martel}, {Martin}, {Martin}, {Martinez}, {Maschmann}, {Masci}, {Masetti}, {Maszkiewicz}, {Matthews}, {Matuskey}, {McBrayer}, {McCarthy}, {McCaughrean}, {McClare}, {McClare}, {McCloskey}, {McClurg}, {McCoy}, {McElwain}, {McGregor}, {McGuffey}, {McKay}, {McKenzie}, {McLean}, {McMaster}, {McNeil}, {De Meester}, {Mehalick}, {Meixner}, {Mel{\'e}ndez}, {Menzel}, {Menzel}, {Merz}, {Mesterharm}, {Meyer}, {Meyett}, {Meza}, {Midwinter}, {Milam}, {Miller}, {Miller}, {Miskey}, {Misselt}, {Mitchell}, {Mohan}, {Montoya}, {Moran}, {Morishita}, {Moro-Mart{\'\i}n}, {Morrison}, {Morrison}, {Morse}, {Moschos}, {Moseley}, {Mosier}, {Mosner}, {Mountain}, {Muckenthaler}, {Mueller}, {Mueller}, {Muhiem}, {M{\"u}hlmann}, {Mullally}, {Mullen}, {Munger}, {Murphy}, {Murray}, {Muzerolle}, {Mycroft}, {Myers}, {Myers}, {Myers}, {Myers}, {Myrick}, {Nagle}, {Nayak}, {Naylor}, {Neff}, {Nelan}, {Nella}, {Nguyen}, {Nguyen}, {Nickson}, {Nidhiry}, {Niedner},
  {Nieto-Santisteban}, {Nikolov}, {Nishisaka}, {Noriega-Crespo}, {Nota}, {O'Mara}, {Oboryshko}, {O'Brien}, {Ochs}, {Offenberg}, {Ogle}, {Ohl}, {Olmsted}, {Osborne}, {O'Shaughnessy}, {{\"O}stlin}, {O'Sullivan}, {Otor}, {Ottens}, {Ouellette}, {Outlaw}, {Owens}, {Pacifici}, {Page}, {Paranilam}, {Park}, {Parrish}, {Paschal}, {Patapis}, {Patel}, {Patrick}, {Pattishall}, {Paul}, {Paul}, {Pauly}, {Pavlovsky}, {Pe{\~n}a-Guerrero}, {Pedder}, {Peek}, {Pelham}, {Penanen}, {Perriello}, {Perrin}, {Perrine}, {Perrygo}, {Peslier}, {Petach}, {Peterson}, {Pfarr}, {Pierson}, {Pietraszkiewicz}, {Pilchen}, {Pipher}, {Pirzkal}, {Pitman}, {Player}, {Plesha}, {Plitzke}, {Pohner}, {Poletis}, {Pollizzi}, {Polster}, {Pontius}, {Pontoppidan}, {Porges}, {Potter}, {Prescott}, {Proffitt}, {Pueyo}, {Quispe Neira}, {Radich}, {Rager}, {Rameau}, {Ramey}, {Ramos Alarcon}, {Rampini}, {Rapp}, {Rashford}, {Rauscher}, {Ravindranath}, {Rawle}, {Rawlings}, {Ray}, {Regan}, {Rehm}, {Rehm}, {Reid}, {Reis}, {Renk}, {Reoch}, {Ressler}, {Rest},
  {Reynolds}, {Richon}, {Richon}, {Ridgaway}, {Riedel}, {Rieke}, {Rieke}, {Rifelli}, {Rigby}, {Riggs}, {Ringel}, {Ritchie}, {Rix}, {Robberto}, {Robinson}, {Robinson}, {Robinson}, {Rock}, {Rodriguez}, {Rodr{\'\i}guez del Pino}, {Roellig}, {Rohrbach}, {Roman}, {Romelfanger}, {Romo}, {Rosales}, {Rose}, {Roteliuk}, {Roth}, {Rothwell}, {Rouzaud}, {Rowe}, {Rowlands}, {Roy}, {Royer}, {Rui}, {Rumler}, {Rumpl}, {Russ}, {Ryan}, {Ryan}, {Saad}, {Sabata}, {Sabatino}, {Sabbi}, {Sabelhaus}, {Sabia}, {Sahu}, {Saif}, {Salvignol}, {Samara-Ratna}, {Samuelson}, {Sanders}, {Sappington}, {Sargent}, {Sauer}, {Savadkin}, {Sawicki}, {Schappell}, {Scheffer}, {Scheithauer}, {Scherer}, {Schiff}, {Schlawin}, {Schmeitzky}, {Schmitz}, {Schmude}, {Schneider}, {Schreiber}, {Schroeven-Deceuninck}, {Schultz}, {Schwab}, {Schwartz}, {Scoccimarro}, {Scott}, {Scott}, {Seaton}, {Seely}, {Seery}, {Seidleck}, {Sembach}, {Shanahan}, {Shaughnessy}, {Shaw}, {Shay}, {Sheehan}, {Sheth}, {Shih}, {Shivaei}, {Siegel}, {Sienkiewicz}, {Simmons}, {Simon},
  {Sirianni}, {Sivaramakrishnan}, {Slade}, {Sloan}, {Slocum}, {Slowinski}, {Smith}, {Smith}, {Smith}, {Smith}, {Smith}, {Smith}, {Smolik}, {Soderblom}, {Sohn}, {Sokol}, {Sonneborn}, {Sontag}, {Sooy}, {Soummer}, {Southwood}, {Spain}, {Sparmo}, {Speer}, {Spencer}, {Sprofera}, {Stallcup}, {Stanley}, {Stansberry}, {Stark}, {Starr}, {Stassi}, {Steck}, {Steeley}, {Stephens}, {Stephenson}, {Stewart}, {Stiavelli}, {}, {Strada}, {Straughn}, {Streetman}, {Strickland}, {Strobele}, {Stuhlinger}, {Stys}, {Such}, {Sukhatme}, {Sullivan}, {Sullivan}, {Sumner}, {Sun}, {Sunnquist}, {Swade}, {Swam}, {Swenton}, {Swoish}, {Tam Litten}, {Tamas}, {Tao}, {Taylor}, {Taylor}, {te Plate}, {Van Tea}, {Teague}, {Telfer}, {Temim}, {Texter}, {Thatte}, {Thompson}, {Thompson}, {Thomson}, {Thronson}, {Tierney}, {Tikkanen}, {Tinnin}, {Tippet}, {Todd}, {Tran}, {Trauger}, {Trejo}, {Vinh Truong}, {Tsukamoto}, {Tufail}, {Tumlinson}, {Tustain}, {Tyra}, {Ubeda}, {Underwood}, {Uzzo}, {Vaclavik}, {Valenduc}, {Valenti}, {Van Campen}, {van de Wetering},
  {Van Der Marel}, {van Haarlem}, {Vandenbussche}, {van Dishoeck}, {Vanterpool}, {Vernoy}, {Vila Costas}, {Volk}, {Voorzaat}, {Voyton}, {Vydra}, {Waddy}, {Waelkens}, {Wahlgren}, {Walker}, {Wander}, {Warfield}, {Warner}, {Wasiak}, {Wasiak}, {Wehner}, {Weiler}, {Weilert}, {Weiss}, {Wells}, {Welty}, {Wheate}, {Wheeler}, {White}, {Whitehouse}, {Whiteleather}, {Whitman}, {Williams}, {Willmer}, {Willott}, {Willoughby}, {Wilson}, {Wilson}, {Wilson}, {Windhorst}, {Wislowski}, {Wolfe}, {Wolfe}, {Wolff}, {Wondel}, {Woo}, {Woods}, {Worden}, {Workman}, {Wright}, {Wu}, {Wu}, {Wun}, {Wymer}, {Yadetie}, {Yan}, {Yang}, {Yates}, {Yeager}, {Yerger}, {Young}, {Young}, {Yu}, {Yu}, {Zak}, {Zeidler}, {Zepp}, {Zhou}, {Zincke}, {Zonak}, \& {Zondag}}]{Gardner2023}
{Gardner}, J.~P., {Mather}, J.~C., {Abbott}, R., {et~al.} 2023, \pasp, 135, 068001, \dodoi{10.1088/1538-3873/acd1b5}

\bibitem[{{Gaskell} {et~al.}(2004){Gaskell}, {Goosmann}, {Antonucci}, \& {Whysong}}]{Gaskell2004}
{Gaskell}, C.~M., {Goosmann}, R.~W., {Antonucci}, R. R.~J., \& {Whysong}, D.~H. 2004, \apj, 616, 147, \dodoi{10.1086/423885}

\bibitem[{{Glikman} {et~al.}(2012){Glikman}, {Urrutia}, {Lacy}, {Djorgovski}, {Mahabal}, {Myers}, {Ross}, {Petitjean}, {Ge}, {Schneider}, \& {York}}]{Glikman2012}
{Glikman}, E., {Urrutia}, T., {Lacy}, M., {et~al.} 2012, \apj, 757, 51, \dodoi{10.1088/0004-637X/757/1/51}

\bibitem[{{Golubchik} {et~al.}(2024){Golubchik}, {Steinhardt}, {Zitrin}, {Meena}, {Furtak}, {Chelouche}, \& {Kaspi}}]{Golubchik2024}
{Golubchik}, M., {Steinhardt}, C.~L., {Zitrin}, A., {et~al.} 2024, arXiv e-prints, arXiv:2408.00073, \dodoi{10.48550/arXiv.2408.00073}

\bibitem[{{Gordon} {et~al.}(2003){Gordon}, {Clayton}, {Misselt}, {Landolt}, \& {Wolff}}]{Gordon2003}
{Gordon}, K.~D., {Clayton}, G.~C., {Misselt}, K.~A., {Landolt}, A.~U., \& {Wolff}, M.~J. 2003, \apj, 594, 279, \dodoi{10.1086/376774}

\bibitem[{{G{\"o}tberg} {et~al.}(2020){G{\"o}tberg}, {de Mink}, {McQuinn}, {Zapartas}, {Groh}, \& {Norman}}]{Gotberg2020}
{G{\"o}tberg}, Y., {de Mink}, S.~E., {McQuinn}, M., {et~al.} 2020, \aap, 634, A134, \dodoi{10.1051/0004-6361/201936669}

\bibitem[{{Goulding} {et~al.}(2023){Goulding}, {Greene}, {Setton}, {Labbe}, {Bezanson}, {Miller}, {Atek}, {Bogd{\'a}n}, {Brammer}, {Chemerynska}, {Cutler}, {Dayal}, {Fudamoto}, {Fujimoto}, {Furtak}, {Kokorev}, {Khullar}, {Leja}, {Marchesini}, {Natarajan}, {Nelson}, {Oesch}, {Pan}, {Papovich}, {Price}, {van Dokkum}, {Wang}, {Weaver}, {Whitaker}, \& {Zitrin}}]{Goulding2023}
{Goulding}, A.~D., {Greene}, J.~E., {Setton}, D.~J., {et~al.} 2023, \apjl, 955, L24, \dodoi{10.3847/2041-8213/acf7c5}

\bibitem[{{Greene} {et~al.}(2020){Greene}, {Strader}, \& {Ho}}]{Greene2020}
{Greene}, J.~E., {Strader}, J., \& {Ho}, L.~C. 2020, \araa, 58, 257, \dodoi{10.1146/annurev-astro-032620-021835}

\bibitem[{{Greene} {et~al.}(2024){Greene}, {Labbe}, {Goulding}, {Furtak}, {Chemerynska}, {Kokorev}, {Dayal}, {Volonteri}, {Williams}, {Wang}, {Setton}, {Burgasser}, {Bezanson}, {Atek}, {Brammer}, {Cutler}, {Feldmann}, {Fujimoto}, {Glazebrook}, {de Graaff}, {Khullar}, {Leja}, {Marchesini}, {Maseda}, {Matthee}, {Miller}, {Naidu}, {Nanayakkara}, {Oesch}, {Pan}, {Papovich}, {Price}, {van Dokkum}, {Weaver}, {Whitaker}, \& {Zitrin}}]{Greene2024}
{Greene}, J.~E., {Labbe}, I., {Goulding}, A.~D., {et~al.} 2024, \apj, 964, 39, \dodoi{10.3847/1538-4357/ad1e5f}

\bibitem[{{Hall} {et~al.}(2002){Hall}, {Anderson}, {Strauss}, {York}, {Richards}, {Fan}, {Knapp}, {Schneider}, {Vanden Berk}, {Geballe}, {Bauer}, {Becker}, {Davis}, {Rix}, {Nichol}, {Bahcall}, {Brinkmann}, {Brunner}, {Connolly}, {Csabai}, {Doi}, {Fukugita}, {Gunn}, {Haiman}, {Harvanek}, {Heckman}, {Hennessy}, {Inada}, {Ivezi{\'c}}, {Johnston}, {Kleinman}, {Krolik}, {Krzesinski}, {Kunszt}, {Lamb}, {Long}, {Lupton}, {Miknaitis}, {Munn}, {Narayanan}, {Neilsen}, {Newman}, {Nitta}, {Okamura}, {Pentericci}, {Pier}, {Schlegel}, {Snedden}, {Szalay}, {Thakar}, {Tsvetanov}, {White}, \& {Zheng}}]{Hall2002}
{Hall}, P.~B., {Anderson}, S.~F., {Strauss}, M.~A., {et~al.} 2002, \apjs, 141, 267, \dodoi{10.1086/340546}

\bibitem[{{Harikane} {et~al.}(2023){Harikane}, {Ouchi}, {Oguri}, {Ono}, {Nakajima}, {Isobe}, {Umeda}, {Mawatari}, \& {Zhang}}]{Harikane2023}
{Harikane}, Y., {Ouchi}, M., {Oguri}, M., {et~al.} 2023, \apjs, 265, 5, \dodoi{10.3847/1538-4365/acaaa9}

\bibitem[{{Hopkins} {et~al.}(2004){Hopkins}, {Strauss}, {Hall}, {Richards}, {Cooper}, {Schneider}, {Vanden Berk}, {Jester}, {Brinkmann}, \& {Szokoly}}]{Hopkins2004}
{Hopkins}, P.~F., {Strauss}, M.~A., {Hall}, P.~B., {et~al.} 2004, \aj, 128, 1112, \dodoi{10.1086/423291}

\bibitem[{{Horne}(1986)}]{Horne1986}
{Horne}, K. 1986, \pasp, 98, 609, \dodoi{10.1086/131801}

\bibitem[{{Inayoshi} \& {Ichikawa}(2024)}]{Inayoshi&Ichikawa2024}
{Inayoshi}, K., \& {Ichikawa}, K. 2024, arXiv e-prints, arXiv:2402.14706, \dodoi{10.48550/arXiv.2402.14706}

\bibitem[{{Inayoshi} \& {Maiolino}(2024)}]{Inayoshi&Maiolino2024}
{Inayoshi}, K., \& {Maiolino}, R. 2024, arXiv e-prints, arXiv:2409.07805.
\newblock \doarXiv{2409.07805}

\bibitem[{{Jakobsen} {et~al.}(2022){Jakobsen}, {Ferruit}, {Alves de Oliveira}, {Arribas}, {Bagnasco}, {Barho}, {Beck}, {Birkmann}, {B{\"o}ker}, {Bunker}, {Charlot}, {de Jong}, {de Marchi}, {Ehrenwinkler}, {Falcolini}, {Fels}, {Franx}, {Franz}, {Funke}, {Giardino}, {Gnata}, {Holota}, {Honnen}, {Jensen}, {Jentsch}, {Johnson}, {Jollet}, {Karl}, {Kling}, {K{\"o}hler}, {Kolm}, {Kumari}, {Lander}, {Lemke}, {L{\'o}pez-Caniego}, {L{\"u}tzgendorf}, {Maiolino}, {Manjavacas}, {Marston}, {Maschmann}, {Maurer}, {Messerschmidt}, {Moseley}, {Mosner}, {Mott}, {Muzerolle}, {Pirzkal}, {Pittet}, {Plitzke}, {Posselt}, {Rapp}, {Rauscher}, {Rawle}, {Rix}, {R{\"o}del}, {Rumler}, {Sabbi}, {Salvignol}, {Schmid}, {Sirianni}, {Smith}, {Strada}, {te Plate}, {Valenti}, {Wettemann}, {Wiehe}, {Wiesmayer}, {Willott}, {Wright}, {Zeidler}, \& {Zincke}}]{Jakobsen2022}
{Jakobsen}, P., {Ferruit}, P., {Alves de Oliveira}, C., {et~al.} 2022, \aap, 661, A80, \dodoi{10.1051/0004-6361/202142663}

\bibitem[{{Jiang} {et~al.}(2013){Jiang}, {Zhou}, {Ji}, {Shu}, {Liu}, {Wang}, {Dong}, {Bai}, {Wang}, \& {Wang}}]{Jiang2013}
{Jiang}, P., {Zhou}, H., {Ji}, T., {et~al.} 2013, \aj, 145, 157, \dodoi{10.1088/0004-6256/145/6/157}

\bibitem[{{Jiang} {et~al.}(2019){Jiang}, {Stone}, \& {Davis}}]{Jiang2019}
{Jiang}, Y.-F., {Stone}, J.~M., \& {Davis}, S.~W. 2019, \apj, 880, 67, \dodoi{10.3847/1538-4357/ab29ff}

\bibitem[{{Jones} {et~al.}(1994){Jones}, {Tielens}, {Hollenbach}, \& {McKee}}]{Jones1994}
{Jones}, A.~P., {Tielens}, A.~G.~G.~M., {Hollenbach}, D.~J., \& {McKee}, C.~F. 1994, \apj, 433, 797, \dodoi{10.1086/174689}

\bibitem[{{Juod{\v{z}}balis} {et~al.}(2024){Juod{\v{z}}balis}, {Ji}, {Maiolino}, {D'Eugenio}, {Scholtz}, {Risaliti}, {Fabian}, {Mazzolari}, {Gilli}, {Prandoni}, {Arribas}, {Bunker}, {Carniani}, {Charlot}, {Curtis-Lake}, {de Graaff}, {Hainline}, {Parlanti}, {Perna}, {P{\'e}rez-Gonz{\'a}lez}, {Robertson}, {Tacchella}, {{\"U}bler}, {Williams}, {Willott}, \& {Witstok}}]{Juodzbalis2024}
{Juod{\v{z}}balis}, I., {Ji}, X., {Maiolino}, R., {et~al.} 2024, arXiv e-prints, arXiv:2407.08643, \dodoi{10.48550/arXiv.2407.08643}

\bibitem[{{Katz} {et~al.}(2024){Katz}, {Cameron}, {Saxena}, {Barrufet}, {Choustikov}, {Cleri}, {de Graaff}, {Ellis}, {Fosbury}, {Heintz}, {Maseda}, {Matthee}, {McConchie}, \& {Oesch}}]{Katz2024}
{Katz}, H., {Cameron}, A.~J., {Saxena}, A., {et~al.} 2024, arXiv e-prints, arXiv:2408.03189, \dodoi{10.48550/arXiv.2408.03189}

\bibitem[{{Killi} {et~al.}(2023){Killi}, {Watson}, {Brammer}, {McPartland}, {Antwi-Danso}, {Newshore}, {Coe}, {Allen}, {Fynbo}, {Gould}, {Heintz}, {Rusakov}, \& {Vejlgaard}}]{Killi2023}
{Killi}, M., {Watson}, D., {Brammer}, G., {et~al.} 2023, arXiv e-prints, arXiv:2312.03065, \dodoi{10.48550/arXiv.2312.03065}

\bibitem[{{King}(2024)}]{King2024}
{King}, A. 2024, \mnras, 531, 550, \dodoi{10.1093/mnras/stae1171}

\bibitem[{{Kocevski} {et~al.}(2023){Kocevski}, {Onoue}, {Inayoshi}, {Trump}, {Arrabal Haro}, {Grazian}, {Dickinson}, {Finkelstein}, {Kartaltepe}, {Hirschmann}, {Aird}, {Holwerda}, {Fujimoto}, {Juneau}, {Amor{\'\i}n}, {Backhaus}, {Bagley}, {Barro}, {Bell}, {Bisigello}, {Calabr{\`o}}, {Cleri}, {Cooper}, {Ding}, {Grogin}, {Ho}, {Hutchison}, {Inoue}, {Jiang}, {Jones}, {Koekemoer}, {Li}, {Li}, {McGrath}, {Molina}, {Papovich}, {P{\'e}rez-Gonz{\'a}lez}, {Pirzkal}, {Wilkins}, {Yang}, \& {Yung}}]{Kocevski2023}
{Kocevski}, D.~D., {Onoue}, M., {Inayoshi}, K., {et~al.} 2023, \apjl, 954, L4, \dodoi{10.3847/2041-8213/ace5a0}

\bibitem[{{Kocevski} {et~al.}(2024){Kocevski}, {Finkelstein}, {Barro}, {Taylor}, {Calabr{\`o}}, {Laloux}, {Buchner}, {Trump}, {Leung}, {Yang}, {Dickinson}, {P{\'e}rez-Gonz{\'a}lez}, {Pacucci}, {Inayoshi}, {Somerville}, {McGrath}, {Akins}, {Bagley}, {Bisigello}, {Bowler}, {Carnall}, {Casey}, {Cheng}, {Cleri}, {Costantin}, {Cullen}, {Davis}, {Donnan}, {Dunlop}, {Ellis}, {Ferguson}, {Fujimoto}, {Fontana}, {Giavalisco}, {Grazian}, {Grogin}, {Hathi}, {Hirschmann}, {Huertas-Company}, {Holwerda}, {Illingworth}, {Juneau}, {Kartaltepe}, {Koekemoer}, {Li}, {Lucas}, {Magee}, {Mason}, {McLeod}, {McLure}, {Napolitano}, {Papovich}, {Pirzkal}, {Rodighiero}, {Santini}, {Wilkins}, \& {Yung}}]{Kocevski2024}
{Kocevski}, D.~D., {Finkelstein}, S.~L., {Barro}, G., {et~al.} 2024, arXiv e-prints, arXiv:2404.03576, \dodoi{10.48550/arXiv.2404.03576}

\bibitem[{{Kokorev} {et~al.}(2023){Kokorev}, {Fujimoto}, {Labbe}, {Greene}, {Bezanson}, {Dayal}, {Nelson}, {Atek}, {Brammer}, {Caputi}, {Chemerynska}, {Cutler}, {Feldmann}, {Fudamoto}, {Furtak}, {Goulding}, {de Graaff}, {Leja}, {Marchesini}, {Miller}, {Nanayakkara}, {Oesch}, {Pan}, {Price}, {Setton}, {Smit}, {Stefanon}, {Wang}, {Weaver}, {Whitaker}, {Williams}, \& {Zitrin}}]{Kokorev2023LRDz8}
{Kokorev}, V., {Fujimoto}, S., {Labbe}, I., {et~al.} 2023, \apjl, 957, L7, \dodoi{10.3847/2041-8213/ad037a}

\bibitem[{{Kokorev} {et~al.}(2024{\natexlab{a}}){Kokorev}, {Caputi}, {Greene}, {Dayal}, {Trebitsch}, {Cutler}, {Fujimoto}, {Labb{\'e}}, {Miller}, {Iani}, {Navarro-Carrera}, \& {Rinaldi}}]{Kokorev2024PhotLRD}
{Kokorev}, V., {Caputi}, K.~I., {Greene}, J.~E., {et~al.} 2024{\natexlab{a}}, \apj, 968, 38, \dodoi{10.3847/1538-4357/ad4265}

\bibitem[{{Kokorev} {et~al.}(2024{\natexlab{b}}){Kokorev}, {Chisholm}, {Endsley}, {Finkelstein}, {Greene}, {Akins}, {Bromm}, {Casey}, {Fujimoto}, {Labb{\'e}}, \& {Larson}}]{Kokorev2024LRDz4}
{Kokorev}, V., {Chisholm}, J., {Endsley}, R., {et~al.} 2024{\natexlab{b}}, arXiv e-prints, arXiv:2407.20320, \dodoi{10.48550/arXiv.2407.20320}

\bibitem[{{Kokubo} \& {Harikane}(2024)}]{KokuboHarikane2024}
{Kokubo}, M., \& {Harikane}, Y. 2024, arXiv e-prints, arXiv:2407.04777, \dodoi{10.48550/arXiv.2407.04777}

\bibitem[{{Kriek} \& {Conroy}(2013)}]{KriekConroy2013}
{Kriek}, M., \& {Conroy}, C. 2013, \apjl, 775, L16, \dodoi{10.1088/2041-8205/775/1/L16}

\bibitem[{{Kriek} {et~al.}(2008){Kriek}, {van Dokkum}, {Franx}, {Illingworth}, {Marchesini}, {Quadri}, {Rudnick}, {Taylor}, {F{\"o}rster Schreiber}, {Gawiser}, {Labb{\'e}}, {Lira}, \& {Wuyts}}]{Kriek2008}
{Kriek}, M., {van Dokkum}, P.~G., {Franx}, M., {et~al.} 2008, \apj, 677, 219, \dodoi{10.1086/528945}

\bibitem[{{Kubota} \& {Done}(2019)}]{Kubota&Done2019}
{Kubota}, A., \& {Done}, C. 2019, \mnras, 489, 524, \dodoi{10.1093/mnras/stz2140}

\bibitem[{{Labbe} {et~al.}(2023){Labbe}, {Greene}, {Bezanson}, {Fujimoto}, {Furtak}, {Goulding}, {Matthee}, {Naidu}, {Oesch}, {Atek}, {Brammer}, {Chemerynska}, {Coe}, {Cutler}, {Dayal}, {Feldmann}, {Franx}, {Glazebrook}, {Leja}, {Marchesini}, {Maseda}, {Nanayakkara}, {Nelson}, {Pan}, {Papovich}, {Price}, {Suess}, {Wang}, {Whitaker}, {Williams}, \& {Zitrin}}]{Labbe2023uncoverLRD}
{Labbe}, I., {Greene}, J.~E., {Bezanson}, R., {et~al.} 2023, arXiv e-prints, arXiv:2306.07320, \dodoi{10.48550/arXiv.2306.07320}

\bibitem[{{Labbe} {et~al.}(2024){Labbe}, {Greene}, {Matthee}, {Treiber}, {Kokorev}, {Miller}, {Kramarenko}, {Setton}, {Ma}, {Goulding}, {Bezanson}, {Naidu}, {Williams}, {Atek}, {Brammer}, {Cutler}, {Chemerynska}, {Cloonan}, {Dayal}, {de Graaff}, {Fudamoto}, {Fujimoto}, {Furtak}, {Glazebrook}, {Heintz}, {Leja}, {Marchesini}, {Nanayakkara}, {Nelson}, {Oesch}, {Pan}, {Price}, {Shivaei}, {Sobral}, {Suess}, {van Dokkum}, {Wang}, {Weaver}, {Whitaker}, \& {Zitrin}}]{Labbe2024Monster}
{Labbe}, I., {Greene}, J.~E., {Matthee}, J., {et~al.} 2024, arXiv e-prints, arXiv:2412.04557.
\newblock \doarXiv{2412.04557}

\bibitem[{{Lambrides} {et~al.}(2024){Lambrides}, {Garofali}, {Larson}, {Ptak}, {Chiaberge}, {Long}, {Hutchison}, {Norman}, {McKinney}, {Akins}, {Berg}, {Chisholm}, {Civano}, {Cloonan}, {Endsley}, {Faisst}, {Gilli}, {Gillman}, {Hirschmann}, {Kartaltepe}, {Kocevski}, {Kokorev}, {Pacucci}, {Richardson}, {Stiavelli}, \& {Whalen}}]{Lambrides2024}
{Lambrides}, E., {Garofali}, K., {Larson}, R., {et~al.} 2024, arXiv e-prints, arXiv:2409.13047, \dodoi{10.48550/arXiv.2409.13047}

\bibitem[{{Langeroodi} {et~al.}(2024){Langeroodi}, {Hjorth}, {Ferrara}, \& {Gall}}]{Langeroodi2024}
{Langeroodi}, D., {Hjorth}, J., {Ferrara}, A., \& {Gall}, C. 2024, arXiv e-prints, arXiv:2410.14671, \dodoi{10.48550/arXiv.2410.14671}

\bibitem[{{Laor} \& {Davis}(2011)}]{LaorDavis2011}
{Laor}, A., \& {Davis}, S.~W. 2011, \mnras, 417, 681, \dodoi{10.1111/j.1365-2966.2011.19310.x}

\bibitem[{{Larson} {et~al.}(2023){Larson}, {Finkelstein}, {Kocevski}, {Hutchison}, {Trump}, {Arrabal Haro}, {Bromm}, {Cleri}, {Dickinson}, {Fujimoto}, {Kartaltepe}, {Koekemoer}, {Papovich}, {Pirzkal}, {Tacchella}, {Zavala}, {Bagley}, {Behroozi}, {Champagne}, {Cole}, {Jung}, {Morales}, {Yang}, {Zhang}, {Zitrin}, {Amor{\'\i}n}, {Burgarella}, {Casey}, {Ch{\'a}vez Ortiz}, {Cox}, {Chworowsky}, {Fontana}, {Gawiser}, {Grazian}, {Grogin}, {Harish}, {Hathi}, {Hirschmann}, {Holwerda}, {Juneau}, {Leung}, {Lucas}, {McGrath}, {P{\'e}rez-Gonz{\'a}lez}, {Rigby}, {Seill{\'e}}, {Simons}, {de La Vega}, {Weiner}, {Wilkins}, {Yung}, \& {Ceers Team}}]{Larson2023}
{Larson}, R.~L., {Finkelstein}, S.~L., {Kocevski}, D.~D., {et~al.} 2023, \apjl, 953, L29, \dodoi{10.3847/2041-8213/ace619}

\bibitem[{{Lauer} {et~al.}(2007){Lauer}, {Tremaine}, {Richstone}, \& {Faber}}]{Lauer2007}
{Lauer}, T.~R., {Tremaine}, S., {Richstone}, D., \& {Faber}, S.~M. 2007, \apj, 670, 249, \dodoi{10.1086/522083}

\bibitem[{{Li} {et~al.}(2024{\natexlab{a}}){Li}, {Silverman}, {Shen}, {Volonteri}, {Jahnke}, {Zhuang}, {Scoggins}, {Ding}, {Harikane}, {Onoue}, \& {Tanaka}}]{Li2024MstarMBH}
{Li}, J., {Silverman}, J.~D., {Shen}, Y., {et~al.} 2024{\natexlab{a}}, arXiv e-prints, arXiv:2403.00074, \dodoi{10.48550/arXiv.2403.00074}

\bibitem[{{Li} {et~al.}(2024{\natexlab{b}}){Li}, {Inayoshi}, {Chen}, {Ichikawa}, \& {Ho}}]{Li2024}
{Li}, Z., {Inayoshi}, K., {Chen}, K., {Ichikawa}, K., \& {Ho}, L.~C. 2024{\natexlab{b}}, arXiv e-prints, arXiv:2407.10760, \dodoi{10.48550/arXiv.2407.10760}

\bibitem[{{Lin} {et~al.}(2024){Lin}, {Wang}, {Fan}, {Cai}, {Champagne}, {Sun}, {Volonteri}, {Yang}, {Hennawi}, {Ba{\~n}ados}, {Barth}, {Eilers}, {Farina}, {Liu}, {Jin}, {Jun}, {Lupi}, {Kakiichi}, {Mazzucchelli}, {Onoue}, {Pan}, {Pizzati}, {Rojas-Ruiz}, {Schindler}, {Trakhtenbrot}, {Shen}, {Trebitsch}, {Zhuang}, {Endsley}, {Meyer}, {Li}, {Li}, {Pudoka}, {Tee}, {Wu}, \& {Zhang}}]{Lin2024}
{Lin}, X., {Wang}, F., {Fan}, X., {et~al.} 2024, arXiv e-prints, arXiv:2407.17570, \dodoi{10.48550/arXiv.2407.17570}

\bibitem[{{Lotz} {et~al.}(2017){Lotz}, {Koekemoer}, {Coe}, {Grogin}, {Capak}, {Mack}, {Anderson}, {Avila}, {Barker}, {Borncamp}, {Brammer}, {Durbin}, {Gunning}, {Hilbert}, {Jenkner}, {Khandrika}, {Levay}, {Lucas}, {MacKenty}, {Ogaz}, {Porterfield}, {Reid}, {Robberto}, {Royle}, {Smith}, {Storrie-Lombardi}, {Sunnquist}, {Surace}, {Taylor}, {Williams}, {Bullock}, {Dickinson}, {Finkelstein}, {Natarajan}, {Richard}, {Robertson}, {Tumlinson}, {Zitrin}, {Flanagan}, {Sembach}, {Soifer}, \& {Mountain}}]{Lotz2017}
{Lotz}, J.~M., {Koekemoer}, A., {Coe}, D., {et~al.} 2017, \apj, 837, 97, \dodoi{10.3847/1538-4357/837/1/97}

\bibitem[{{Lupi} {et~al.}(2024){Lupi}, {Trinca}, {Volonteri}, {Dotti}, \& {Mazzucchelli}}]{Lupi2024}
{Lupi}, A., {Trinca}, A., {Volonteri}, M., {Dotti}, M., \& {Mazzucchelli}, C. 2024, arXiv e-prints, arXiv:2406.17847, \dodoi{10.48550/arXiv.2406.17847}

\bibitem[{{Ma} {et~al.}(2024){Ma}, {Goulding}, {Greene}, {Zakamska}, {Wylezalek}, \& {Jiang}}]{Ma2024}
{Ma}, Y., {Goulding}, A., {Greene}, J.~E., {et~al.} 2024, arXiv e-prints, arXiv:2401.04177, \dodoi{10.48550/arXiv.2401.04177}

\bibitem[{{Maiolino} {et~al.}(2001){Maiolino}, {Marconi}, {Salvati}, {Risaliti}, {Severgnini}, {Oliva}, {La Franca}, \& {Vanzi}}]{Maiolino2001}
{Maiolino}, R., {Marconi}, A., {Salvati}, M., {et~al.} 2001, \aap, 365, 28, \dodoi{10.1051/0004-6361:20000177}

\bibitem[{{Maiolino} {et~al.}(2023){Maiolino}, {Scholtz}, {Curtis-Lake}, {Carniani}, {Baker}, {de Graaff}, {Tacchella}, {{\"U}bler}, {D'Eugenio}, {Witstok}, {Curti}, {Arribas}, {Bunker}, {Charlot}, {Chevallard}, {Eisenstein}, {Egami}, {Ji}, {Jones}, {Lyu}, {Rawle}, {Robertson}, {Rujopakarn}, {Perna}, {Sun}, {Venturi}, {Williams}, \& {Willott}}]{Maiolino2023}
{Maiolino}, R., {Scholtz}, J., {Curtis-Lake}, E., {et~al.} 2023, arXiv e-prints, arXiv:2308.01230, \dodoi{10.48550/arXiv.2308.01230}

\bibitem[{{Maiolino} {et~al.}(2024{\natexlab{a}}){Maiolino}, {Risaliti}, {Signorini}, {Trefoloni}, {Juodzbalis}, {Scholtz}, {Uebler}, {D'Eugenio}, {Carniani}, {Fabian}, {Ji}, {Mazzolari}, {Bertola}, {Brusa}, {Bunker}, {Charlot}, {Comastri}, {Cresci}, {DeCoursey}, {Egami}, {Fiore}, {Gilli}, {Perna}, {Tacchella}, \& {Venturi}}]{Maiolino2024LRDGeometry}
{Maiolino}, R., {Risaliti}, G., {Signorini}, M., {et~al.} 2024{\natexlab{a}}, arXiv e-prints, arXiv:2405.00504, \dodoi{10.48550/arXiv.2405.00504}

\bibitem[{{Maiolino} {et~al.}(2024{\natexlab{b}}){Maiolino}, {Scholtz}, {Witstok}, {Carniani}, {D'Eugenio}, {de Graaff}, {{\"U}bler}, {Tacchella}, {Curtis-Lake}, {Arribas}, {Bunker}, {Charlot}, {Chevallard}, {Curti}, {Looser}, {Maseda}, {Rawle}, {Rodr{\'\i}guez del Pino}, {Willott}, {Egami}, {Eisenstein}, {Hainline}, {Robertson}, {Williams}, {Willmer}, {Baker}, {Boyett}, {DeCoursey}, {Fabian}, {Helton}, {Ji}, {Jones}, {Kumari}, {Laporte}, {Nelson}, {Perna}, {Sandles}, {Shivaei}, \& {Sun}}]{Maiolino2024GNz11}
{Maiolino}, R., {Scholtz}, J., {Witstok}, J., {et~al.} 2024{\natexlab{b}}, \nat, 627, 59, \dodoi{10.1038/s41586-024-07052-5}

\bibitem[{{Markov} {et~al.}(2024){Markov}, {Gallerani}, {Ferrara}, {Pallottini}, {Parlanti}, {Di Mascia}, {Sommovigo}, \& {Kohandel}}]{Markov2024}
{Markov}, V., {Gallerani}, S., {Ferrara}, A., {et~al.} 2024, arXiv e-prints, arXiv:2402.05996, \dodoi{10.48550/arXiv.2402.05996}

\bibitem[{{Matthee} {et~al.}(2024{\natexlab{a}}){Matthee}, {Naidu}, {Brammer}, {Chisholm}, {Eilers}, {Goulding}, {Greene}, {Kashino}, {Labbe}, {Lilly}, {Mackenzie}, {Oesch}, {Weibel}, {Wuyts}, {Xiao}, {Bordoloi}, {Bouwens}, {van Dokkum}, {Illingworth}, {Kramarenko}, {Maseda}, {Mason}, {Meyer}, {Nelson}, {Reddy}, {Shivaei}, {Simcoe}, \& {Yue}}]{Matthee2024}
{Matthee}, J., {Naidu}, R.~P., {Brammer}, G., {et~al.} 2024{\natexlab{a}}, \apj, 963, 129, \dodoi{10.3847/1538-4357/ad2345}

\bibitem[{{Matthee} {et~al.}(2024{\natexlab{b}}){Matthee}, {Naidu}, {Kotiwale}, {Furtak}, {Kramarenko}, {Mackenzie}, {Greene}, {Adamo}, {Bouwens}, {Di Cesare}, {Eilers}, {de Graaff}, {Heintz}, {Kashino}, {Maseda}, {Tacchella}, \& {Torralba}}]{Matthee2024LRDoverdensity}
{Matthee}, J., {Naidu}, R.~P., {Kotiwale}, G., {et~al.} 2024{\natexlab{b}}, arXiv e-prints, arXiv:2412.02846, \dodoi{10.48550/arXiv.2412.02846}

\bibitem[{{Naab} {et~al.}(2009){Naab}, {Johansson}, \& {Ostriker}}]{Naab2009}
{Naab}, T., {Johansson}, P.~H., \& {Ostriker}, J.~P. 2009, \apjl, 699, L178, \dodoi{10.1088/0004-637X/699/2/L178}

\bibitem[{{Narayanan} {et~al.}(2018){Narayanan}, {Conroy}, {Dav{\'e}}, {Johnson}, \& {Popping}}]{Narayanan2018a}
{Narayanan}, D., {Conroy}, C., {Dav{\'e}}, R., {Johnson}, B.~D., \& {Popping}, G. 2018, \apj, 869, 70, \dodoi{10.3847/1538-4357/aaed25}

\bibitem[{{Newman} {et~al.}(2013){Newman}, {Treu}, {Ellis}, {Sand}, {Nipoti}, {Richard}, \& {Jullo}}]{Newman2013a}
{Newman}, A.~B., {Treu}, T., {Ellis}, R.~S., {et~al.} 2013, \apj, 765, 24, \dodoi{10.1088/0004-637X/765/1/24}

\bibitem[{{Noll} {et~al.}(2009){Noll}, {Burgarella}, {Giovannoli}, {Buat}, {Marcillac}, \& {Mu{\~n}oz-Mateos}}]{Noll2009dustlaw}
{Noll}, S., {Burgarella}, D., {Giovannoli}, E., {et~al.} 2009, \aap, 507, 1793, \dodoi{10.1051/0004-6361/200912497}

\bibitem[{{Osterbrock} \& {Ferland}(2006)}]{OsterbrockFerland2006}
{Osterbrock}, D.~E., \& {Ferland}, G.~J. 2006, {Astrophysics of gaseous nebulae and active galactic nuclei}

\bibitem[{{Pacucci} \& {Narayan}(2024)}]{PacucciNarayan2024}
{Pacucci}, F., \& {Narayan}, R. 2024, arXiv e-prints, arXiv:2407.15915, \dodoi{10.48550/arXiv.2407.15915}

\bibitem[{{Pacucci} {et~al.}(2023){Pacucci}, {Nguyen}, {Carniani}, {Maiolino}, \& {Fan}}]{Pacucci2023}
{Pacucci}, F., {Nguyen}, B., {Carniani}, S., {Maiolino}, R., \& {Fan}, X. 2023, \apjl, 957, L3, \dodoi{10.3847/2041-8213/ad0158}

\bibitem[{{P{\'e}rez-Gonz{\'a}lez} {et~al.}(2024){P{\'e}rez-Gonz{\'a}lez}, {Barro}, {Rieke}, {Lyu}, {Rieke}, {Alberts}, {Williams}, {Hainline}, {Sun}, {Pusk{\'a}s}, {Annunziatella}, {Baker}, {Bunker}, {Egami}, {Ji}, {Johnson}, {Robertson}, {Rodr{\'\i}guez Del Pino}, {Rujopakarn}, {Shivaei}, {Tacchella}, {Willmer}, \& {Willott}}]{Perez-Gonzalez2024}
{P{\'e}rez-Gonz{\'a}lez}, P.~G., {Barro}, G., {Rieke}, G.~H., {et~al.} 2024, \apj, 968, 4, \dodoi{10.3847/1538-4357/ad38bb}

\bibitem[{{Price} {et~al.}(2024){Price}, {Bezanson}, {Labbe}, {Furtak}, {de Graaff}, {Greene}, {Kokorev}, {Setton}, {Suess}, {Brammer}, {Cutler}, {Leja}, {Pan}, {Wang}, {Weaver}, {Whitaker}, {Atek}, {Burgasser}, {Chemerynska}, {Dayal}, {Feldmann}, {F{\"o}rster Schreiber}, {Fudamoto}, {Fujimoto}, {Glazebrook}, {Goulding}, {Khullar}, {Kriek}, {Marchesini}, {Maseda}, {Miller}, {Muzzin}, {Nanayakkara}, {Nelson}, {Oesch}, {Shipley}, {Smit}, {Taylor}, {van Dokkum}, {Williams}, \& {Zitrin}}]{Price2024Spectra}
{Price}, S.~H., {Bezanson}, R., {Labbe}, I., {et~al.} 2024, arXiv e-prints, arXiv:2408.03920, \dodoi{10.48550/arXiv.2408.03920}

\bibitem[{{Reinoso} {et~al.}(2023){Reinoso}, {Klessen}, {Schleicher}, {Glover}, \& {Solar}}]{Reinoso2023}
{Reinoso}, B., {Klessen}, R.~S., {Schleicher}, D., {Glover}, S. C.~O., \& {Solar}, P. 2023, \mnras, 521, 3553, \dodoi{10.1093/mnras/stad790}

\bibitem[{{Salim} \& {Narayanan}(2020)}]{Salim&Narayanan2020}
{Salim}, S., \& {Narayanan}, D. 2020, \araa, 58, 529, \dodoi{10.1146/annurev-astro-032620-021933}

\bibitem[{{S{\'a}nchez-Bl{\'a}zquez} {et~al.}(2006){S{\'a}nchez-Bl{\'a}zquez}, {Peletier}, {Jim{\'e}nez-Vicente}, {Cardiel}, {Cenarro}, {Falc{\'o}n-Barroso}, {Gorgas}, {Selam}, \& {Vazdekis}}]{Sanchez-Blazquez2006}
{S{\'a}nchez-Bl{\'a}zquez}, P., {Peletier}, R.~F., {Jim{\'e}nez-Vicente}, J., {et~al.} 2006, \mnras, 371, 703, \dodoi{10.1111/j.1365-2966.2006.10699.x}

\bibitem[{{Schaerer} {et~al.}(2024){Schaerer}, {Guibert}, {Marques-Chaves}, \& {Martins}}]{Schaerer2024a}
{Schaerer}, D., {Guibert}, J., {Marques-Chaves}, R., \& {Martins}, F. 2024, arXiv e-prints, arXiv:2407.12122, \dodoi{10.48550/arXiv.2407.12122}

\bibitem[{{Sesar} {et~al.}(2007){Sesar}, {Ivezi{\'c}}, {Lupton}, {Juri{\'c}}, {Gunn}, {Knapp}, {DeLee}, {Smith}, {Miknaitis}, {Lin}, {Tucker}, {Doi}, {Tanaka}, {Fukugita}, {Holtzman}, {Kent}, {Yanny}, {Schlegel}, {Finkbeiner}, {Padmanabhan}, {Rockosi}, {Bond}, {Lee}, {Stoughton}, {Jester}, {Harris}, {Harding}, {Brinkmann}, {Schneider}, {York}, {Richmond}, \& {Vanden Berk}}]{Sesar2007}
{Sesar}, B., {Ivezi{\'c}}, {\v{Z}}., {Lupton}, R.~H., {et~al.} 2007, \aj, 134, 2236, \dodoi{10.1086/521819}

\bibitem[{{Setton} {et~al.}(2024){Setton}, {Greene}, {de Graaff}, {Ma}, {Leja}, {Matthee}, {Bezanson}, {Boogaard}, {Cleri}, {Katz}, {Labbe}, {Maseda}, {McConachie}, {Miller}, {Price}, {Suess}, {van Dokkum}, {Wang}, {Weibel}, {Whitaker}, \& {Williams}}]{Setton2024LRD3600}
{Setton}, D.~J., {Greene}, J.~E., {de Graaff}, A., {et~al.} 2024, arXiv e-prints, arXiv:2411.03424, \dodoi{10.48550/arXiv.2411.03424}

\bibitem[{{Shakura} \& {Sunyaev}(1973)}]{ShakuraSunyaev1973}
{Shakura}, N.~I., \& {Sunyaev}, R.~A. 1973, \aap, 24, 337

\bibitem[{{Stasi{\'n}ska} \& {Izotov}(2001)}]{Stasinska&Izotov2001}
{Stasi{\'n}ska}, G., \& {Izotov}, Y. 2001, \aap, 378, 817, \dodoi{10.1051/0004-6361:20011303}

\bibitem[{{Suess} {et~al.}(2023){Suess}, {Williams}, {Robertson}, {Ji}, {Johnson}, {Nelson}, {Alberts}, {Hainline}, {D'Eugenio}, {{\"U}bler}, {Rieke}, {Rieke}, {Bunker}, {Carniani}, {Charlot}, {Eisenstein}, {Maiolino}, {Stark}, {Tacchella}, \& {Willott}}]{Suess2023MinorMergerInQG}
{Suess}, K.~A., {Williams}, C.~C., {Robertson}, B., {et~al.} 2023, \apjl, 956, L42, \dodoi{10.3847/2041-8213/acf5e6}

\bibitem[{{Suess} {et~al.}(2024){Suess}, {Weaver}, {Price}, {Pan}, {Wang}, {Bezanson}, {Brammer}, {Cutler}, {Labbe}, {Leja}, {Williams}, {Whitaker}, {Dayal}, {de Graaff}, {Feldmann}, {Franx}, {Fudamoto}, {Fujimoto}, {Furtak}, {Goulding}, {Greene}, {Khullar}, {Kokorev}, {Kriek}, {Lorenz}, {Marchesini}, {Maseda}, {Matthee}, {Miller}, {Mitsuhashi}, {Mowla}, {Muzzin}, {Naidu}, {Nanayakkara}, {Nelson}, {Oesch}, {Setton}, {Shipley}, {Smit}, {Spilker}, {van Dokkum}, \& {Zitrin}}]{Suess2024MEGASCIENCE}
{Suess}, K.~A., {Weaver}, J.~R., {Price}, S.~H., {et~al.} 2024, arXiv e-prints, arXiv:2404.13132, \dodoi{10.48550/arXiv.2404.13132}

\bibitem[{{Sun} {et~al.}(2024){Sun}, {Ho}, {Zhuang}, {Ma}, {Chen}, \& {Li}}]{Sun2024}
{Sun}, W., {Ho}, L.~C., {Zhuang}, M.-Y., {et~al.} 2024, \apj, 960, 104, \dodoi{10.3847/1538-4357/acf1f6}

\bibitem[{{Szomoru} {et~al.}(2012){Szomoru}, {Franx}, \& {van Dokkum}}]{Szomoru2012}
{Szomoru}, D., {Franx}, M., \& {van Dokkum}, P.~G. 2012, \apj, 749, 121, \dodoi{10.1088/0004-637X/749/2/121}

\bibitem[{{Tee} {et~al.}(2024){Tee}, {Fan}, {Wang}, \& {Yang}}]{Tee2024}
{Tee}, W.~L., {Fan}, X., {Wang}, F., \& {Yang}, J. 2024, arXiv e-prints, arXiv:2412.05242, \dodoi{10.48550/arXiv.2412.05242}

\bibitem[{{Temple} {et~al.}(2021){Temple}, {Hewett}, \& {Banerji}}]{Temple2021}
{Temple}, M.~J., {Hewett}, P.~C., \& {Banerji}, M. 2021, \mnras, 508, 737, \dodoi{10.1093/mnras/stab2586}

\bibitem[{{Thomas} {et~al.}(2016){Thomas}, {Groves}, {Sutherland}, {Dopita}, {Kewley}, \& {Jin}}]{Thomas2016}
{Thomas}, A.~D., {Groves}, B.~A., {Sutherland}, R.~S., {et~al.} 2016, \apj, 833, 266, \dodoi{10.3847/1538-4357/833/2/266}

\bibitem[{{Thompson} {et~al.}(2005){Thompson}, {Quataert}, \& {Murray}}]{Thompson2005}
{Thompson}, T.~A., {Quataert}, E., \& {Murray}, N. 2005, \apj, 630, 167, \dodoi{10.1086/431923}

\bibitem[{{Trayford} {et~al.}(2020){Trayford}, {Lagos}, {Robotham}, \& {Obreschkow}}]{Trayford2020}
{Trayford}, J.~W., {Lagos}, C. d.~P., {Robotham}, A. S.~G., \& {Obreschkow}, D. 2020, \mnras, 491, 3937, \dodoi{10.1093/mnras/stz3234}

\bibitem[{{van Dokkum} \& {Conroy}(2024)}]{vanDokkum&Conroy2024}
{van Dokkum}, P., \& {Conroy}, C. 2024, \apjl, 973, L32, \dodoi{10.3847/2041-8213/ad77b8}

\bibitem[{{van Dokkum} {et~al.}(2008){van Dokkum}, {Franx}, {Kriek}, {Holden}, {Illingworth}, {Magee}, {Bouwens}, {Marchesini}, {Quadri}, {Rudnick}, {Taylor}, \& {Toft}}]{vanDokkum2008MassiveQ}
{van Dokkum}, P.~G., {Franx}, M., {Kriek}, M., {et~al.} 2008, \apjl, 677, L5, \dodoi{10.1086/587874}

\bibitem[{{van Dokkum} {et~al.}(2010){van Dokkum}, {Whitaker}, {Brammer}, {Franx}, {Kriek}, {Labb{\'e}}, {Marchesini}, {Quadri}, {Bezanson}, {Illingworth}, {Muzzin}, {Rudnick}, {Tal}, \& {Wake}}]{vanDokkum2010}
{van Dokkum}, P.~G., {Whitaker}, K.~E., {Brammer}, G., {et~al.} 2010, \apj, 709, 1018, \dodoi{10.1088/0004-637X/709/2/1018}

\bibitem[{{Vanzella} {et~al.}(2023){Vanzella}, {Claeyssens}, {Welch}, {Adamo}, {Coe}, {Diego}, {Mahler}, {Khullar}, {Kokorev}, {Oguri}, {Ravindranath}, {Furtak}, {Hsiao}, {Abdurro'uf}, {Mandelker}, {Brammer}, {Bradley}, {Brada{\v{c}}}, {Conselice}, {Dayal}, {Nonino}, {Andrade-Santos}, {Windhorst}, {Pirzkal}, {Sharon}, {de Mink}, {Fujimoto}, {Zitrin}, {Eldridge}, \& {Norman}}]{Vanzella2023}
{Vanzella}, E., {Claeyssens}, A., {Welch}, B., {et~al.} 2023, \apj, 945, 53, \dodoi{10.3847/1538-4357/acb59a}

\bibitem[{{Voggel} {et~al.}(2019){Voggel}, {Seth}, {Baumgardt}, {Mieske}, {Pfeffer}, \& {Rasskazov}}]{Voggel2019}
{Voggel}, K.~T., {Seth}, A.~C., {Baumgardt}, H., {et~al.} 2019, \apj, 871, 159, \dodoi{10.3847/1538-4357/aaf735}

\bibitem[{{Volonteri} {et~al.}(2008){Volonteri}, {Haardt}, \& {G{\"u}ltekin}}]{Volonteri2008}
{Volonteri}, M., {Haardt}, F., \& {G{\"u}ltekin}, K. 2008, \mnras, 384, 1387, \dodoi{10.1111/j.1365-2966.2008.12911.x}

\bibitem[{{Volonteri} {et~al.}(2017){Volonteri}, {Reines}, {Atek}, {Stark}, \& {Trebitsch}}]{Volonteri2017}
{Volonteri}, M., {Reines}, A.~E., {Atek}, H., {Stark}, D.~P., \& {Trebitsch}, M. 2017, \apj, 849, 155, \dodoi{10.3847/1538-4357/aa93f1}

\bibitem[{{Volonteri} {et~al.}(2024){Volonteri}, {Trebitsch}, {Dubois}, {Greene}, {Dong-Paez}, {Habouzit}, {Lupi}, {Ma}, {Beckmann}, \& {Dayal}}]{Volonteri2024}
{Volonteri}, M., {Trebitsch}, M., {Dubois}, Y., {et~al.} 2024, arXiv e-prints, arXiv:2408.12854, \dodoi{10.48550/arXiv.2408.12854}

\bibitem[{{Wang} {et~al.}(2024{\natexlab{a}}){Wang}, {de Graaff}, {Davies}, {Greene}, {Leja}, {Goulding}, {Williams}, {Brammer}, {Suess}, {Weibel}, {Bezanson}, {Boogaard}, {Cleri}, {Hirschmann}, {Katz}, {Labbe}, {Maseda}, {Matthee}, {McConachie}, {Naidu}, {Oesch}, {Rix}, {Setton}, \& {Whitaker}}]{Wang2024BRD}
{Wang}, B., {de Graaff}, A., {Davies}, R.~L., {et~al.} 2024{\natexlab{a}}, arXiv e-prints, arXiv:2403.02304.
\newblock \doarXiv{2403.02304}

\bibitem[{{Wang} {et~al.}(2024{\natexlab{b}}){Wang}, {Leja}, {de Graaff}, {Brammer}, {Weibel}, {van Dokkum}, {Baggen}, {Suess}, {Greene}, {Bezanson}, {Cleri}, {Hirschmann}, {Labb{\'e}}, {Matthee}, {McConachie}, {Naidu}, {Nelson}, {Oesch}, {Setton}, \& {Williams}}]{Wang2024RUBIESz78MassGal}
{Wang}, B., {Leja}, J., {de Graaff}, A., {et~al.} 2024{\natexlab{b}}, \apjl, 969, L13, \dodoi{10.3847/2041-8213/ad55f7}

\bibitem[{{Weaver} {et~al.}(2024){Weaver}, {Cutler}, {Pan}, {Whitaker}, {Labb{\'e}}, {Price}, {Bezanson}, {Brammer}, {Marchesini}, {Leja}, {Wang}, {Furtak}, {Zitrin}, {Atek}, {Chemerynska}, {Coe}, {Dayal}, {van Dokkum}, {Feldmann}, {F{\"o}rster Schreiber}, {Franx}, {Fujimoto}, {Fudamoto}, {Glazebrook}, {de Graaff}, {Greene}, {Juneau}, {Kassin}, {Kriek}, {Khullar}, {Maseda}, {Mowla}, {Muzzin}, {Nanayakkara}, {Nelson}, {Oesch}, {Pacifici}, {Papovich}, {Setton}, {Shapley}, {Shipley}, {Smit}, {Stefanon}, {Taylor}, {Weibel}, \& {Williams}}]{Weaver2024}
{Weaver}, J.~R., {Cutler}, S.~E., {Pan}, R., {et~al.} 2024, \apjs, 270, 7, \dodoi{10.3847/1538-4365/ad07e0}

\bibitem[{{Weibel} {et~al.}(2024){Weibel}, {de Graaff}, {Setton}, {Miller}, {Oesch}, {Brammer}, {Lagos}, {Whitaker}, {Williams}, {Baggen}, {Bezanson}, {Boogaard}, {Cleri}, {Greene}, {Hirschmann}, {Hviding}, {Kuruvanthodi}, {Labb{\'e}}, {Leja}, {Maseda}, {Matthee}, {McConachie}, {Naidu}, {Roberts-Borsani}, {Schaerer}, {Suess}, {Valentino}, {van Dokkum}, \& {Wang}}]{Weibel2024}
{Weibel}, A., {de Graaff}, A., {Setton}, D.~J., {et~al.} 2024, arXiv e-prints, arXiv:2409.03829, \dodoi{10.48550/arXiv.2409.03829}

\bibitem[{{Williams} {et~al.}(2024){Williams}, {Alberts}, {Ji}, {Hainline}, {Lyu}, {Rieke}, {Endsley}, {Suess}, {Sun}, {Johnson}, {Florian}, {Shivaei}, {Rujopakarn}, {Baker}, {Bhatawdekar}, {Boyett}, {Bunker}, {Cameron}, {Carniani}, {Charlot}, {Curtis-Lake}, {DeCoursey}, {de Graaff}, {Egami}, {Eisenstein}, {Gibson}, {Hausen}, {Helton}, {Maiolino}, {Maseda}, {Nelson}, {P{\'e}rez-Gonz{\'a}lez}, {Rieke}, {Robertson}, {Saxena}, {Tacchella}, {Willmer}, \& {Willott}}]{Williams2024}
{Williams}, C.~C., {Alberts}, S., {Ji}, Z., {et~al.} 2024, \apj, 968, 34, \dodoi{10.3847/1538-4357/ad3f17}

\bibitem[{{Wright} {et~al.}(2024){Wright}, {Whitaker}, {Weaver}, {Cutler}, {Wang}, {Carnall}, {Suess}, {Bezanson}, {Nelson}, {Miller}, {Ito}, \& {Valentino}}]{Wright2024}
{Wright}, L., {Whitaker}, K.~E., {Weaver}, J.~R., {et~al.} 2024, \apjl, 964, L10, \dodoi{10.3847/2041-8213/ad2b6d}

\bibitem[{{Yue} {et~al.}(2024){Yue}, {Eilers}, {Ananna}, {Panagiotou}, {Kara}, \& {Miyaji}}]{Yue2024}
{Yue}, M., {Eilers}, A.-C., {Ananna}, T.~T., {et~al.} 2024, arXiv e-prints, arXiv:2404.13290, \dodoi{10.48550/arXiv.2404.13290}

\bibitem[{{Zakamska} {et~al.}(2005){Zakamska}, {Schmidt}, {Smith}, {Strauss}, {Krolik}, {Hall}, {Richards}, {Schneider}, {Brinkmann}, \& {Szokoly}}]{Zakamska2005}
{Zakamska}, N.~L., {Schmidt}, G.~D., {Smith}, P.~S., {et~al.} 2005, \aj, 129, 1212, \dodoi{10.1086/427543}

\bibitem[{{Zakamska} {et~al.}(2006){Zakamska}, {Strauss}, {Krolik}, {Ridgway}, {Schmidt}, {Smith}, {Heckman}, {Schneider}, {Hao}, \& {Brinkmann}}]{Zakamska2006}
{Zakamska}, N.~L., {Strauss}, M.~A., {Krolik}, J.~H., {et~al.} 2006, \aj, 132, 1496, \dodoi{10.1086/506986}

\bibitem[{{Zhang} {et~al.}(2024){Zhang}, {Jiang}, {Liu}, \& {Ho}}]{Zhang2024}
{Zhang}, Z., {Jiang}, L., {Liu}, W., \& {Ho}, L.~C. 2024, arXiv e-prints, arXiv:2411.02729, \dodoi{10.48550/arXiv.2411.02729}

\bibitem[{{Zitrin} {et~al.}(2015){Zitrin}, {Fabris}, {Merten}, {Melchior}, {Meneghetti}, {Koekemoer}, {Coe}, {Maturi}, {Bartelmann}, {Postman}, {Umetsu}, {Seidel}, {Sendra}, {Broadhurst}, {Balestra}, {Biviano}, {Grillo}, {Mercurio}, {Nonino}, {Rosati}, {Bradley}, {Carrasco}, {Donahue}, {Ford}, {Frye}, \& {Moustakas}}]{Zitrin2015SLcode}
{Zitrin}, A., {Fabris}, A., {Merten}, J., {et~al.} 2015, \apj, 801, 44, \dodoi{10.1088/0004-637X/801/1/44}

\end{thebibliography}
\bibliographystyle{aasjournal}

\end{CJK*}
\end{document}